\documentclass[aps,showpacs,twocolumn,dvips,prb]{revtex4}
\usepackage{feynmp}
\usepackage{amssymb}
\usepackage{epsfig}
\usepackage{graphicx}
\usepackage{subfigure}
\usepackage{mathrsfs}
\usepackage{hyperref}
\usepackage{cleveref}
\usepackage{cases}
\usepackage{bbm}
\usepackage{xcolor}
\usepackage{tabularx}
\usepackage{array}
\usepackage{multirow}
\usepackage{CJK}
\usepackage{lmodern}
\usepackage{textcomp}
\usepackage{makecell}
\usepackage{threeparttable}

\begin{document}



\title{Quantum criticality of fermion velocities and critical temperature nearby a
putative quantum phase transition in the $d$-wave superconductors}

\date{\today}

\author{Xiao-Yue Ren}
\affiliation{Department of Physics, Tianjin University, Tianjin 300072, P.R. China}
\author{Ya-Hui Zhai}
\affiliation{Department of Physics, Tianjin University, Tianjin 300072, P.R. China}
\author{Jing Wang}
\altaffiliation{Corresponding author: jing$\textunderscore$wang@tju.edu.cn}
\affiliation{Department of Physics, Tianjin University, Tianjin 300072, P.R. China}

\begin{abstract}
Quantum critical behaviors induced by a putative
quantum phase transition are vigilantly investigated, which separates
a $d$-wave superconducting state and $d$-wave superconducting+$X$
state below the superconducting dome of the $d-$wave superconductors
with tuning the non-thermal doping variable. Within the framework of
renormalization group approach, we start with a phenomenological effective
theory originated from the Landau-Ginzburg-Wilson theory and practice
one-loop calculations to construct a set of coupled flows of all
interaction parameters. After extracting related physical information
from these coupled evolutions, we address that both fermion velocities
and critical temperatures exhibit critical behaviors, which are
robust enough against the initial conditions due to strong
quantum fluctuations. At first, the evolution of Yukawa coupling
between $X$-state order parameter and nodal fermions in tandem
with quantum fluctuations heavily renormalize fermion velocities and
generally drive them into certain finite anisotropic fixed point at the
lowest-energy limit, whose concrete value relies upon the very quantum
phase transition. In addition, these unique properties of fermion
velocities largely reshape the fate of superfluid density,
giving rise to either an enhancement or a dip of critical temperature.
Moreover, we find that fermion-fermion interactions bring
non-ignorable quantitative corrections to quantum critical
behaviors despite they are subordinate to
quantum fluctuations of order parameters.
\end{abstract}


\maketitle


\section{Introduction}

A plethora of both theoretical and experimental research efforts have been
devoted to the $d$-wave cuprate superconductors in the last three
decades owing to their unique pairing mechanisms and
anomalous properties in the normal states~\cite{Lee2006RMP,Vojta2000PRL,
Vojta2000PRB,Vojta2000IJMPB,Sachdev2000Science,Sachdev2003RMP,Sachdev2008PRB,Sachdev2011PT,
Wang2011PRB,Fradkin2012NPhys,Kivelson2014PNAS,Fradkin2015RMP,
Dagotto1994RMP,Dagotto2005Science,Kivelson1995Nature,Kivelson1998Nature,
Kivelson2003RMP_DFS,Sigrist1991RMP,Sigrist1995RMP,
Tinkham1996Book,Anderson1997Book,Phillips2020NPhys,
Kim-Kivelson2008PRB,She2010PRB,She2015PRB,Xu2008PRB,Larkin2005Book}.
Compared to their $s$-wave counterparts~\cite{Tinkham1996Book,Anderson1997Book,
Larkin2005Book}, it is noteworthy that such superconductors own a $d_{x^2-y^2}$
superconducting gap~\cite{Lee2006RMP,Ding1996Nature,Loeser1996Science,
Valla1999Science,Orenstein2000Science,Yoshida2003PRL,
Dagotto1994RMP}, which vanishes at four nodes $(\pm \pi/2,\pm \pi/2)$
in the first Brillouin zone~\cite{Lee2006RMP,Fradkin2015RMP,
Dagotto1994RMP}. This indicates that the gapless nodal quasiparticles (QPs) can be
excited from these nodal points and present even at the lowest
energy in the superconducting phase~\cite{Orenstein2000Science,
Sachdev2000Science,Sachdev2003RMP,Lee2006RMP,Sachdev2011PT}.
Generally, these nodal QPs are nearly non-interacting~\cite{Orenstein2000Science}. However,
this feature can be significantly changed once the nodal fermions
interact with certain critical bosonic mode accompanied by
a quantum phase transition (QPT)~\cite{Vojta2003RPP,Sachdev2011Book,Coleman2005Nature},
around which the quantum fluctuations couple strongly
to the nodal fermions, giving rise to severe fermion damping~\cite{Vojta2000PRB,Vojta2000PRL,Vojta2000IJMPB,Paaske2001PRL,Kim-Kivelson2008PRB}
and other striking properties~\cite{Sachdev2008PRB,Xu2008PRB,Sachdev2009PRB,
Liu2012PRB,Liu2013NJP,She2015PRB}. It is therefore reasonably
expected that these nodal QPs together with quantum critical degrees of freedom
would be responsible for unusual behaviors around
the QPT~\cite{Orenstein2000Science,Coleman2005Nature,Lee2006RMP,Fradkin2012NPhys,
Kivelson2014PNAS,Fradkin2015RMP,Sachdev2011Book,Vojta2003RPP,
Moon2010PRB,Moon2012PRB,Moon2016PRB,Moon2016SRep,Wang-EM2014PRB,
Yoshida2003PRL,Paaske2001PRL,Sachdev2009PRB,Liu2012PRB,Liu2013NJP}.

On the basis of diversity and complexity of realistic systems,
a series of stimulated frameworks are proposed~\cite{Lee1993PRL,
Vojta2000PRL,Vojta2000PRB,Coleman2005Nature,Dagotto2005Science,
Sachdev2000Science,Sachdev2011PT,Castellani1997ZPB,She2011PRL}
to explore and unravel the intimate connection
between quantum criticality and unusual properties associated with
nodal QPs in the $d$-wave superconductors. One of the most well-known
pioneering scenarios was put forward by Vojta \emph{et al}. in 2000~\cite{Vojta2000PRL,Vojta2000PRB,Vojta2000IJMPB}.
Within their strategy, a putative quantum critical point
(QCP)~\cite{Sachdev2011Book,Vojta2003RPP} exists somewhere in the superconducting dome
accompanied by certain QPT from a $d_{x^{2}-y^{2}}$ superconducting
state to another $d_{x^{2}-y^{2}}+X$ superconducting state as schematically presented
in Fig.~\ref{fig1} due to the topological changes of nodal positions~\cite{Vojta2000PRL,Vojta2000PRB,Vojta2000IJMPB}.
Hereby, the $X$ state is developed by the $C_4$ symmetry breaking of nodal positions
and owns seven potential candidates based upon the group-theory analysis~\cite{Vojta2000PRL,Vojta2000PRB,Vojta2000IJMPB}.
These states can be effectively reduced to four distinct types~\cite{Wang2013PRB},
which are associated with the QPTs denominated by Type-$\tau_{0,x,y,z}$
in this work. 
In particular, the Type-$\tau_{x}$ QCP dubbed the nematic QCP
has been suggested~\cite{Kivelson1998Nature} and indirectly
detected~\cite{Keimer2008Science} below the superconducting
dome of $d$-wave high-$T_{c}$ superconductor, which is expected to be
associated with several non-Fermi-liquid behaviors~\cite{Sachdev2008PRB,Wang2011PRB,
Kim-Kivelson2008PRB,Xu2008PRB,She2015PRB}. This accordingly
stimulates us to systematically investigate the critical
consequences and differences of these distinct kinds of QCPs on the physics of
related quantum critical regions owing to the combination of ferocious quantum
fluctuations of order parameters and their interplay with other degrees of freedom,
which as far as we know have not been yet sufficiently studied.

Specifically, the quantum fluctuation of $X$
order parameter nearby a QCP strongly couples to gapless nodal QPs, which
then leads to nontrivial critical effects on two fermion velocities of
nodal QPs consisting of the Fermi velocity $v_F$ and the gap velocity $v_\Delta$~\cite{Durst2000PRB,Lee1993PRL}.
Principally, their ratio $v_{\Delta}/v_{F}$ plays an important role
in pining down the low-energy fates of physical quantities in that
it always enters into a number of important observable quantities including
the superfluid density and critical temperature~\cite{Lee1997PRL}
as well as electric and thermal conductivities~\cite{Lee1993PRL,Durst2000PRB,
Mesot1999PRL,Vojta2009AP}. This signals any unusual renormalization
of this velocity ratio will give rise to certain enhancement or suppression
of these observable quantities. It is therefore of considerable
necessity to explore the low-energy tendency of $v_{\Delta}/v_{F}$.
Stimulated by this, Huh and Sachdev~\cite{Sachdev2008PRB} carefully
examined the Type-$\tau_x$ QPT, which is so-called nematic QPT with
spontaneously breaking $C_4$ symmetry down to $C_2$ symmetry of the
system~\cite{Sachdev2008PRB,Kim-Kivelson2008PRB,Xu2008PRB,Vojta2009AP,
Keimer2008Science,Kim2010Nature,Sachdev2011Book,Metzner2007PRB,
Sachdev2002PRB,Kivelson2009PRB,Sachdev2010PRB,Fradkin2010ARCMP,
Kivelson1998Nature,Metzner2000PRL,Kivelson2001PRB,Vojta2000PRB,Vojta2000PRL,
Sachdev2009PRB,Kim2010PRB,Wang2011PRB}, and obtain a fixed
point $v_{\Delta}/v_{F}\rightarrow0$ at the lowest-energy limit.
In addition, Wang \emph{et al.}~\cite{Wang2013PRB,Wang2015PLA,Wang2013NJP}
addressed two distinct fixed points corresponding to
$v_{\Delta}/v_{F}\rightarrow1$ and $v_{\Delta}/v_{F}\rightarrow\infty$ for Type-$\tau_{y}$ and
Type-$\tau_{z}$ QPTs, respectively. Further, the consequences of
these fixed points on the physical implications are subsequently
investigated in Refs.~\cite{Kim-Kivelson2008PRB,Xu2008PRB,Wang2011PRB,She2015PRB,
Wang2015PLA,Wang2013NJP,Wang2013PRB}.

Despite of these considerable progresses on
the behaviors of fermion velocities nearby the QPTs~\cite{Sachdev2008PRB,Wang2013PRB,
Kim-Kivelson2008PRB,Xu2008PRB,Wang2011PRB,Wang2015PLA,Wang2013NJP,Liu2012PRB,She2015PRB},
several quantum critical degrees of freedom are insufficiently taken into account,
which may be essential to dictate the low-energy behaviors of the system.
On one hand, the Yukawa coupling between nodal
QPs and certain order parameter is fixed as an energy-independent constant
to approximately collect the physical ingredients nearby the QPTs
in these works~\cite{Sachdev2008PRB,Wang2013PRB,Kim-Kivelson2008PRB,Xu2008PRB,
Wang2011PRB,Wang2015PLA,Wang2013NJP,Liu2012PRB,She2015PRB}.
Going beyond this fixed-coupling assumption,
much more physical information would be captured and hence
the low-energy fates of fermion velocities may be partially or
heavily modified by the coupled entanglements of all interaction
parameters due to quantum criticality.
On the other hand, although the nodal fermions are always excited~\cite{Lee2006RMP},
they own a long lifetime and can coexist with the superconducting
state~\cite{Orenstein2000Science}. This implies the fermion-fermion
interactions can be safely neglected away from the QCP. However, quantum
criticality would coax these nodal QPs to mutually
intertwine with each other and influence fermion velocities plus Yukawa coupling.
Consequently, fermion-fermion interactions may play important roles
in determining critical behaviors around certain QCP~\cite{Vafek2014PRB,Vafek2010PRB,Vafek2012PRB,
Roy2016SR,Roy2017PRB,Roy2019PRL,Wang2017PRB,Wang2017PRB-2,Wang2018JPCM,Moon2017PRB,Wang2020PRB,
Wang2020NPB,Wang2021NPB,Roy-Sau2016PRB,Mandal2018PRB, Roy2018PRX,
Roy-Saram2016PRB, Nandkishore2017PRB, Roy-Sau2017PRL,
Roy-Slager2018PRX, Roy2004.13043,Roy2021JHEP,
Roy2021PRB,Chubukov2010PRB,Chubukov2012NPhys_chiral_SC,
Khodas2016PRX,Nandkishore2013PRB,Nandkishore2016NJP_RG-shell,
Herbut2016JHEP,Herbut2018Science,Moon2016SRep-2,Yao2017PRB,
Yao2021PRB,Wang2019JPCM,Hui2020EPJB}.
Consequently, one can expect that uncovering the contributions from
these two quantum critical ingredients may well improve our understandings
on the quantum criticality of certain QCP in the $d$-wave superconductor.

In order to encapsulate more physical
information driven by the QCP, it is therefore
imperative to systematically investigate the
effects of fermion-order parameter couplings and fermion-fermion
interactions as well as their interplay on the
low-energy fates of fermion velocities and related
observable quantities. To this purpose, we within this work employ the
momentum-shell renormalization group (RG) approach~\cite{Shankar1994RMP,Wilson1975RMP,Polchinski1992}
to unbiasedly treat all these critical physical degrees of
freedom nearby a putative QPT from the $d$-wave superconducting to
$d$-wave superconducting+$X$ state as illustrated in Fig.~\ref{fig1}.
After collecting all one-loop corrections, a set of coupled RG flows of
all interaction parameters are derived to characterize the quantum criticality
nearby all four types of potential QPTs dubbed Type-$\tau_{0,x,y,z}$
that are explicitly clarified in Sec.~\ref{sub_phen-model}.

Decoding the physical information contained in
the coupled RG equations yields a number of quantum critical
properties in the vicinity of all QCPs.
At first, we find that the fermion velocities
exhibit several interesting fixed points. With respect to
the Type-$\tau_{0}$ QPT, the Yukawa interplay designated as $\lambda$
between nodal QPs and related order parameter is marginal to one-loop level and
the ratio of fermion velocities flows towards either fixed point $(v_{\Delta}/v_{F})^*\approx0.3478$ or $(v_{\Delta}/v_{F})^*\approx0.0942$
at the low-energy limit caused by the quantum criticality. Concerning
Type-$\tau_{x,y,z}$ QPTs, the evolution of Yukawa coupling $\lambda$ and
quantum fluctuations heavily reshape three fixed points $v_{\Delta}/v_{F}
\rightarrow0,1,\infty$ for Type-$\tau_{x,y,z}$ under the fixed-coupling assumptions~\cite{Sachdev2008PRB,Wang2013PRB} to evolve towards finite anisotropies
as approaching the QPTs. To be specific, the extreme anisotropies
of fermion velocities are changed to
finite anisotropies for both Type-$\tau_{x,z}$ QPTs but instead the
isotropic fermion velocities for Type-$\tau_{y}$ QPT are broken and
attracted by a finite anisotropic fixed point.
In addition, we notice that the unusual behaviors of fermion
velocities considerably modify the fates of superfluid density
and critical temperature around the underlying four types of QPTs.
As approaching the Type-$\tau_{0,x}$ QPTs, the critical temperatures are largely
suppressed. Conversely, both Type-$\tau_{y}$ and Type-$\tau_{z}$
QPTs are in favor of the superconductivity. Furthermore, the roles
of fermion-fermion interactions that have not yet been adequately
considered are also inspected in quantum criticality. We realize that they can
give rise to quantitative contributions to quantum critical
behaviors in the vicinity of all putative QPTs.
Last but not the least important, it is worth
pointing out that our qualitative results are considerably
robust enough with the variation of initial conditions.

The rest of paper is organized as follows. In Sec.~\ref{Sec_eff-theory},
we establish our low-energy effective field theory that
includes the most of key physical ingredients to describe
the main physics around the QPT.
On the basis of the effective theory, we within Sec.~\ref{Sec_RG}
perform one-loop momentum-shell RG analysis to deliver the coupled
RG equations of all interaction parameters. After combining
both the tentatively analytical discussions and vigilant numerical calculations,
Sec.~\ref{Sec_velocity} and Sec.~\ref{Sec_rho-s} are followed to
present the critical behaviors of fermion velocities and superfluid
density together with critical temperature nearby the QCP, respectively.
Finally, we provide a brief summary in Sec.~\ref{Sec_summary}.

\section{Effective theory}\label{Sec_eff-theory}

\begin{figure}
\hspace{-0.68cm}
\includegraphics[width=3.5in]{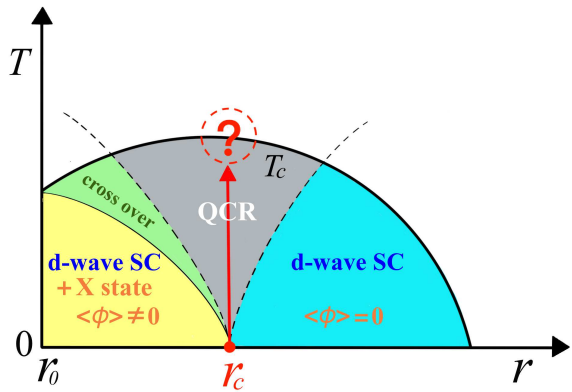}
\vspace{-0.9cm}
\caption{(Color online) Schematic illustration for some potential
quantum phase transition (QPT) from a $d$-wave SC to $d$-wave SC+$X$ state
beneath the superconducting dome of the $d-$wave superconductor with tuning
the non-thermal doping variable~\cite{Vojta2000PRL}. Hereby, $T_c$ denotes
the critical temperature of $d$-wave superconductor and the field $\phi$
characterizes the order parameter of $X$ state, which depends
upon the specific symmetry breaking accompanied by the QPT.
In addition, the very value $r_{c}$ that roughly locates at
the optimal doping is the so-called quantum critical point (QCP),
which separates the disordered ($\langle\phi\rangle=0$)
and ordered ($\langle\phi\rangle\neq0$) $X$ phases at $T=0$.
As to the finite-temperature region around the QPT,
critical behaviors are expected to be induced in the
quantum critical region (QCR) due to the strong quantum
fluctuations. The fate of critical temperature circled by the dashed line
will be explicitly addressed in Sec.~\ref{Sec_rho-s}.}\label{fig1}
\end{figure}

In this work, our focus is put on a putative QPT in the $d$-wave
superconductor as schematically displayed in Fig.~\ref{fig1} as well as
the associated critical behaviors of fermion velocities and physical quantities.
To begin with, we within this section are going to construct
the low-energy effective field theory around the QPT and
defer the one-loop RG analysis to the next section~\ref{Sec_RG}.

\subsection{Phenomenological model}\label{sub_phen-model}

As the $d$-wave superconductor is pushed closer to the QCP depicted in
Fig.~\ref{fig1}, the possible quantum criticality in the quantum critical
region (QCR) can be principally ascribed to three major distinct
types of physical ingredients that are gapless fermionic
quasiparticles (QP) excited from the nodal points and
quantum fluctuation for order parameter $\phi$ of $X$ state
in tandem with their intimate interplay~\cite{Vojta2000PRB,Vojta2000PRL,
Vojta2000IJMPB,Sachdev2008PRB,Wang2013PRB}.
It is of importance to address that the quantum fluctuations are so
ferocious that the QCR presented in Fig.~\ref{fig1}
inherits the strong fluctuations of QCP and hence the quantum
fluctuations dominate over the thermal fluctuations
within such region~\cite{Vojta2003RPP,Sachdev2011Book}.
This indicates that the quantum fluctuations are in charge of
the singular physical behaviors and henceforth the thermal
fluctuations can be ignored.

\begin{figure}
\centering
\includegraphics[width=1.3in]{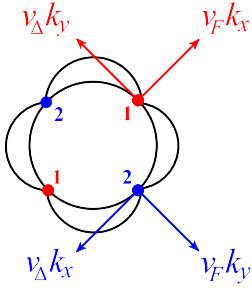}
\vspace{-0.1cm}
\caption{(Color online) Illustrations for the momenta
($k_{x,y}$) and fermion velocities ($v_{F,\Delta}$) of two
pairs of nodal QPs excited from four nodal points of
the $d-$wave superconductor.}\label{fig2}
\end{figure}

Without loss of generality, we within this
work put our focus on the QCR. The phenomenological model is therefore introduced to
capture the physical information nearby the QCP~\cite{Sachdev2008PRB,Wang2011PRB,Wang2013PRB},
\begin{eqnarray}
S=S_{\Psi}+S_{\phi_0}+S_{\Psi\phi_0},\label{Eq_S-phenomen}
\end{eqnarray}
where $S_{\Psi}$, $S_{\phi_0}$, and $S_{\Psi\phi_0}$ serve as the
degrees of fermionic QPs, order parameter, and their couplings, respectively.
To be concrete, the gapless fermions with linear dispersion are allowed to
be freely excited from four nodes on the Fermi surface and this fermionic part
can be expressed as follows~\cite{Sachdev2008PRB,Wang2011PRB,Wang2013PRB},
\begin{eqnarray}
S_{\Psi}\!\!&=&\!\!\int\!\!\frac{d^{2}\mathbf{k}}{(2\pi)^{2}}\frac{d\omega}{2\pi}\Psi^{\dag}
_{1a}(-i\omega\!+\!v_{F}k_{x}\tau^{z}\!+\!v_{\Delta}k_{y}\tau^{x})\Psi_{1a}\nonumber\\
&&\!\!\!\!\!\!+\!\!\int\!\!\frac{d^{2}\mathbf{k}}{(2\pi)^{2}}\frac{d\omega}{2\pi}\Psi^{\dag}
_{2a}(-i\omega\!+\!v_{F}k_{y}\tau^{z}\!\!+\!v_{\Delta}k_{x}\tau^{x})\Psi_{2a},\label{Eq_Psi}
\end{eqnarray}
with $\tau^{x,y,z}$ denoting the Pauli matrices~\cite{Sachdev2008PRB,Wang2011PRB,Wang2013PRB}. Hereby, the spinors $\Psi^{\dagger}_{1a}$ and $\Psi^{\dagger}_{2a}$ with the
repeated spin index $a$ being summed from $1$ to the number of fermion flavor
$N$ are employed to specify the fermionic QPs stemming from nodal points at
$(\frac{\pi}{2},\frac{\pi}{2})$ plus $(-\frac{\pi}{2},-\frac{\pi}{2})$ and
$(-\frac{\pi}{2},\frac{\pi}{2})$ plus $(\frac{\pi}{2},-\frac{\pi}{2})$,
respectively~\cite{Vojta2000PRB,Vojta2000PRL,Vojta2000IJMPB,Sachdev2008PRB}.
Besides, as displayed in Fig.~\ref{fig2},
the $k_{x,y}$ describe the momenta with relative to
the nodal points and $v_{F,\Delta}$ correspondingly serve as the Fermi velocity
and gap velocity.

\begin{figure}
\centering
\includegraphics[width=3.0in]{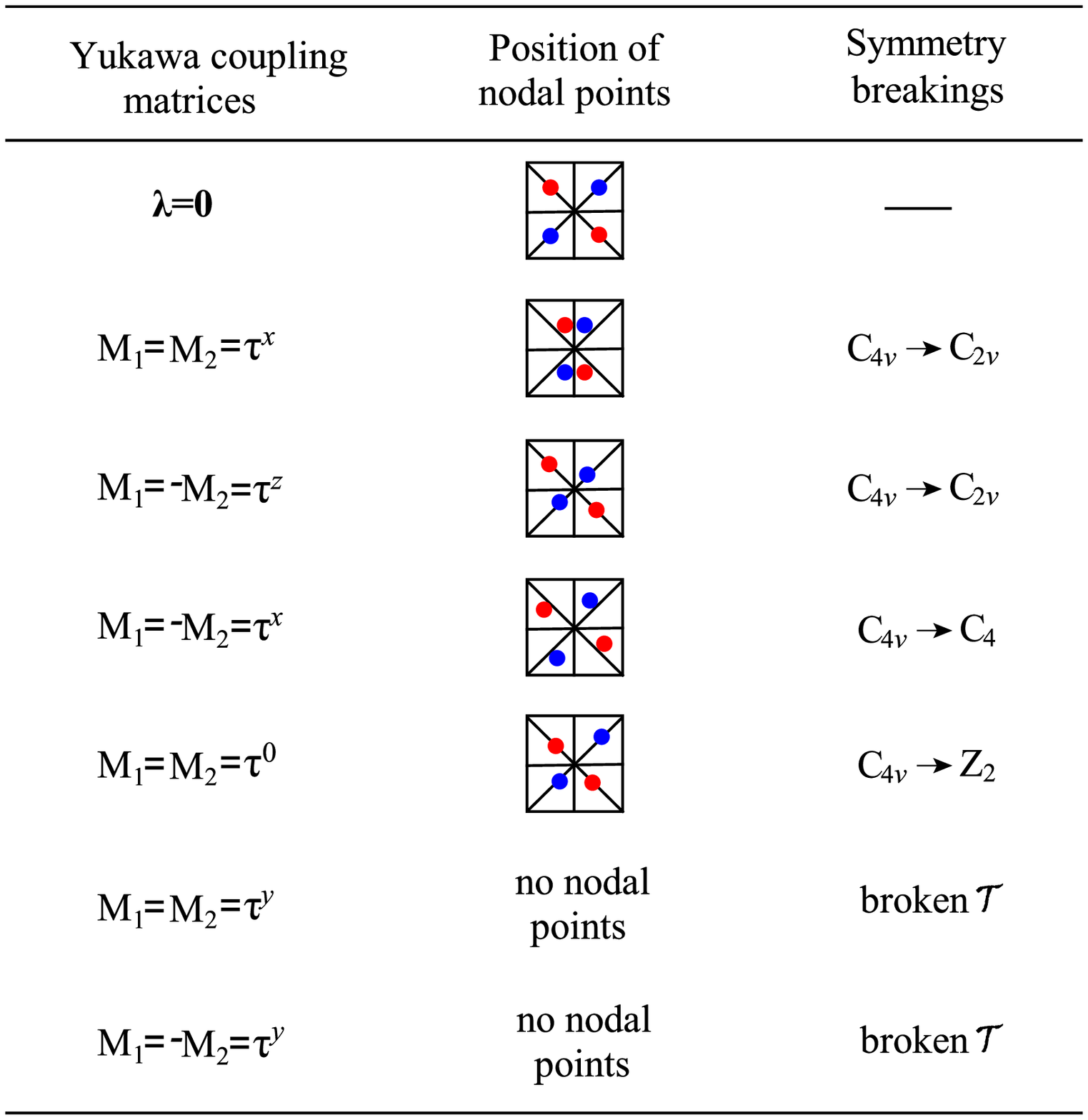}
\vspace{0.1cm}
\caption{(Color online) Seven distinct types of
symmetry breakings associated with positions of nodal points and
the accompanied fermion-order parameter Yukawa couplings with $\mathcal{T}$ representing
the time-reversal symmetry~\cite{Vojta2000IJMPB,Vojta2000PRL}.}\label{Fig_R1}
\end{figure}

With respect to the order-parameter part, there in all exist
seven different sorts of order parameters for the state $X$ in Fig.~\ref{fig1}~\cite{Vojta2000PRB,Vojta2000PRL,Vojta2000IJMPB},
which are solely determined by the specific symmetry breaking of nodal
positions as collected in Fig.~\ref{Fig_R1}.
It is of particular significance to point out that the couplings
between gapless QPs and order parameters are heavily dependent on
the symmetry breaking~\cite{Vojta2000PRL}. As a result, it is
convenient to bring about the Yukawa couplings before presenting the
$S_{\phi_0}$, which are written as~\cite{Vojta2000PRB,Vojta2000PRL,Vojta2000IJMPB,
Sachdev2008PRB,Wang2011PRB,Wang2013PRB}
\begin{eqnarray}
S_{\Psi\phi_0}=\int d^{2}\mathbf{x}d\tau[\lambda_{0}
\phi_0(\Psi^{\dag}_{1}\mathcal{M}_{1}\Psi_{1}+
\Psi^{\dag}_{2}\mathcal{M}_{2}\Psi_{2})],\label{S_phi_S_psi}
\end{eqnarray}
where the matrices $\mathcal{M}_{1,2}$ are directly associated with the
distinct types of order parameters with $\lambda_0$ designating the
coupling strength, which are explicitly classified as
follows~\cite{Vojta2000PRB,Vojta2000PRL,Vojta2000IJMPB}:
Case-I with $\mathcal{M}_{1}=\tau_{y},\mathcal{M}_{2}=\tau_{y}$,
Case-II with $\mathcal{M}_{1}=\tau_{y},\mathcal{M}_{2}=-\tau_{y}$,
Case-III with $\lambda_{0}=0$ (such situation is trivial and not discussed further),
Case-IV with $\mathcal{M}_{1}=\tau_{x},\mathcal{M}_{2}=\tau_{x}$,
Case-V with $\mathcal{M}_{1}=\tau_{z},\mathcal{M}_{2}=-\tau_{z}$,
and Case-VI with $\mathcal{M}_{1}=\tau_{x},\mathcal{M}_{2}=-\tau_{x}$.
In distinction to such six sorts, two real components $\phi_{0A}$
and $\phi_{0B}$ constitute the order parameter of Case-VII, which respectively
interact with $\Psi_{1}$ and $\Psi_{2}$, yielding to~\cite{Vojta2000IJMPB}
\begin{eqnarray}
S_{\Psi\phi_0}=\int d^{2}\mathbf{x}d\tau[\lambda_{0}(\phi_{0A}\Psi^{\dag}_{1}\mathcal{M}_1\Psi_{1}+
\phi_{0B}\Psi^{\dag}_{2}\mathcal{M}_2\Psi_{2})],\label{S_phi_0_S_psi}
\end{eqnarray}
with $\mathcal{M}_1=\mathcal{M}_2=\tau_{0}$.

To proceed, one can figure out that the matrices $\mathcal{M}_{1}$
and $\mathcal{M}_{2}$ always appear in pairs during the calculations
of one-loop corrections and henceforth the results are insensitive
to their signs~\cite{Vojta2000PRB,Vojta2000PRL,Vojta2000IJMPB,Wang2013PRB}.
As a corollary, these seven types of Yukawa couplings can be reduced to
another four simplified categories of phase transitions, which are accordingly
designated as Type-$\tau_x$ with $\mathcal{M}_{1,2}=\tau_{x}$,
Type-$\tau_y$ with $\mathcal{M}_{1,2}=\tau_{y}$, Type-$\tau_z$ with
$\mathcal{M}_{1,2}=\tau_{z}$, and Type-$\tau_0$
with $\mathcal{M}_{1,2}=\tau_{0}$, respectively.
In order to be consistent with such version of classification,
the corresponding order-parameter part $S_{\phi_0}$ can be cast as~\cite{Vojta2000PRB,Vojta2000PRL,Vojta2000IJMPB,Wang2013PRB,Sachdev2008PRB,Wang2013NJP}
\begin{eqnarray}
S_{\phi_0}=\frac{1}{2}\int\frac{d^{3}q}
{(2\pi)^{3}}[-2(r-r_c)+q^{2}]\phi^{2}_0,\label{Eq_S-phi}
\end{eqnarray}
for Type-$\tau_x$, Type-$\tau_y$, and Type-$\tau_z$.
In comparison, one needs to replace the $\phi^{2}_0$ in Eq.~(\ref{Eq_S-phi})
with $(\phi_{0A}^{2}+\phi_{0B}^{2})$ to obtain their Type-$\tau_0$ counterpart.
For convenience, we hereby neglect the order-parameter self-interaction terms ($\phi^4_0$)
in that they are irrelevant with the support of the power countering~\cite{Sachdev2008PRB}.
It is worth highlighting that the term $(r-r_c)$ is equivalent to the mass parameter,
which is tuned to be zero as the QCP is accessed at $r=r_c$ displayed in Fig.~\ref{fig1}.

\subsection{Renormalized order-parameter action and effective theory}

\begin{figure}
\centering
\includegraphics[width=3.2in]{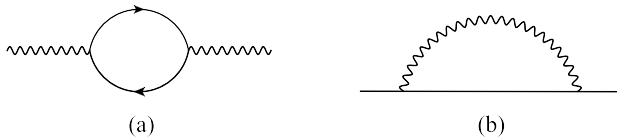}\hspace{0.9cm}
\vspace{-0.5cm}
\caption{One-loop corrections due to the Yukawa coupling between nodal QPs and
order parameter to: (a) the order-parameter propagator (polarization function)
and (b) the fermionic propagator (self-energy), where the solid and wavy lines
represent the fermionic and order-parameter propagators, respectively.}\label{fig3}
\end{figure}

Before proceeding further, it is of particular importance to
point out that the free order-parameter action~(\ref{Eq_S-phi}) would be
qualitatively renormalized by one-loop corrections due to
switching on the Yukawa couplings between nodal QPs and order parameter.
In order to evaluate such effects, we are forced to compute the
polarization function of order parameter depicted in Fig.~\ref{fig3}(a),
which can be formally expressed
as~\cite{Vojta2000PRB,Vojta2000PRL,Vojta2000IJMPB,Wang2013PRB,Sachdev2008PRB,Wang2013NJP}
\begin{eqnarray}
\Pi(\mathbf{q},\epsilon)\!=\!\!\int\!\frac{d^{2}\mathbf{k}}{(2\pi)^{2}}
\frac{d\omega}{2\pi}\mathrm{Tr}[\mathcal{M}G^{0}_{\Psi}(\mathbf{k},\omega)
\mathcal{M}G^{0}_{\Psi}(\mathbf{k\!+\!q},\omega\!+\!\epsilon)],
\end{eqnarray}
with the vertex matrix $\mathcal{M}$ being designated in Eq.~(\ref{S_phi_0_S_psi}).

Hereby, the free fermionic propagator can be forwardly derived from Eq.~(\ref{Eq_Psi})~\cite{Vojta2000PRL,Sachdev2008PRB}. Specifically, it reads
\begin{eqnarray}
G^{0}_{\Psi}(\mathbf{k},\omega)=
\frac{1}{-i\omega+v_{F}k_{x}\tau^{z}+v_{\Delta}k_{y}\tau^{x}}, \label{Eq_6}
\end{eqnarray}
for nodal QPs $\Psi_{1}$ and its $\Psi_{2}$ counterpart would be analogously
obtained via exchanging the positions of momenta $k_{x}$ and $k_{y}$ in
Eq.~(\ref{Eq_6}).

After performing long but straightforward
calculations~\cite{Sachdev2008PRB,Wang2011PRB,Wang2013PRB},
we are left with one-loop polarization functions for four different types
of phase transitions as follows, namely
\begin{eqnarray}
\!\!\!\!\!\!\Pi^{\tau_x}(\mathbf{q},\epsilon)
\!\!&=&\!\! \frac{1}{16v_{F}v_{\Delta}}
\frac{\epsilon^{2}+v_{F}^{2}q_{x}^{2}}{\sqrt{\epsilon^{2}
+v_{F}^{2}q_{x}^{2}+v_{\Delta}^{2}q_{y}^{2}}}\!+\!(q_{x}\!\rightarrow \!q_{y}),\label{Eq_Pi-x}\\
\!\!\!\!\!\!\Pi^{\tau_y}(\mathbf{q},\epsilon)
\!\!&=&\!\!\frac{1}{16v_{F}v_{\Delta}}
\sqrt{\epsilon^{2}+v_{F}^{2}q_{x}^{2}+v_{\Delta}^{2}q_{y}^{2}}\!+\!(q_{x}\!\rightarrow \!q_{y}),\\
\!\!\!\!\!\!\Pi^{\tau_z}(\mathbf{q},\epsilon)
\!\!&=&\!\!\frac{1}{16v_{F}v_{\Delta}}\!
\frac{\epsilon^{2}+v_{\Delta}^{2}q_{y}^{2}}{\sqrt{\epsilon^{2}
+v_{F}^{2}q_{x}^{2}+v_{\Delta}^{2}q_{y}^{2}}}\!+\!(q_{x}\!\rightarrow \!q_{y}),
\end{eqnarray}
for Type-$\tau_{x}$, Type-$\tau_{y}$, and Type-$\tau_{z}$, in tandem with
Type-$\tau_{0}$ that are listed by
\begin{eqnarray}
\!\!\!\!\!\!\Pi^{\tau_0}_{A}(q_{x},q_{y},\epsilon)
\!\!&=&\!\!-\frac{1}{16v_{F}v_{\Delta}}
\sqrt{\epsilon^{2}+v_{F}^{2}q_{x}^{2}+v_{\Delta}^{2}q_{y}^{2}},\\
\!\!\!\!\!\!\Pi^{\tau_0}_{B}(q_{y},q_{x},\epsilon)
\!\!&=&\!\!-\frac{1}{16v_{F}v_{\Delta}}
\sqrt{\epsilon^{2}+v_{F}^{2}q_{y}^{2}+v_{\Delta}^{2}q_{x}^{2}}.\label{Eq_Pi-0}
\end{eqnarray}

Inserting these polarization functions~(\ref{Eq_Pi-x})-(\ref{Eq_Pi-0})
into the free order-parameter action~(\ref{Eq_S-phi}) by virtue of Dyson equation
reformulates the quadratic part of $S_{\phi0}$ into~\cite{Sachdev2008PRB,Wang2013NJP}
\begin{eqnarray}
[-2(r-r_c)+q^{2})\phi^{2}_0\!\rightarrow\![-2(r-r_c)+q^{2}\!
+\!\Pi^{\mathcal{M}}(q)]\phi^{2}_0.
\end{eqnarray}
At the low-energy regime, one can realize that the
term $\Pi^{\mathcal{M}}(q)$ is proportional to $q$ and such additional linear-$q$ term
dominates over the $q^{2}$ term which henceforth can be neglected~\cite{Sachdev2008PRB,Wang2013NJP}.
This manifestly indicates that the incorporation of $q$ term
qualitatively alters the dynamical nature of the order-parameter action.
Additionally, it is the polarization term that substantially
modifies the action via involving two significant quantities
including the nodal QPs' Fermi velocity $v_F$ and the gap
velocity $v_\Delta$. In other words, it is now suitable to
designate a renormalized order-parameter
field $\phi$ to replace the bare one. As a result, we are left with the
following renormalized order-parameter action
\begin{small}
\begin{numcases}{\!\!\!\!S_{\phi}\!\!=\!\!\!}
\!\!\frac{1}{2}\int\frac{d^{3}q}{(2\pi)^{3}}
[-2(r-r_{c})\!+\!\Pi^{\mathcal{M}}\!(q)]\phi^{2}\!\!,\!\!\!\! \!\!\!\!\!\!\!\!\!\!\!\!\!\!\!
\!\!\!\!\!\!\!\!\!\!\!\!\!&
$\mathcal{M}\!\!=\!\tau_j$,\label{Eq_phi-21}\\
~\nonumber\\
\!\!\frac{1}{2}\!\!\!\sum_{\sigma=A,B}\!\int\!\frac{d^{3}q}{(2\pi)^{3}}
[-2(r-r_{c})\!+\!\Pi^{\mathcal{M}}_{\sigma}\!(q)]\phi^{2}_{\sigma},
\ \!\!\!\!\!\!\!&$\!\mathcal{M}\!\!=\!\tau_0$.\label{Eq_phi-22}
\end{numcases}
\end{small}
\hspace{-0.15cm} And the order-parameter propagator for Type-$\mathcal{M}$
is then given by~\cite{Vojta2000PRL,Sachdev2008PRB,Wang2011PRB,Wang2013PRB}
\begin{eqnarray}
G^\mathcal{M}_{\phi}(\mathbf{q},\epsilon)
=\frac{1}{\Pi^\mathcal{M}(\mathbf{q},\epsilon)},\label{Eq_phi-propagator}
\end{eqnarray}
as approaching the QCP shown in Fig.~\ref{fig1},
where $\mathcal{M}=\tau_{0,x,y,z}$ correspond to
the four potential distinct types of phase transitions delivered
in Sec.~\ref{sub_phen-model}. In addition, the Yukawa
coupling~(\ref{S_phi_S_psi})-(\ref{S_phi_0_S_psi})
between the nodal QPs and order parameter would be accordingly reshaped as
\begin{small}
\begin{numcases}{\!\!\!\! \!\!\!\!\!\!\!\!\!S_{\Psi\phi\!\!}=\!\!\!}
\!\!\!\!\frac{1}{2}\!\int\!\!\! d^{2}\mathbf{x}d\tau[\lambda
\phi(\Psi^{\dag}_{1}\mathcal{M}\Psi_{1}\!\!+\!
\Psi^{\dag}_{2}\mathcal{M}\Psi_{2})],\!\!\!\!\! \!\!\!\!&
$\mathcal{M}\!=\!\tau_{j}$,\label{Eq_Psi-phi-21}\\
~\nonumber\\
\!\!\!\!\frac{1}{2}\!\int\!\!\! d^{2}\mathbf{x}d\tau[\lambda(\phi_{A}\mathcal{M}\Psi^{\dag}_{1}\Psi_{1}\!\!+\!
\phi_{B}\mathcal{M}\Psi^{\dag}_{2}\Psi_{2})],\!\!\!\!\! \!\!\!\!&
$\mathcal{M}\!=\!\tau_{0}$, \label{Eq_Psi-phi-22}
\end{numcases}
\end{small}
\hspace{-0.25cm} with the coupling strength $\lambda_0$ being also
adjusted to $\lambda$ for consistence.


In order to capture more critical information influenced by the QCP
shown in Fig.~\ref{fig1}, we hereafter reformulate the nodal QPs' part $S_{\Psi}$~(\ref{Eq_Psi}) via supplementing the interaction between
nodal QPs themselves dubbed by $S_{\mathrm{ff}}$ in conjunction with
the renormalized order-parameter
action~(\ref{Eq_phi-21})-(\ref{Eq_phi-22}) and Yukawa
couplings~(\ref{Eq_Psi-phi-21})-(\ref{Eq_Psi-phi-22})
to establish our effective action as follows,
\begin{eqnarray}
S_{\mathrm{eff}}=S_{\Psi}+S_{\phi}+S_{\Psi\phi}+S_{\mathrm{ff}},\label{Eq_S-eff}
\end{eqnarray}
where the fermion-fermion interactions
$S_{\mathrm{ff}}$ can be expressed as~\cite{Vafek2014PRB,Wang2017PRB}
\begin{eqnarray}
S_{\mathrm{ff}}&=&\sum_{i=0}^{3}u_{i}\int d^{2}\mathbf{x}[\Psi^{\dag}(\mathbf{x})\tau_{i}\Psi(\mathbf{x})]^{2},
\end{eqnarray}
with the indexes $i=0,1,2,3$ corresponding to four types of fermion-fermion
interactions and the parameter $u_{i}$ measuring their coupling strengths.

Compared to other physical ingredients involved in the phenomenological model,
we here would like to address more comments on the fermion-fermion couplings,
which as far as we know have not yet been seriously investigated.
In the absence of a potential QCP or with the focus on the regions far away from the QCP,
it is in principle sensible to drop the fermion-fermion interactions in
the superconducting dome of $d-$wave superconductors as the nodal QPs are known
to coexist harmoniously with the SC state~\cite{Orenstein2000Science}.
In a sharp contrast, as it concerns the question on the critical behaviors neighboring the
QCP shown in Fig.~\ref{fig1}, we therefore ought to take discreetly into account the
contributions from the interplay between the nodal QPs.
On one hand, these nodal QPs themselves may mutually intertwine with each other
owing to the strong fluctuations and become one of the major elements at the lowest-energy
limit~\cite{Vafek2014PRB,Vafek2010PRB,Vafek2012PRB,
Roy2016SR,Roy2017PRB,Roy2019PRL,Wang2017PRB,Wang2018JPCM,Moon2017PRB,Wang2020PRB,
Wang2020NPB,Wang2021NPB,Roy-Sau2016PRB,Mandal2018PRB, Roy2018PRX,
Roy-Saram2016PRB, Nandkishore2017PRB, Roy-Sau2017PRL,
Roy-Slager2018PRX, Roy2004.13043,Roy2021JHEP,
Roy2021PRB,Chubukov2010PRB,Chubukov2012NPhys_chiral_SC,
Khodas2016PRX,Nandkishore2013PRB,Nandkishore2016NJP_RG-shell,
Herbut2016JHEP,Herbut2018Science,Moon2016SRep-2,Yao2017PRB,
Yao2021PRB,Wang2019JPCM,Hui2020EPJB}.
On the other hand, fermionic couplings can also impact other interaction parameters
including fermion velocities $v_{F,\Delta}$ and Yukawa coupling via participating in the
coupled RG evolutions, which will be established in Sec.~\ref{Sec_RG} based upon
the strong quantum fluctuations connecting various types of degrees of freedom.
In this sense, they can indirectly influence and may play important roles in
determining the critical behaviors induced by the QCP.


Before going further, it is of necessity to highlight that
the nodal QPs are always assumed to be well-defined in the
QCR within $T< T_{c}$ as aforementioned in Sec.~\ref{sub_phen-model}
and also advocated in many previous efforts~\cite{Lee2006RMP,Vojta2000PRL,
Vojta2000PRB,Vojta2000IJMPB,Kim-Kivelson2008PRB,
Xu2008PRB,Larkin2005Book,Sachdev2008PRB,She2010PRB,She2015PRB,Fradkin2012NPhys}.
This implies that our effective theory can capture the core physics of
quantum criticality nearby the QCP although the nodal QPs may survive not very long~\cite{Ong1995PRL,Valla1999Science,Vojta2000IJMPB}.
Afterwards, we adopt the effective action~(\ref{Eq_S-eff}) as our starting point to
derive the coupled flow equations of all associated parameters in the
frame of one-loop RG approach~\cite{Shankar1994RMP,Wilson1975RMP,Polchinski1992}
and explore the physical behaviors of fermion velocities as well as their effects
on superfluid densities and critical temperatures nearby all four types
of QCPs illustrated in Fig.~\ref{fig1}.

\begin{figure}
\centering
\includegraphics[width=3in]{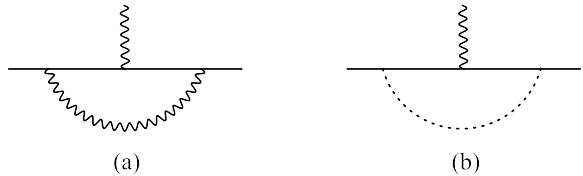}\hspace{0.9cm}
\vspace{-0.1cm}
\caption{One-loop corrections to the Yukawa coupling between
nodal QPs and  order parameter owing to (a) fermion-order parameter
interaction and (b) fermion-fermion interaction, where the
solid, dashed and wavy lines represent the fermion,
fermion-fermion interaction and order parameter, respectively.}\label{fig-vertex}
\end{figure}

\section{RG analysis}\label{Sec_RG}

To proceed, we within this section endeavor to perform the one-loop RG
analysis~\cite{Shankar1994RMP,Wilson1975RMP,Polchinski1992} for
our effective action~(\ref{Eq_S-eff}) to obtain the coupled RG
equations of all interaction parameters, from which the singular
properties induced by the QCP are expected to be extracted.
To this end, we from now on put our focus on the QCP, namely
assuming $r\rightarrow r_c$ in Eq.~(\ref{Eq_S-eff}), and then
compute all one-loop Feynman diagrams to carry out
the standard momentum-shell RG procedures
from the field theory perspective.

\subsection{One-loop corrections}\label{Sec_correction}

We commence with the one-loop corrections to fermionic propagator.
As depicted in Fig.~\ref{fig3}(b), the free fermionic propagator would
receive one-loop correction $\Sigma^{\mathcal{M}}$, which originates from
the Yukawa coupling between the nodal QPs and Type-$\mathcal{M}$
order parameter with $\mathcal{M}=\tau_{0,x,y,z}$.
After paralleling the strategy put forward in
Refs.~\cite{Shankar1994RMP,Wilson1975RMP,Polchinski1992},
we integrate out the momentum shell within $b\Lambda-\Lambda$, where
$\Lambda$ is associated with the lattice constant to
characterize the cutoff of energy scale and the variable parameter is
designated as $b=e^{-l}$ with $l>0$ being a running energy scale~
\cite{Sachdev2008PRB,Vafek2014PRB,Wang2017PRB,Wang2011PRB,Wang2013PRB,Vafek2012PRB,
She2010PRB, Kim-Kivelson2008PRB, She2015PRB, Roy-Sau2016PRB}, and eventually obtain
\begin{eqnarray}
\!\!\!\!\!\!\!\Sigma^{\tau_x}(\mathbf{k},\omega)\!\!\!
&=&\!\!\!\lambda^{2}\left[\mathcal{A}_{1}(-i\omega)
\!+\!\mathcal{A}_{2}v_{F}k_{x}\tau^{z}\!
+\!\mathcal{A}_{3}v_{\Delta}k_{y}\tau^{x}\right]l,\label{Eq_Sigma-tau-x}
\end{eqnarray}
for Type-$\tau_{x}$ order parameter. As to the other three
types with $\mathcal{M}=\tau_{y},\tau_{z},\tau_{0}$,
the structures of their results are analogous to Eq.~(\ref{Eq_Sigma-tau-x})
but the coefficients $A_{1,2,3}$ are substituted respectively by
$\mathcal{B}_{1,2,3}$, $\mathcal{C}_{1,2,3}$, and $\mathcal{D}_{1,2,3}$,
whose expressions are presented in Eqs.~(\ref{Eq_Sigma-y})-(\ref{Eq_Sigma-0-B})
of Appendix~\ref{Appendix_Sigma-Gamma} and
Appendix~\ref{Appendix_Coefficients}. Accordingly, this
gives rise to the renormalized fermionic propagator with the help
of the Dyson equation~\cite{Sachdev2008PRB,Wang2011PRB,Wang2013PRB},
\begin{eqnarray}
G_{\Psi}^{-1}(\mathbf{k},\omega)=
-i\omega+v_{F}k_{x}\tau^{z}+v_{\Delta}k_{y}\tau^{x}
-\Sigma^{\mathcal{M}}(\mathbf{k},\omega),\label{Eq_G_full}
\end{eqnarray}
where $\Sigma^{\mathcal{M}}(\mathbf{k},\omega)$ with $\mathcal{M}=\tau_{0,x,y,z}$ specifies
the self-energy owing to the Type-$\mathcal{M}$ QPT, which will be
one of the critical factors to derive the RG equations.

Next, we take into account the one-loop corrections to the Yukawa coupling and
fermion-fermion interactions. At first, we consider the former, which is marginal
at the tree level. It is therefore of particular importance to examine its fate
after including the one-loop corrections. To this end, we read
off Fig.~\ref{fig-vertex} and realize there exist two sorts of contributions, namely
\begin{eqnarray}
\Xi^{\mathcal{M}}=\Xi^{\mathcal{M}}_{\mathrm{Y}}
+\Xi^{\mathcal{M}}_{\mathrm{ff}},\label{Eq_Xi}
\end{eqnarray}
where $\Xi^{\mathcal{M}}_{\mathrm{Y}}$ and $\Xi^{\mathcal{M}}_{\mathrm{ff}}$
with $\mathcal{M}=\tau_{0,x,y,z}$ labeling the Type-$\mathcal{M}$ QPT
represent the corrections stemming from order-parameter fluctuations
and fermion-fermion interactions, respectively. By borrowing the tactic employed in
Refs.~\cite{Sachdev2008PRB,Wang2011PRB,Wang2013PRB,Vafek2014PRB,Wang2017PRB,
Wang2018JPCM,Wang2020PRB,Wang2021NPB}, we carry out the similarly long but
straightforward calculations and finally are left with the following results,
\begin{numcases}{\Xi^{\mathcal{M}}_{Y}=}
-\mathcal{A}_{3}\lambda^{3}\tau_{x}l,& $\mathcal{M}=\tau_{x}$,\\
(\mathcal{B}_{1}+\mathcal{B}_{2}+\mathcal{B}_{3})\lambda^{3}\tau_{y}l,& $\mathcal{M}=\tau_{y}$,\\
-\mathcal{C}_{2}\lambda^{3}\tau_{z}l,&$\mathcal{M}=\tau_{z}$,\\
-\mathcal{D}^{A,B}_{1}\lambda^{3}\tau_{0}l,&$\mathcal{M}=\tau_{0}$,
\end{numcases}
for the order-parameter part and
\begin{numcases}{\Xi^{\mathcal{M}}_{\mathrm{ff}}=}
\frac{u_{2}^{2}
+u_{3}^{2}-u_{0}^{2}-u_{1}^{2}}{8\pi v_{F}v_{\Delta}}\lambda\tau_{x}l,& $\mathcal{M}=\tau_{x}$,\\
\frac{u_{3}^{2}-u_{0}^{2}-u_{2}^{2}}{4\pi v_{F}v_{\Delta}}\lambda\tau_{y}l,& $\mathcal{M}=\tau_{y}$,\\
\frac{u_{1}^{2}
+u_{2}^{2}-u_{0}^{2}-u_{3}^{2}}{8\pi v_{F}v_{\Delta}}\lambda\tau_{z}l,&$\mathcal{M}=\tau_{z}$,
\end{numcases}
for fermion-fermion part, respectively.

Then, we turn our focus to the fermion-fermion couplings.
In analogy to the Yukawa vertex, both Yukawa couplings and
fermion-fermion interactions can contribute to the fermion-fermion vertex
dubbed by $\Gamma$, which leads to
\begin{eqnarray}
\Gamma^{\mathcal{M}}=\Gamma^{\mathcal{M}}_{\mathrm{Y}}
+\Gamma^{\mathcal{M}}_{\mathrm{ff}},\label{Eq_1L-ff-interaction}
\end{eqnarray}
where the indexes $\mathcal{M}$, $\mathrm{Y}$, and $\mathrm{ff}$
share the same meanings with the notations appearing in Eq.~(\ref{Eq_Xi}).
We again parallel the approaches
adopted in Refs.~\cite{Vafek2010PRB,Vafek2012PRB,Vafek2014PRB,Wang2017PRB,
Wang2020PRB,Mandal2018PRB,Roy2018PRX}
and arrive at the final results, which are attached
in Appendix~\ref{Appendix_Sigma-Gamma} for convenience.


\subsection{Coupled RG equations}\label{Sec_RG-Eqs}

With one-loop corrections in hand, we are now in a suitable position to derive
the coupled RG flows of all interaction parameters that dictate the critical behaviors
around the QCP. In the spirit of momentum-shell RG~\cite{Shankar1994RMP,Wilson1975RMP,
Polchinski1992}, we select the quadratic terms of effective action~(\ref{Eq_S-eff})
as the ``free fixed point" to deliver the RG rescaling transformations of momenta,
energy, and fields in the following~\cite{Sachdev2008PRB,Vafek2014PRB,Wang2017PRB,
Wang2011PRB,Wang2013PRB,Vafek2012PRB, She2010PRB, Kim-Kivelson2008PRB,
She2015PRB, Roy-Sau2016PRB,Xu2008PRB},
\begin{eqnarray}
k&\rightarrow&k'e^{-l},\label{Eq_RG-Scaling-1}\\
\omega&\rightarrow'&\omega e^{-l},\\
\Psi_{1,2}(\mathbf{k},\omega)&\rightarrow&\Psi'_{1,2}(\mathbf{k}',\omega')
e^{\frac{1}{2}\int_{0}^{l}
(4-\eta_{f})dl},\\
\phi(\mathbf{k},\omega)&\rightarrow&\phi'(\mathbf{k}',\omega')e^{\frac{1}{2}\int_{0}^{l}
(4-\eta_{b})dl}, \label{Eq_RG-Scaling-4}
\end{eqnarray}
where the variable parameter $l>0$ is adopted to specify a running energy scale
and delimit the momentum-shell for every RG transformation, which is confined to
$b\Lambda-\Lambda$ with $\Lambda$ specifying a cutoff of energy
scale~\cite{Shankar1994RMP,Wang2011PRB,Sachdev2008PRB}.
The anomalous dimensions $\eta_{f}$ and $\eta_{b}$ are
determined by the one-loop corrections in Sec.~\ref{Sec_correction}.
To one-loop level, we figure out that $\eta_b=0$
and $\eta_f=-\lambda^2 \mathcal{Z}$ which is
inherited from Eq.~(\ref{Eq_G_full}) with $\mathcal{Z}=\mathcal{A},
\mathcal{B},\mathcal{C},\mathcal{D}$ for four distinct types
of phase transitions classified in Sec.~\ref{sub_phen-model}.

Subsequently, we gather all one-loop corrections to
interaction parameters in Sec.~\ref{Sec_correction} and
the RG transformation scalings~(\ref{Eq_RG-Scaling-1})-(\ref{Eq_RG-Scaling-4})
together to derive the coupled RG equations by carrying out the standard procedures
of RG approach~\cite{Shankar1994RMP,Wilson1975RMP,Polchinski1992}.
In the following, we list the coupled RG equations for Type-$\tau_x$
phase transition, consisting of energy-dependent evolutions of
fermion velocities and Yukawa coupling,
\begin{eqnarray}
\frac{dv_{\Delta}}{dl}\!\!&=&\!\!\lambda^{2}(A_{1}-A_{3})v_{\Delta},\label{Eq_x_vD}\\
\frac{dv_{F}}{dl}\!\!&=&\!\!\lambda^{2}(A_{1}-A_{2})v_{F}\vspace{1ex},\label{Eq_x_vF} \\
\frac{d\frac{v_{\Delta}}{v_{F}}}{dl}
\!\!&=&\!\!\lambda^{2}(A_{2}-A_{3})\frac{v_{\Delta}}{v_{F}}, \label{Eq_x_vD/vF}\\
\frac{d\lambda}{dl}\!\!&=&\!\!\left[A_{1}-A_{3}+\frac{u_{2}^{2}
+u_{3}^{2}-u_{0}^{2}-u_{1}^{2}}{8\pi v_{F}v_{\Delta}}\right]\lambda^{3},\label{Eq_x_lambda}
\end{eqnarray}
as well as fermion-fermion strengths,
\begin{widetext}
\begin{eqnarray}
\frac{du_{0}}{dl}
\!\!\!&=&\!\!\!\Bigl\{\!-1+2A_1-\frac{(u_{1}u_{2}
+u_{2}u_{3})}{4\pi v_{F}v_{\Delta}u_{0}}
+\frac{2\lambda^{2}}{3}\Bigl[
\frac{u_{2}}{u_{0}}(A_{3}-A_{1})-4A_{1}\Bigr]\Bigr\}u_{0},\\
\frac{du_{1}}{dl}
\!\!\!&=&\!\!\!\left\{\!-1+2A_1
\!\!+\frac{1}{4\pi v_{F}v_{\Delta}}(u_{0}-u_{1}-u_{2}-2u_{3}
+\frac{2u_{2}u_{3}}{u_{1}} )+
\frac{2\lambda^{2}}{3}\left[
\frac{u_{3}}{u_{1}}(A_{3}-A_{1})
-4A_{3}\right]\right\}u_{1},\\
\frac{du_{2}}{dl}
\!\!\!&=&\!\!\!\left\{\!-1+2A_1
+\frac{1}{4\pi v_{F}v_{\Delta}}\left[(2u_{0}-3u_{1}-2u_{2}-3u_{3})+\frac{2u_{1}u_{3}
}{u_{2}}\right]
\!+\!\frac{2\lambda^{2}}{3}\!\!\left[\frac{u_{3}}{u_{2}}(A_{2}\!
-\!A_{3})\!+\!4A_{3}\!-\!5A_{1}\!-\!5A_{2}
\right]\!\right\}u_{2},\\
\frac{du_{3}}{dl}
\!\!\!&=&\!\!\!\left\{\!-1+2A_1
+\frac{1}{4\pi v_{F}v_{\Delta}}\left[(u_{0}-u_{3}-u_{1}-2u_{2})
+\frac{2u_{1}u_{2}}{u_{3}}
\right]
+\frac{2\lambda^{2}}{3}\left[
\frac{u_{2}}{u_{3}}(A_{2}-A_{3})
-A_{1}-5A_{2}\right]\right\}u_{3}.\label{Eq_x_u_3}
\end{eqnarray}
\end{widetext}
In order to make our presentations more compact, we collect and present
the related coupled RG equations in Appendix~\ref{Appendix_RG} with respective
to the rest three types of phase transitions. Specifically, Eqs.~(\ref{Eq_y_vF})-(\ref{Eq_y_u3})
correspond to Type-$\tau_y$, Eqs.~(\ref{Eq_z_vF})-(\ref{Eq_z_u3}) to Type-$\tau_z$,
and Eqs.~(\ref{Eq_0_vF})-(\ref{Eq_0_lambda}) to Type-$\tau_0$, respectively.

\begin{figure}
\centering
\includegraphics[width=4in]{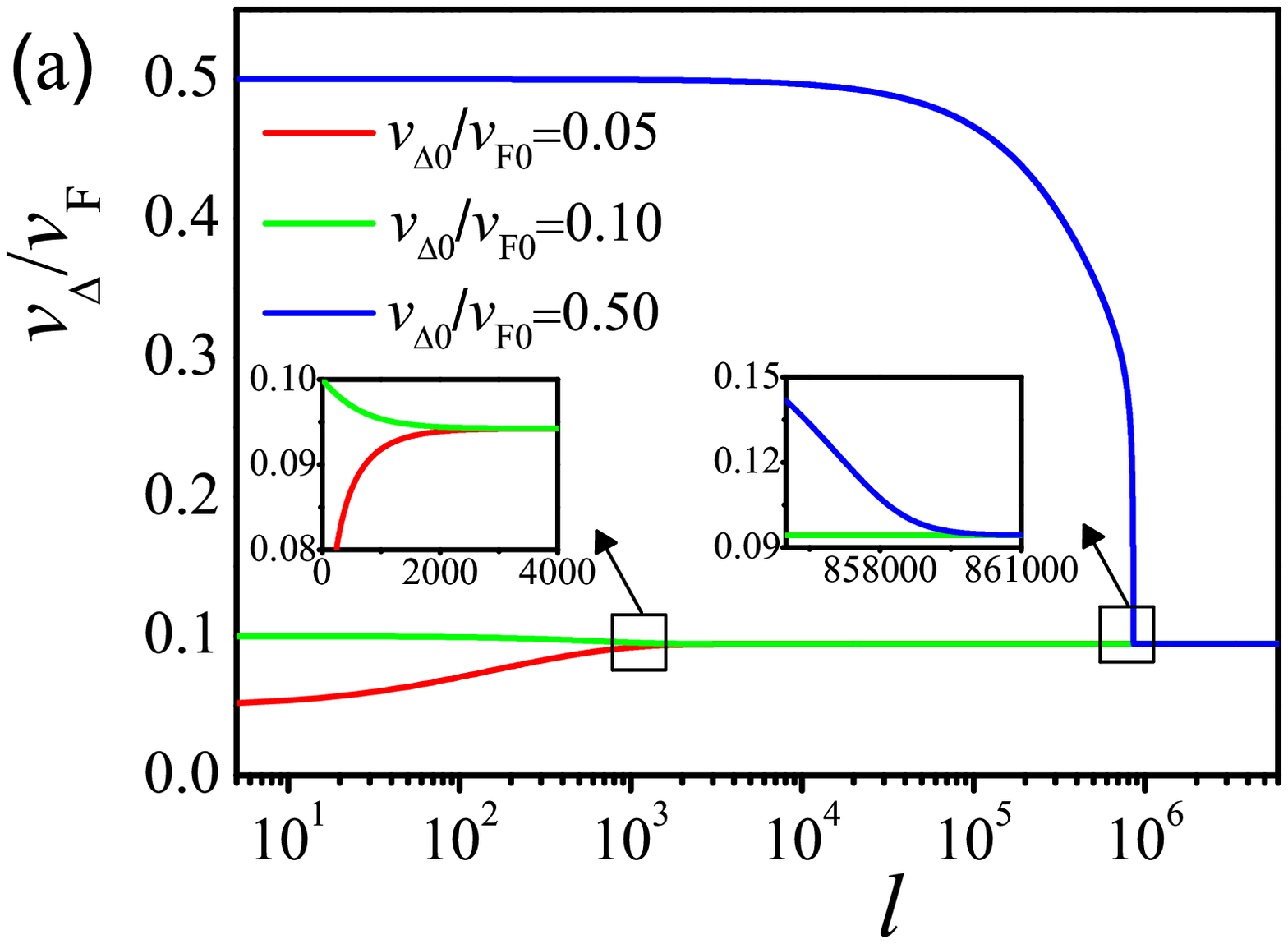}\vspace{-2.6cm}
\includegraphics[width=4in]{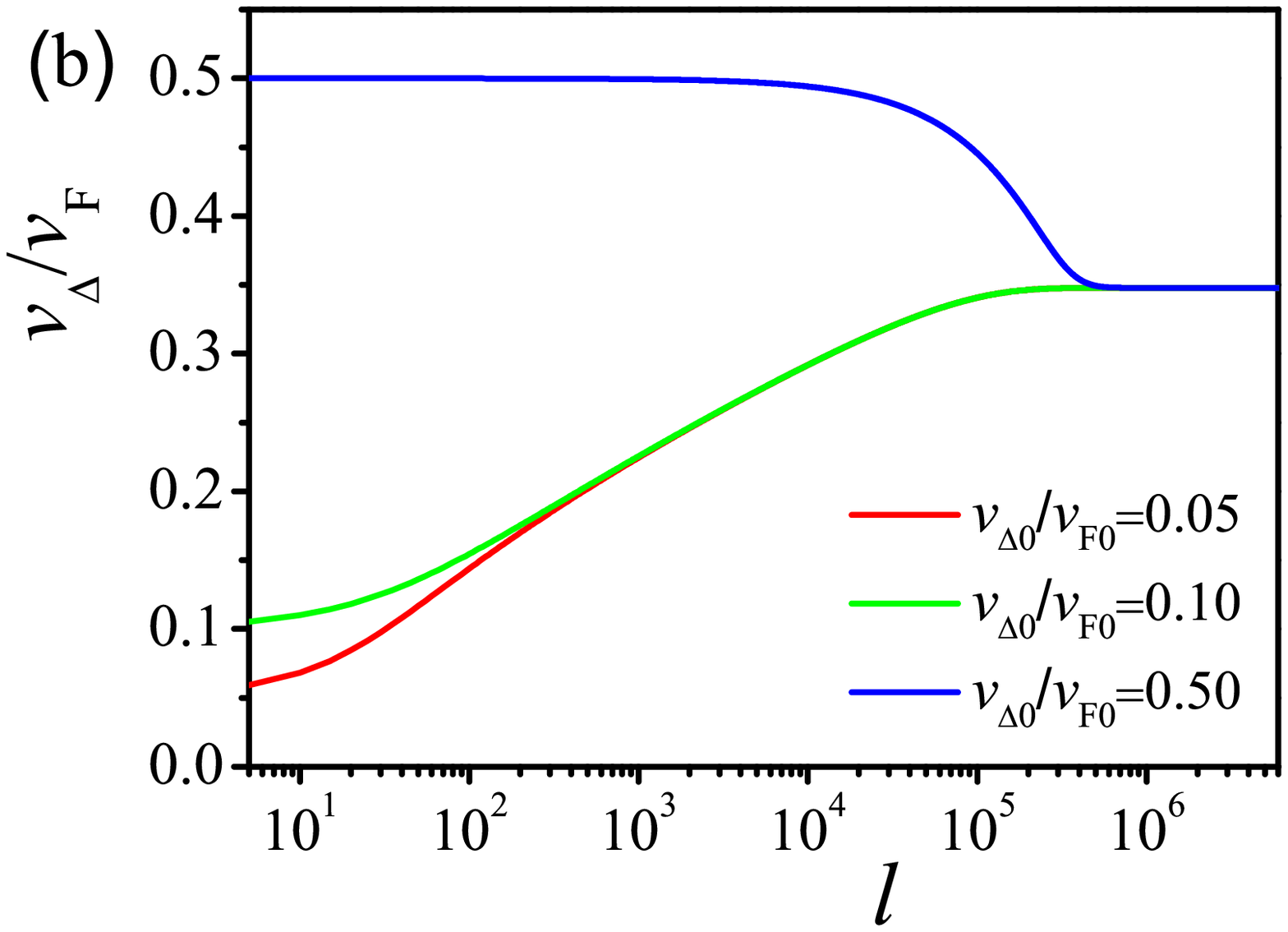}
\vspace{-2.7cm}
\caption{(Color online) Flows of $v_{\Delta}/v_{F}$ with decreasing
the energy scales (enlarging $l$) under three representative initial values
$v_{\Delta0}/v_{F0}=0.05, 0.1, 0.5$ for (a) Type-$\tau_{0A}$ and (b)
Type-$\tau_{0B}$ components of Type-$\tau_{0}$ QPT at a fixed Yukawa coupling $\lambda=1$.}\label{fig_tau_0}
\end{figure}

\begin{figure}
\centering
\includegraphics[width=4in]{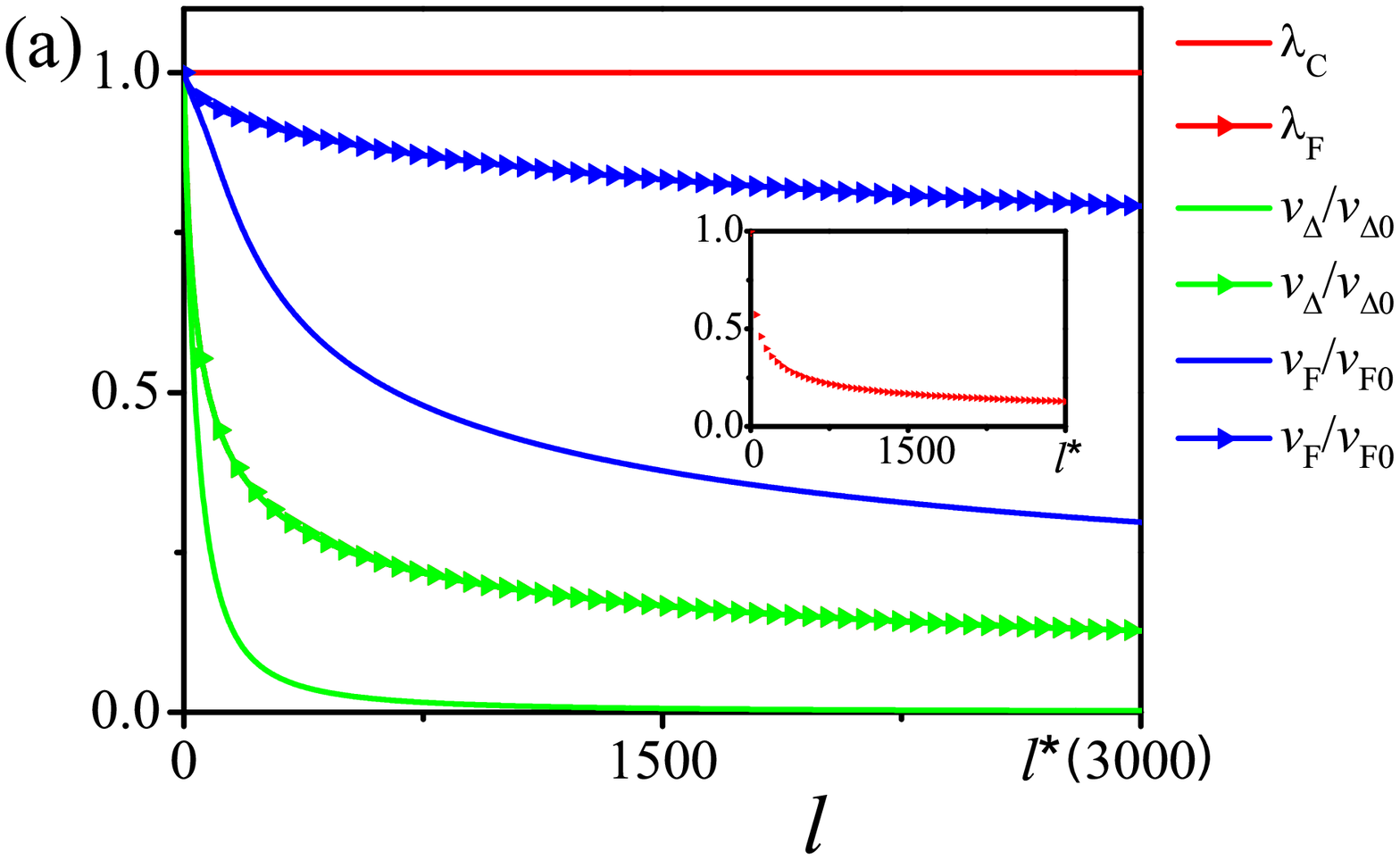}\vspace{-2.7cm}
\includegraphics[width=4in]{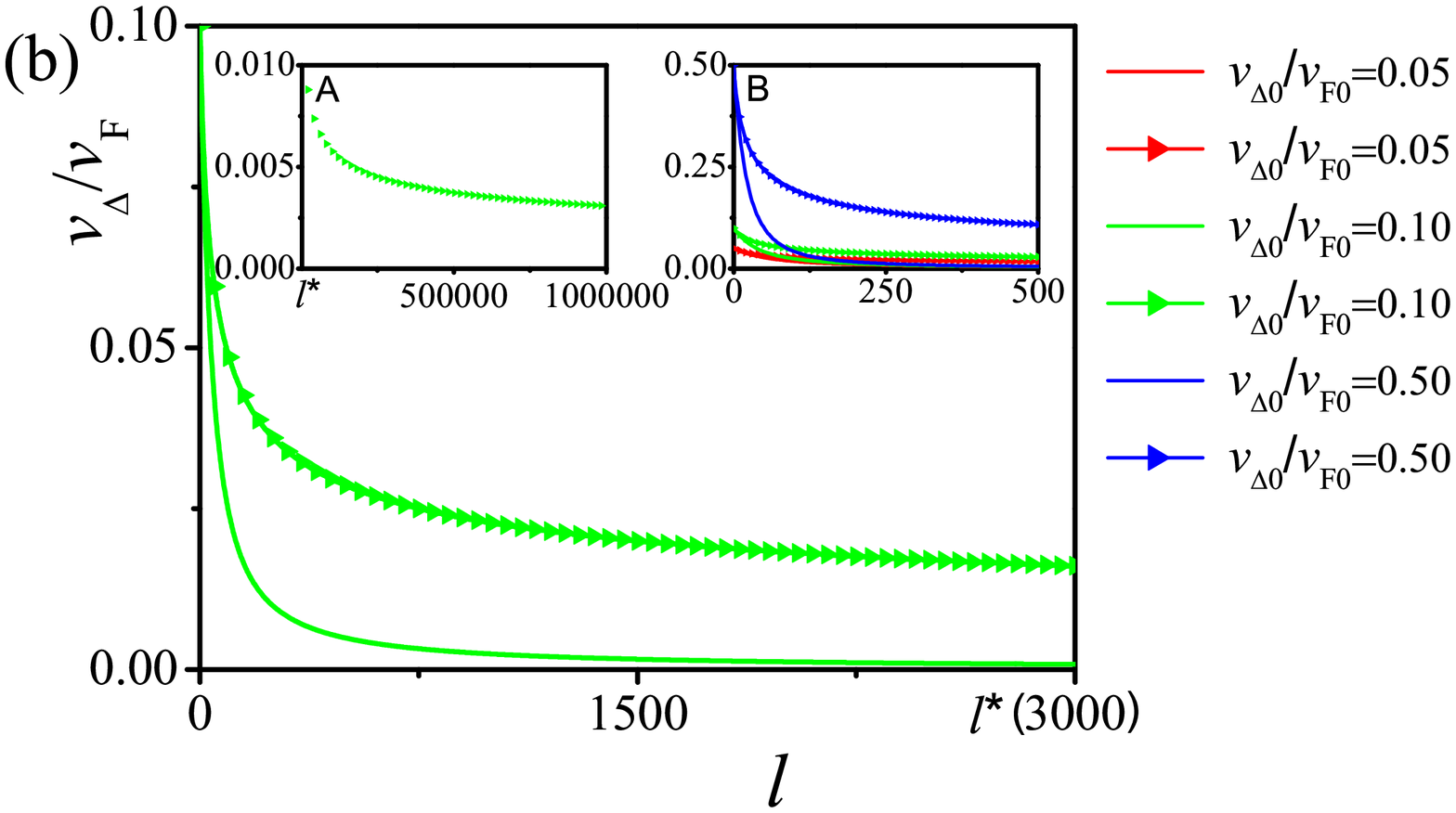}
\vspace{-2.7cm}
\caption{(Color online) Energy-dependent flows of interaction parameters for both
the fixed (bare curves) and flowing (arrowed curves) Yukawa couplings nearby the Type-$\tau_{x}$ QPT (the critical energy scale $l^{*}$ is designated
as the saturated point for $\lambda=1$ case): (a) evolutions of $\lambda$, $v_{\Delta}/v_{\Delta0}$, $v_{F}/v_{F0}$ at a representative initial
value $v_{\Delta0}/v_{F0}=0.1$ and (b) fates
of $v_{\Delta}/v_{F}$ with Inset A displaying the low-energy limit
at $l>l^{*}$ for the running-coupling case
and Inset B presenting its flows at three
representative initial values.}\label{fig_tau_x}
\end{figure}

In the scenario of RG framework, these evolutions encode intimate entanglements of
all interaction parameters~\cite{Shankar1994RMP,Vojta2003RPP,Sachdev2011Book},
which usually enter into the physical implications, and
henceforth are expected to be of particular relevance and significance to dictate
the low-energy fates of critical properties in the vicinity of certain
QCP in Fig.~\ref{fig1}. We are about to attentively investigate and address
the physical consequences of them in the two looming sections.

\section{Low-energy behaviors of fermion velocities}\label{Sec_velocity}

\begin{figure*}
\hspace{-1.8cm}
\includegraphics[width=1.9in]{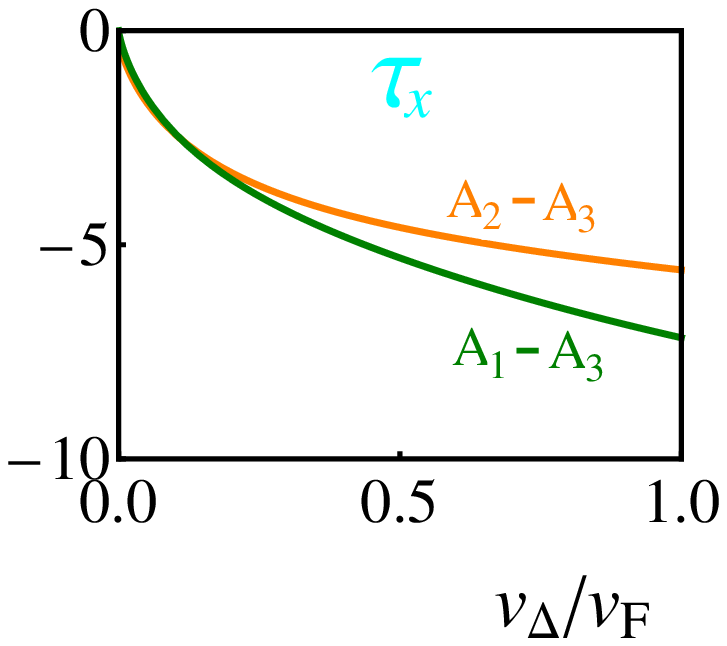}\hspace{0.3cm}
\includegraphics[width=1.9in]{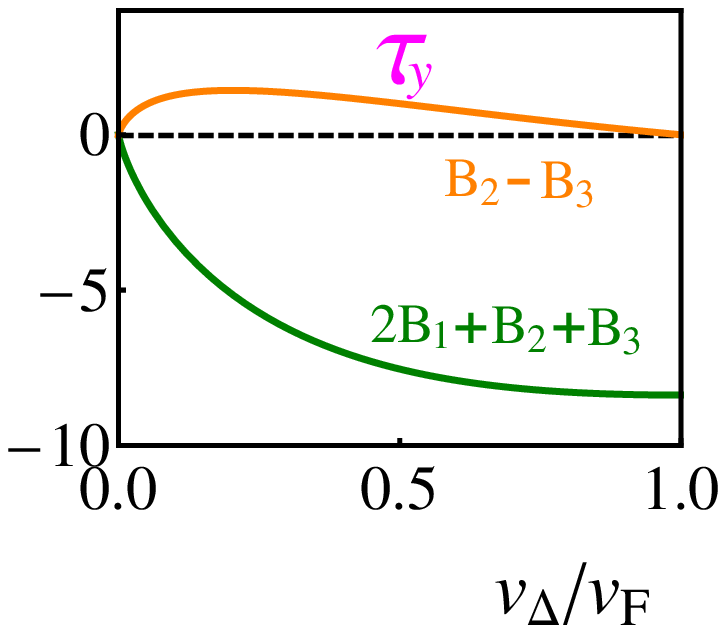}\hspace{0.3cm}
\includegraphics[width=1.9in]{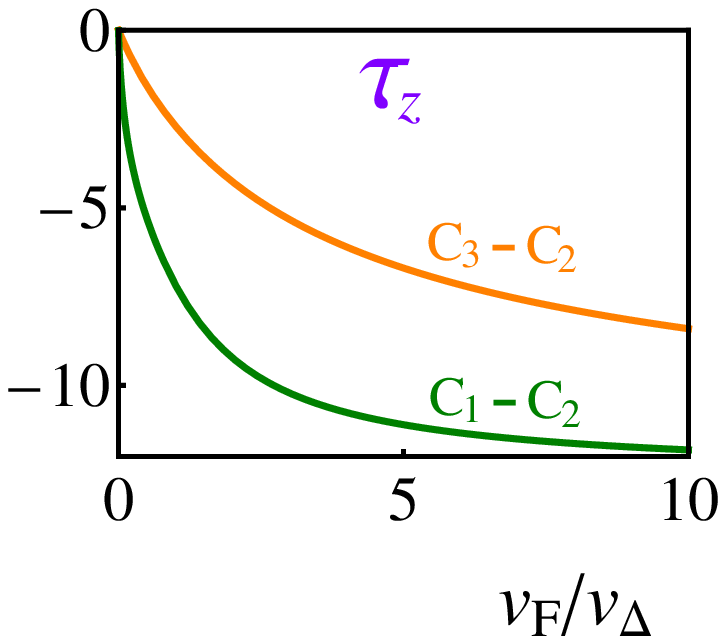}\vspace{-0.05cm}\\
\hspace{-0.9cm}
\includegraphics[width=4.4in]{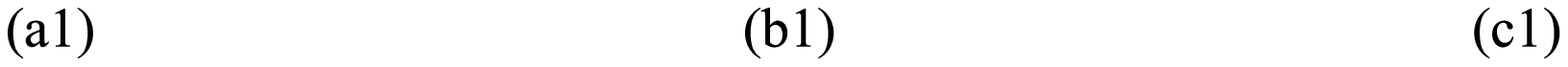}\vspace{0.7cm}\\
\hspace{-1.7cm}
\includegraphics[width=1.8in]{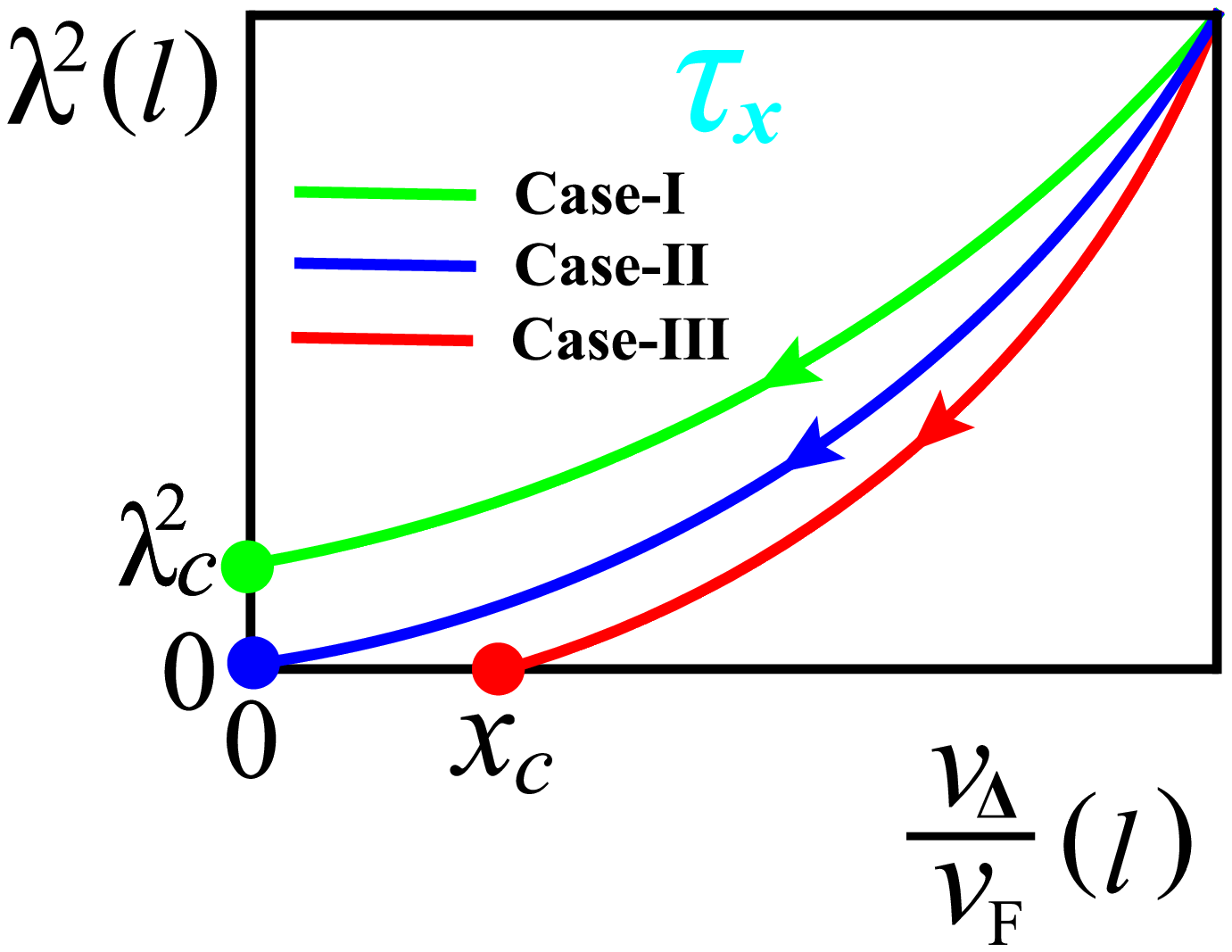} \hspace{0.5cm}
\includegraphics[width=1.8in]{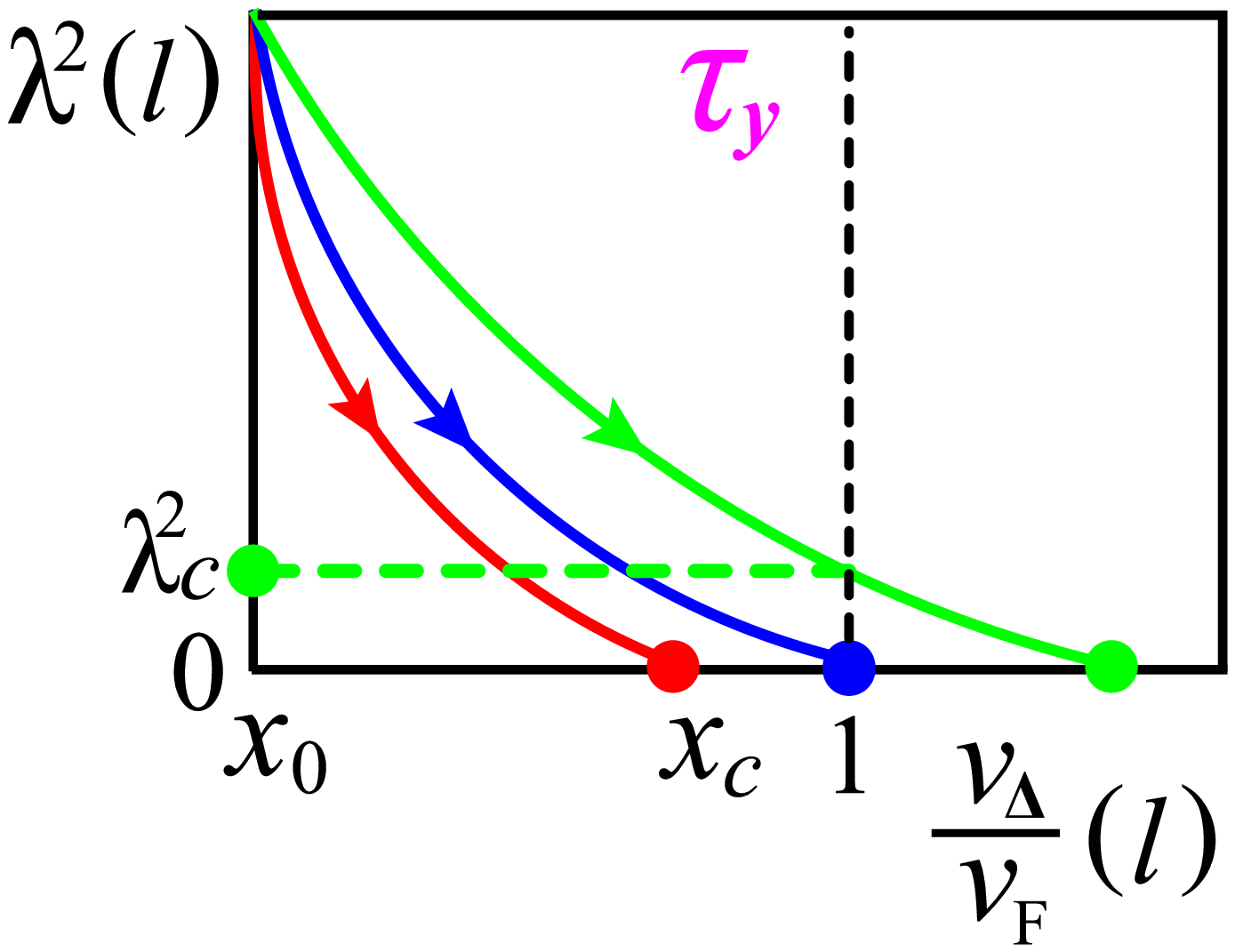} \hspace{0.5cm}
\includegraphics[width=1.8in]{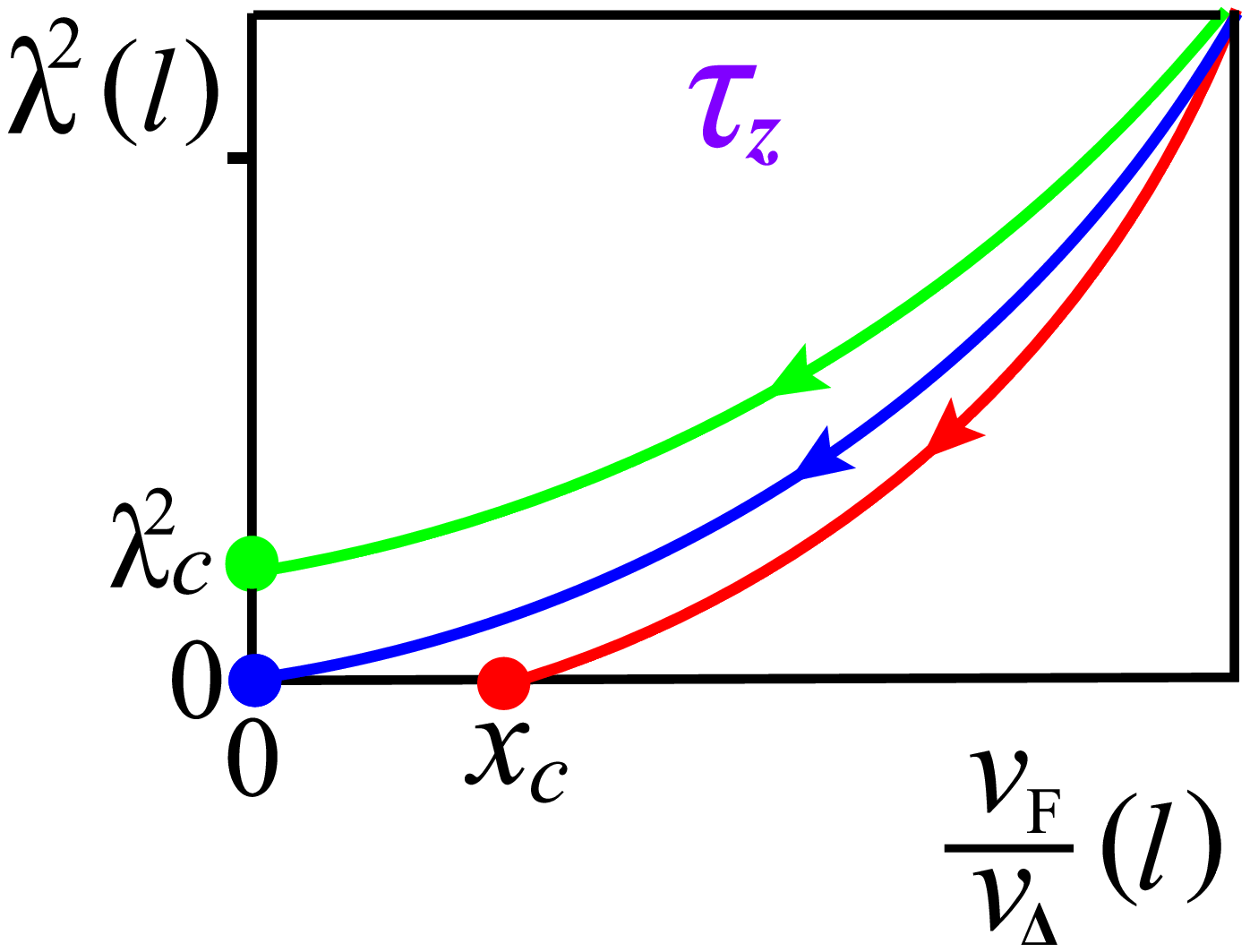} \vspace{-0.1cm}\\
\hspace{-0.9cm}
\includegraphics[width=4.4in]{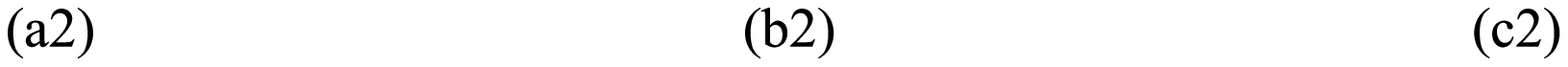}\vspace{0.6cm}\\
\hspace{-0.8cm}
\includegraphics[width=3.2in]{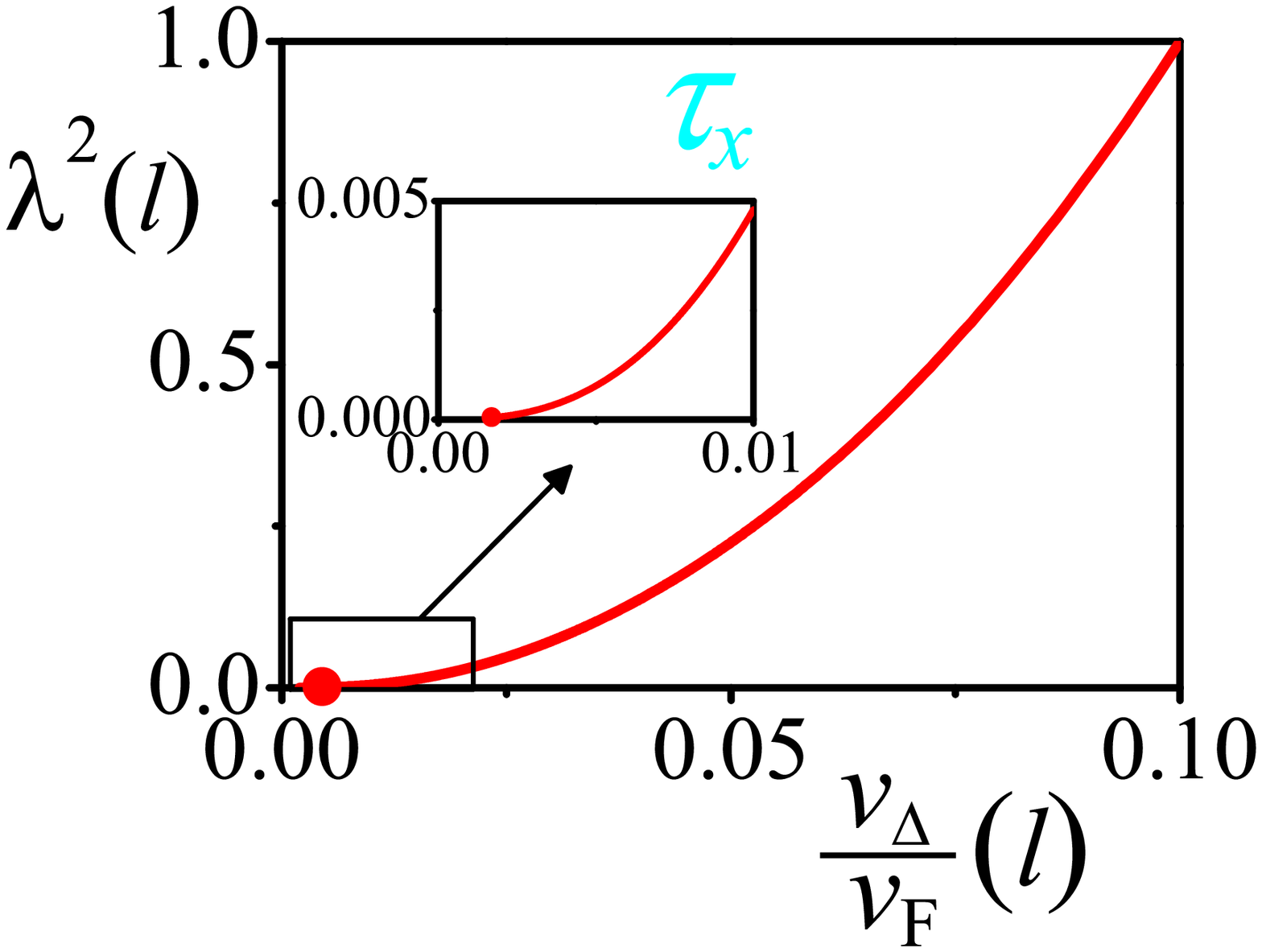} \hspace{-3.05cm}
\includegraphics[width=3.2in]{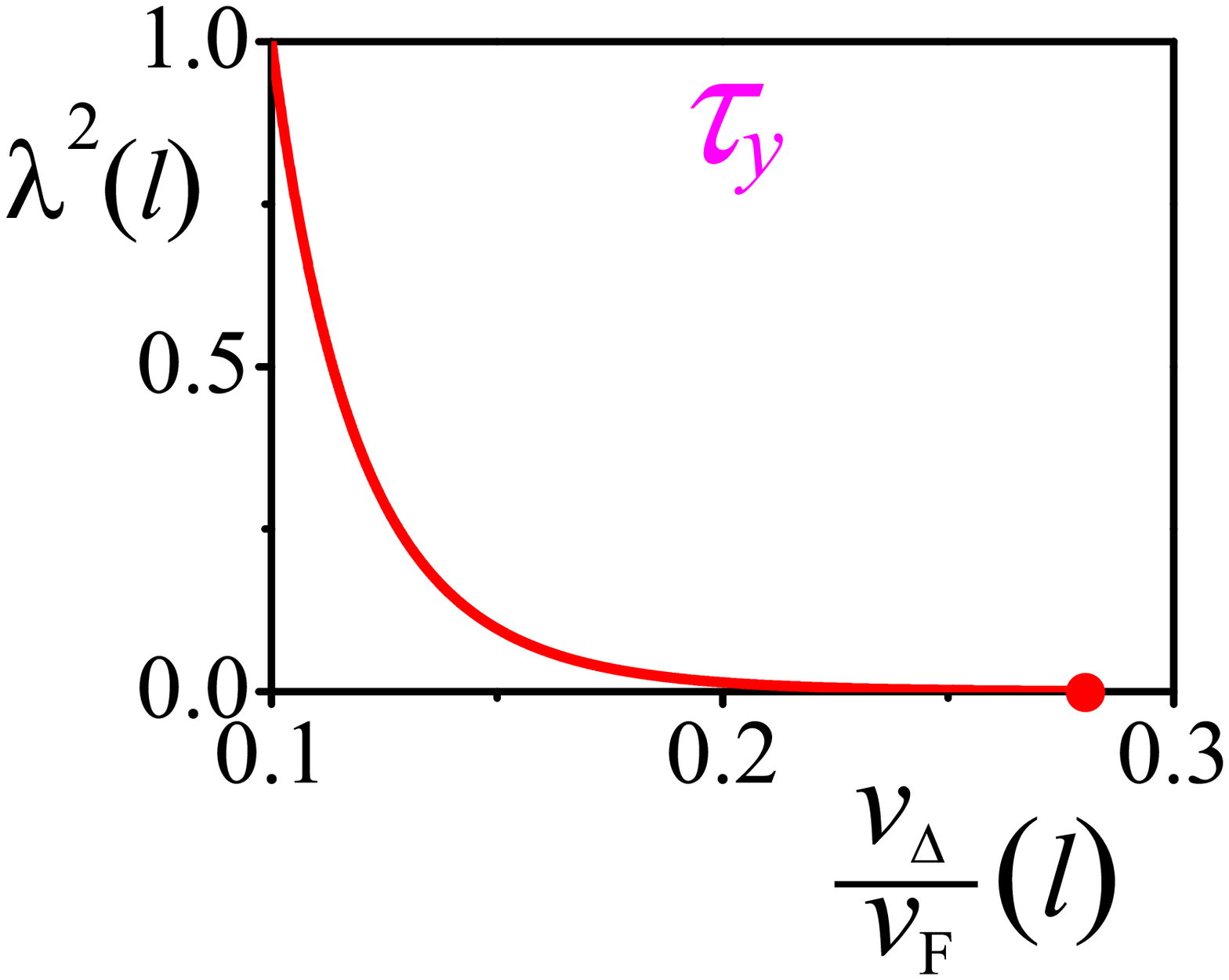} \hspace{-3.1cm}
\includegraphics[width=3.2in]{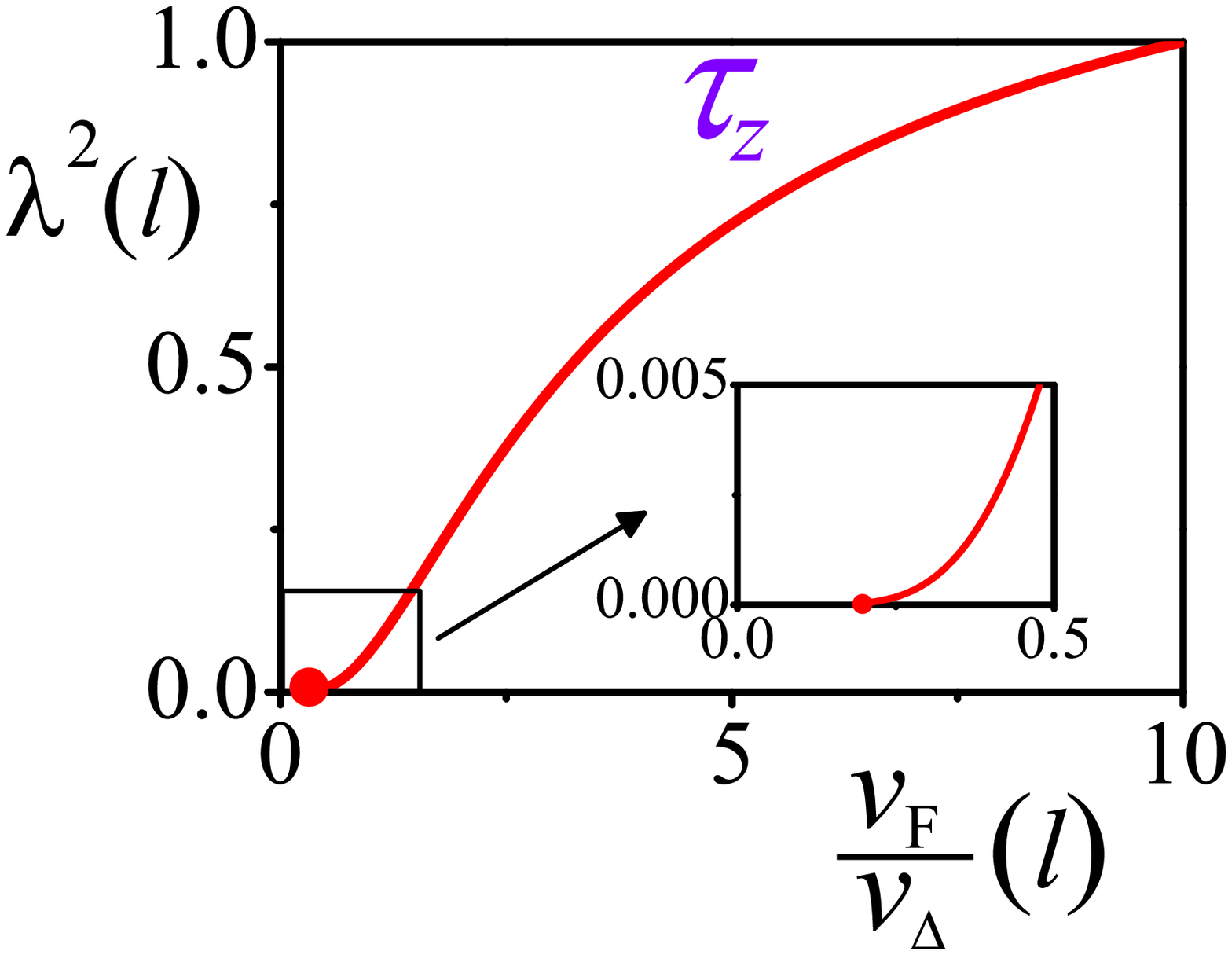}\vspace{-1.8cm}\\
\hspace{-1.1cm}
\includegraphics[width=4.4in]{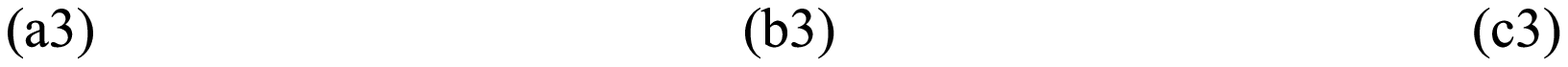}\vspace{0.2cm}\\
\caption{(Color online) Relationships between the coefficients in RG equations
and Yukawa coupling as well as fermion velocities for the potential QPTs.
The left column serves as the Type-$\tau_x$ QPT: (a1) the $v_\Delta/v_F$ dependence
of functions $A_{1}-A_{3}$ and $A_{2}-A_{3}$ in Eq.~(\ref{Eq_x_vD/vF})-(\ref{Eq_x_lambda});
(a2) three underlying fates of $\lambda^{2}(l)$ and $\frac{v_{\Delta}}{v_{F}}(l)$,
which are named as Case-I with $\frac{v_{\Delta}}{v_{F}}=0$ and Case-II with
$\lambda^{2}=\frac{v_{\Delta}}{v_{F}}=0$ as well as Case-III with
$\lambda^{2}=0$; and (a3) the evolution of
$\lambda^{2}(l)$-$\frac{v_{\Delta}}{v_{F}}(l)$ extracted
from the coupled RG equations.
The middle column denotes the Type-$\tau_y$ QPT: (b1) the $v_\Delta/v_F$ dependence
of functions $2B_{1}+B_{2}+B_{3}$ and $B_{2}-B_{3}$ in Eq.~(\ref{Eq_y_vD/vF})-(\ref{Eq_y_lambda});
(b2) three underlying fates of $\lambda^{2}(l)$ and $\frac{v_{\Delta}}{v_{F}}(l)$,
which are dubbed Case-I with $\frac{v_{\Delta}}{v_{F}}>1$, Case-II with $\frac{v_{\Delta}}{v_{F}}=1$ plus Case-III with
$\frac{v_{\Delta}}{v_{F}}<1$; and (b3) the evolution of $\lambda^{2}(l)$-$\frac{v_{\Delta}}{v_{F}}(l)$ extracted
from the coupled RG equations.
The right column corresponds to the Type-$\tau_z$ QPT: (c1) the
$v_{F}/v_{\Delta}$ dependence of functions  $C_{1}-C_{2}$ and $C_{3}-C_{2}$
in Eq.~(\ref{Eq_z_vD/vF})-(\ref{Eq_z_lambda});
(c2) three underlying fates of $\lambda^{2}(l)$ and $\frac{v_{F}}{v_{\Delta}}(l)$,
which are classified as Case-I with $\frac{v_{F}}{v_{\Delta}}=0$, Case-II with $\lambda^{2}=\frac{v_{F}}{v_{\Delta}}=0$ and Case-III with $\lambda^{2}=0$;
and (c3) the evolution of $\lambda^{2}(l)$-$\frac{v_{F}}{v_{\Delta}}(l)$ extracted
from the coupled RG equations.}\label{fig_cofficient}
\end{figure*}

\begin{figure}
\centering
\includegraphics[width=4in]{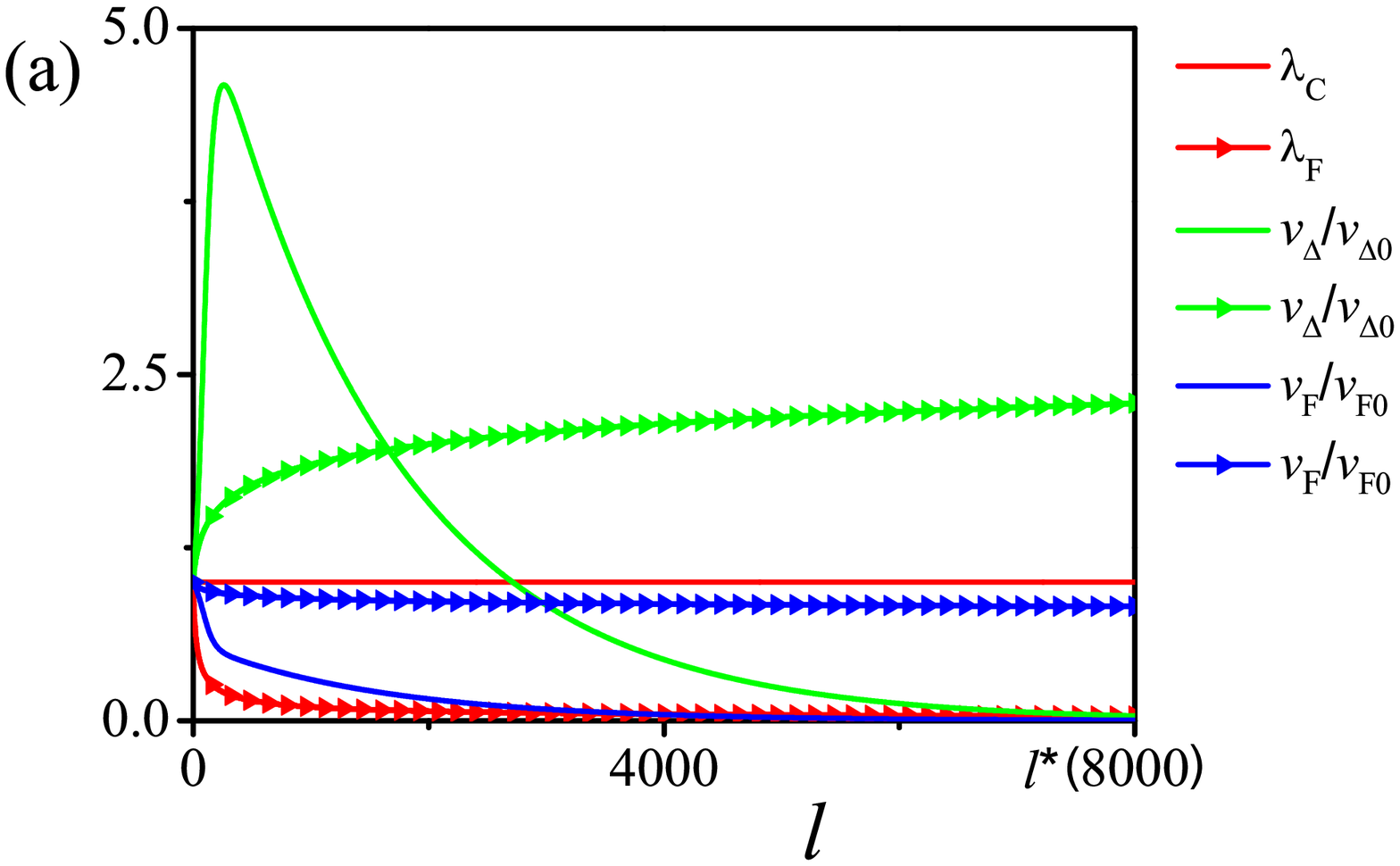}\vspace{-2.7cm}
\includegraphics[width=4in]{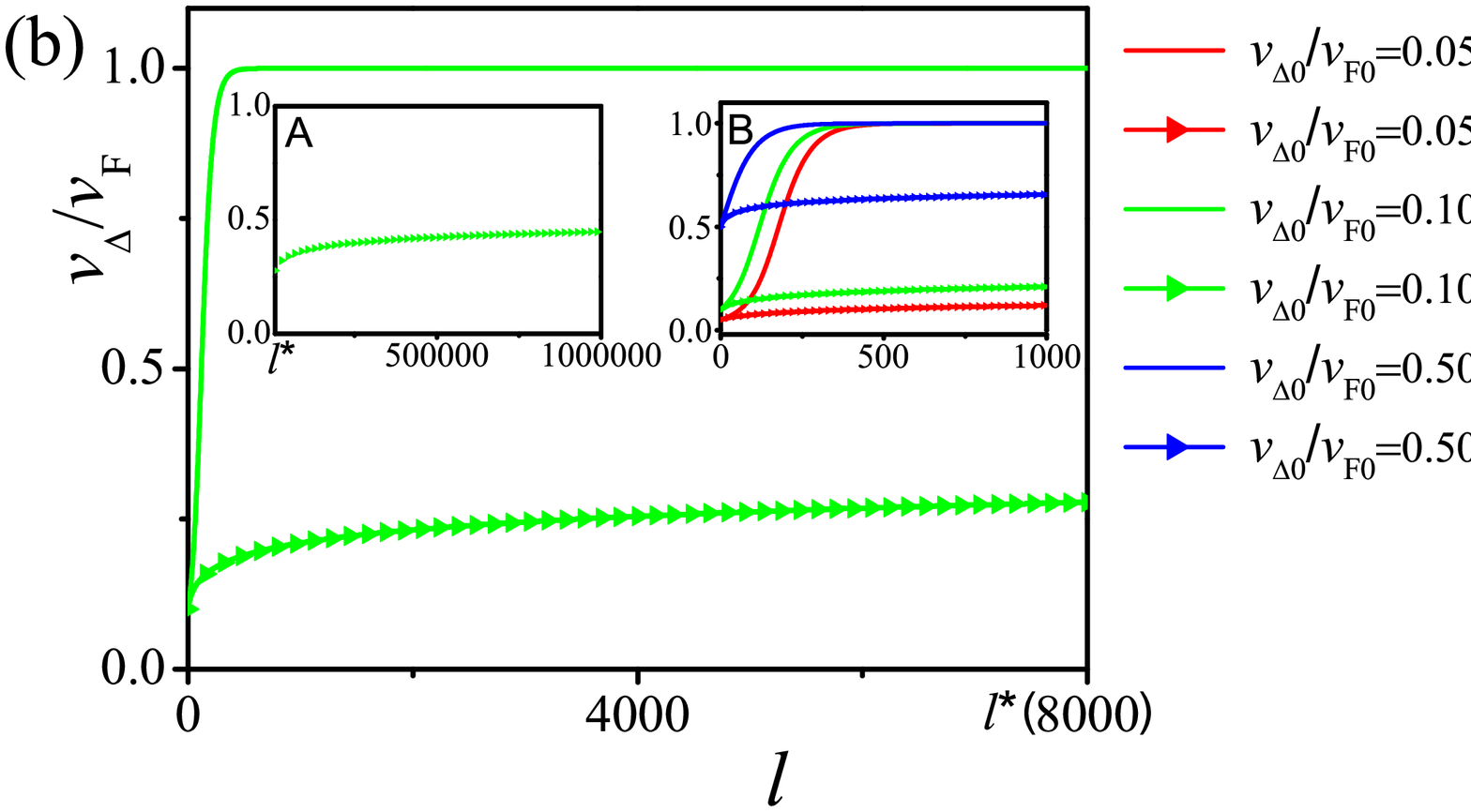}
\vspace{-2.7cm}
\caption{(Color online) Energy-dependent flows for both
the fixed (bare curves) and flowing (arrowed curves) Yukawa couplings
nearby the Type-$\tau_{y}$ QPT (the critical energy scale $l^{*}$ is
designated as the saturated point
for $\lambda=1$ case): (a) evolutions of $\lambda$, $v_{\Delta}/v_{\Delta0}$,
$v_{F}/v_{F0}$ at a representative initial
value $v_{\Delta0}/v_{F0}=0.1$ and (b) fates
of $v_{\Delta}/v_{F}$ with Inset A displaying the low-energy limit
at $l>l^{*}$ for the running-coupling case
and Inset B presenting its flows at three
representative initial values.}\label{fig_tau_y}
\end{figure}

Considering gapless nodal QPs intimately
couple with quantum critical fluctuations around
certain QPT shown in Fig.~\ref{fig1}~\cite{Vojta2000PRL,Lee2006RMP},
two very fermion velocities $v_{F}$ and $v_{\Delta}$ as well as their
ratio $v_{\Delta}/v_{F}$ would be substantially
renormalized. Given the behaviors of fermion velocities are of close
relevance to the low-energy physical observables,
they are henceforth expected to play an important role
in determining the low-energy fates of $d$-wave
superconductors~\cite{Orenstein2000Science,Lee2006RMP,Fradkin2012NPhys,
Kivelson2014PNAS,Fradkin2015RMP,Sachdev2011Book}.
Accordingly, it is of particular
importance to inspect the critical behaviors of
fermion velocities triggered by the QCP. To this end, we within
this section are going to study the energy-dependent
coupled RG flow equations addressed in Sec.~\ref{Sec_RG-Eqs},
which are assumed to contain the critical information of
certain QCP.

\subsection{In the absence of fermion-fermion interactions}\label{Subsec_absence-ff}

Despite both fermion-fermion interactions
and quantum fluctuations of order parameters are involved in
our RG equations in Sec~\ref{Sec_RG}, we at first
switch off the fermion-fermion interactions to explicitly investigate
the effects of all four sorts of order parameters with the reduced
coupled RG equations of $v_F$, $v_\Delta$, plus
$\lambda$, and then defer the contribution from fermionic
couplings to next subsection~\ref{Subsec_ff}.

\subsubsection{Fixed Yukawa coupling}\label{Subsubsec_fixed-lambda}

As aforementioned in Sec.~\ref{Sec_eff-theory}, there exist four
reduced types of QPTs at $r_{c}$, which are schematically shown in Fig.~\ref{fig1}
on the basis of the group theory analysis~\cite{Vojta2000PRL,Vojta2000PRB,Vojta2000IJMPB}.
It is worth pointing out that the critical behaviors of fermion velocities
with approaching the Type-$\tau_{x}$, -$\tau_y$ or -$\tau_z$
QCP were carefully studied by several researchers in the absence of
fermion-fermion interactions~\cite{Sachdev2008PRB,Wang2013PRB,
Kim-Kivelson2008PRB,Xu2008PRB,Wang2011PRB,Liu2012PRB,She2015PRB}. In order to
simplify the analysis, the Yukawa coupling between nodal QPs and order
parameter is regarded as a fixed constant and consequently three distinct fixed points
are driven by the quantum criticality, namely $(v_{\Delta}/v_{F})^*\rightarrow0$~\cite{Sachdev2008PRB}, $(v_{\Delta}/v_{F})^*\rightarrow1$~\cite{Wang2013PRB}, and $(v_{F}/v_{\Delta})^*\rightarrow0$~\cite{Wang2013PRB} for
Type-$\tau_x$, -$\tau_y$ and -$\tau_z$, respectively.

As to Type-$\tau_{0}$ QPT, it has not yet been seriously investigated to the
best of our knowledge. For the sake of completeness, we hereby examine the
fate of fermion velocities for such QPT. Learning from the RG analysis
in Sec.~\ref{Sec_RG-Eqs}, it is of particular interest to figure out that the Yukawa
coupling $\lambda$ is marginal to one-loop level. In other words, this is
equivalent to the situation of fixed Yukawa coupling.
Performing numerical evaluation of coupled RG equations for
Type-$\tau_{0}$ QPT~(\ref{Eq_0_vF})-(\ref{Eq_0_lambda}) gives rise to the main results
delineated in Fig.~\ref{fig_tau_0}. It manifestly shows that
the trajectories of $v_{\Delta}/v_{F}$ with variation of
initial values eventually converge to the same finite value at the
lowest-energy limit. To be specific, with lowering the energy scale,
$v_{\Delta}/v_{F}$ is attracted by either fixed point
$(v_{\Delta}/v_{F})^*\approx0.0942$ or $(v_{\Delta}/v_{F})^*\approx0.3478$,
which corresponds to Type-$\tau_{0A}$ or Type-$\tau_{0B}$ component
and is insensitive to initial conditions. As a consequence, Type-$\tau_{0}$ QPT,
in marked contrast to extreme anisotropies caused by its Type-$\tau_{x,z}$
counterparts, prefers to induce some finite anisotropy of fermion velocities.

\subsubsection{Flowing Yukawa coupling}\label{subsub-flow-lambda}

Compared to the fixed-coupling assumption~\cite{Sachdev2008PRB,Wang2013PRB,Kim-Kivelson2008PRB,Xu2008PRB,
Wang2011PRB,Wang2015PLA,Wang2013NJP,Liu2012PRB,She2015PRB}, much more physical information
would be captured after seriously taking into account the potential evolution
of Yukawa coupling $\lambda$ appearing in Eq.~(\ref{S_phi_0_S_psi}).
As apparently exhibited in Sec.~\ref{Sec_RG-Eqs}, the coupled RG equations are
jointly dictated by both the flow of $\lambda$ and its entanglement
with other interaction parameters. In this context, one can expect that
the low-energy properties of fermion velocities
may be partially or heavily modified by the evolution of coupling $\lambda$
around the putative QCP. In order to clarify these intriguing and
significant issues, we hereby place our focus on whether and how
the tendencies of fermion velocities can be reshaped for all types of
QPTs. With respect to the Type-$\tau_{0}$ QPT, it is worthwhile to highlight that
the coupling $\lambda$ is marginal as depicted in Eq.~(\ref{Eq_0_lambda}),
indicating an effective fixed-coupling case which is
studied in Sec.~\ref{Subsubsec_fixed-lambda}. As to the other
three types of QPTs, we subsequently address one by one in the following.

At the outset, we inspect how fermion velocities behave as approaching
the Type-$\tau_{x}$ QPT. The corresponding coupled RG evolutions are provided in
Eqs.~(\ref{Eq_x_vD})-(\ref{Eq_x_lambda}), which are indicative of the close
interplay between parameter $\lambda$ and fermion velocities $v_{F}$
and $v_{\Delta}$. After choosing several representative initial conditions
to perform numerical calculations, we realize that flowing
$\lambda$ plays an important role in the energy-dependent
tendencies of all related parameters as displayed in
Fig.~\ref{fig_tau_x}. Before going further, two helpful points
need to be clarified. For the sake of comparison,
we from now on would also supplement the corresponding results of fixed-coupling
cases with $\lambda=1$ for simplicity in our numerical results~\cite{Sachdev2008PRB,Wang2013PRB,Kim-Kivelson2008PRB,Xu2008PRB,Wang2011PRB}.
Additionally, a critical energy scale denoted by $l^*$ is designated
to serve as the saturated point for $\lambda=1$ case.
Learning from Fig.~\ref{fig_tau_x}, we find that
both $v_{\Delta}$ and $v_{F}$ for case $\lambda=1$ rapidly decrease upon
lowering the energy scales. In particular, $v_{\Delta}$ falls
down more quickly than $v_{F}$ and thus their ratio $v_\Delta/v_F$
goes towards zero at the lowest-energy limit.
This implies that $v_\Delta$ vanishes but instead
$v_F$ still acquires a finite value at $l\ge l^*$.
In striking comparison, once the Yukawa coupling is
involved in the coupled RG evolutions as well, $\lambda$
itself gradually descends and evolves towards zero at
the lowest-energy limit. As a result, the downtrends of
fermion velocities are much slower than their $\lambda=1$
counterparts. Specifically, both $v_{\Delta}$
and $v_{F}$ gently decrease and evolve towards finite
values at $l^{*}$. Concerning their ratio $v_\Delta/v_F$,
Fig.~\ref{fig_tau_x}(b) shows that the extreme anisotropy
is broken and replaced by a finite anisotropy at $l^{*}$
under the influence of the running coupling $\lambda$.
Therefore, we come to a conclusion that the low-energy fates
of fermion velocities are heavily affected by the participation of
energy-dependent $\lambda$. In particular, the destruction of
extreme anisotropy of $v_{\Delta}/v_{F}$ would impose a direct or indirect
impact on the physical quantities nearby the putative QCP.

\begin{figure}
\centering
\includegraphics[width=4in]{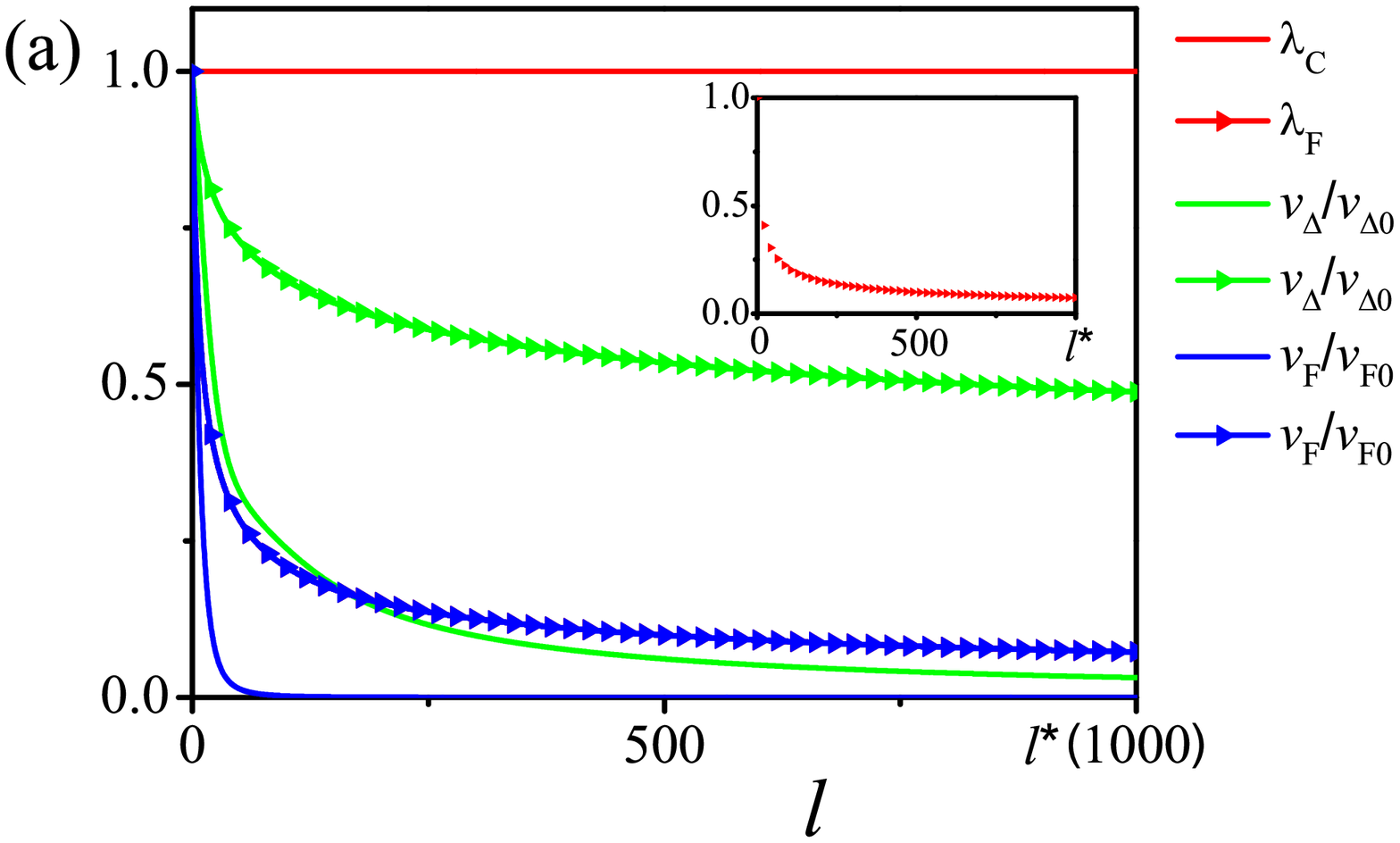}\vspace{-2.7cm}
\includegraphics[width=4in]{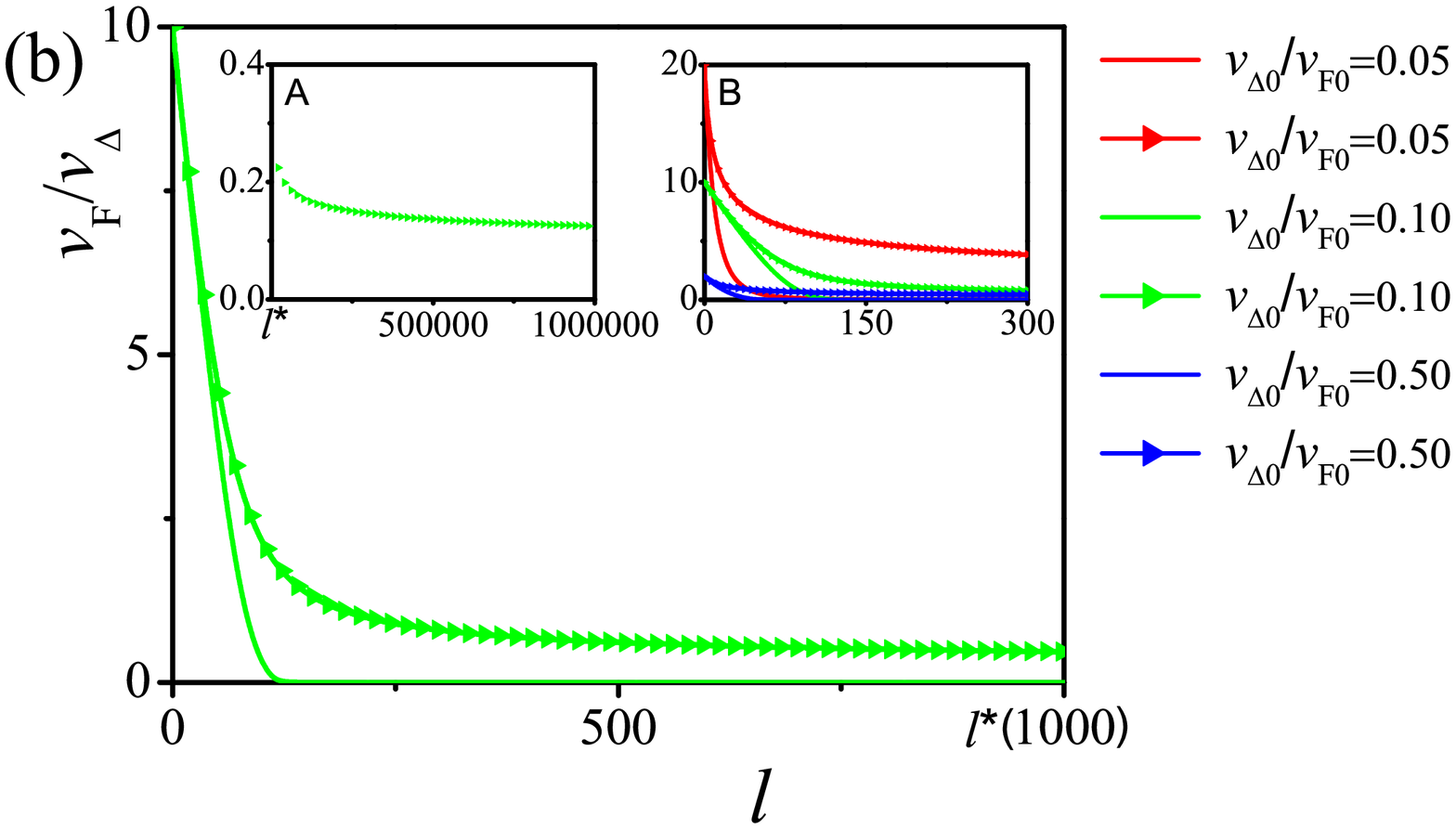}
\vspace{-2.7cm}
\caption{(Color online) Energy-dependent flows for both
the fixed (bare curves) and flowing (arrowed curves) Yukawa
couplings nearby the Type-$\tau_{z}$ QPT (the critical
energy scale $l^{*}$ is designated as the saturated point
for $\lambda=1$ case): (a) evolutions of $\lambda$, $v_{\Delta}/v_{\Delta0}$,
$v_{F}/v_{F0}$ at a representative initial
value $v_{\Delta0}/v_{F0}=0.1$ and (b) fates
of $v_{F}/v_{\Delta}$ with Inset A displaying the low-energy limit
at $l>l^{*}$ for the running-coupling case
and Inset B presenting its flows at three
representative initial values.}\label{fig_tau_z}
\end{figure}

Prior to investigating the Type-$\tau_y$ QPT,
we endeavor to examine the stability of $v_\Delta/v_F$ nearby
the Type-$\tau_x$ QPT at $l=l^*$ and pinpoint its final
fate at $l>l^*$. It is manifestly shown in
Fig.~\ref{fig_tau_x}(b) that $v_{\Delta}/v_{F}$ in the fixed-coupling case
is nearly saturated with the extreme anisotropy at the critical energy scale.
This is apparent distinction to the running-coupling situation,
in which the ratio reduces to certain finite value at $l=l^*$.
With an aim to explore the tendency of $v_{\Delta}/v_{F}$
in the lowest-energy limit, we enlarge the variable $l>l^*$ to
obtain the inset A of Fig.~\ref{fig_tau_x}(b), displaying that
the ratio is still unsaturated. In light of the technical deficiency
of numerical evaluation, it seems unrealistic to determine
whether $v_{\Delta}/v_{F}$ vanishes or reaches a finite value
at $l\rightarrow \infty$. As a consequence, we resort to
tentatively analytical analysis of RG equations in association
with the numerical results. In principle, the behavior of
$v_{\Delta}/v_{F}$ is directly determined by the coefficients
in Eq.~(\ref{Eq_x_vD/vF}) including $\lambda^{2}$ and $A_{2}-A_{3}$.
Each of them going towards zero hints to the stop of RG flow with a stable
$v_{\Delta}/v_{F}$. Once the coupling $\lambda$ goes towards
zero before the latter, the ratio $v_{\Delta}/v_{F}$ can either
be a finite value or zero. However, it is worth emphasizing that the vanishment of
$A_{2}-A_{3}$ is tantamount to $v_{\Delta}/v_{F}=0$ as delineated in
Fig.~\ref{fig_cofficient}(a1). Accordingly, as illustrated in
Fig.~\ref{fig_cofficient}(a2), there are three distinct
circumstances in all for the ratio of fermion velocities at the
lowest-energy limit, which correspond
to Case-I with $\lambda^2\neq0$ and
Case-II with $\lambda^2=v_\Delta/v_F=0$ as well as Case-III
with $v_\Delta/v_F\neq0$. With the help of numerical results,
Fig.~\ref{fig_cofficient}(a3) recapitulates
the tendencies of $\lambda^2$ and $v_\Delta/v_F$ with decreasing the energy scale.
On the basis of these, we figure out that Case-III is selected by the coupled RG
evolutions and hence $v_{\Delta}/v_{F}$ for Type-$\tau_{x}$ QPT
is eventually attracted by a finite fixed point at the
lowest-energy limit. This is consistent with the previous analysis at $l\le l^*$
and therefore hints to the destruction of extreme anisotropy
of fermion velocities due to the evolution of Yukawa coupling.

Subsequently, we move to examine the
low-energy fates of fermion velocities
as accessing the Type-$\tau_{y}$ QPT.
Initially, let us aim at the regime $l\in[0,l^*]$
in which the numerical results of the
associated RG equations are provided
in Fig.~\ref{fig_tau_y}. As for the fixed-coupling
$\lambda=1$~\cite{Sachdev2008PRB,Wang2013PRB,Kim-Kivelson2008PRB,Xu2008PRB,
Wang2011PRB,Wang2015PLA,Wang2013NJP,Liu2012PRB,She2015PRB},
it can be seen from Fig.~\ref{fig_tau_y}(a) that $v_{\Delta}$ quickly
climbs up and then keeps decreasing
until it vanishes at $l=l^*$. Meanwhile,
$v_{F}$ monotonically falls down to zero.
These make the fermion velocities isotropic with
$(v_{\Delta}/v_{F})^*=1$ at $l=l^*$, which is insensitive
to the starting condition as displayed in Fig.~\ref{fig_tau_y}(b).
In sharp contrast, the coupled RG equations~(\ref{Eq_y_vF})-(\ref{Eq_y_lambda})
force the coupling $\lambda$ to interact with other parameters
and descend with lowering the energy scale. As a result, fermion
velocities present distinct behaviors compared to $\lambda=1$
as approaching the critical energy scale. In Fig.~\ref{fig_tau_y}(a),
$v_{\Delta}$ gradually goes up to certain finite values, but
instead $v_{F}$ decreases to nonzero values. Accordingly,
Fig.~\ref{fig_tau_y}(b) shows that $v_{\Delta}/v_{F}$
slowly climbs up and flows towards a finite value,
which is smaller than $(v_{\Delta}/v_{F})^*=1$ and
susceptible to the initial conditions. This suggests that
the evolution of coupling $\lambda$ prevents fermion velocities
being isotropic but rather results in weak anisotropy
as accessing the critical energy scale. Next, we go to
judge the final fate at $l>l^*$ under the
influence of energy-dependent Yukawa coupling in that $v_{\Delta}/v_{F}$
is not saturated at $l=l^{*}$ and even a much larger $l$ as shown in
Inset A of Fig.~\ref{fig_tau_y}(b). In analogy with Type-$\tau_x$ case,
the final fate of $v_{\Delta}/v_{F}$ for the Type-$\tau_{y}$ QPT
depends upon which one of two coefficients $B_{2}-B_{3}$ and
$\lambda^{2}$ in Eq.~(\ref{Eq_y_vD/vF}) is driven to the fixed point more quickly.
To respond this, we realize the fate of $B_2-B_3=0$ amounts to
$v_\Delta/v_F=1$ and then parallel the strategy for Type-$\tau_x$ QPT to
present three potential circumstances for $\lambda^2$ and $v_\Delta/v_F$
in Fig.~\ref{fig_cofficient}(b2), consisting of Case-I with $v_{\Delta}/v_{F}>1$,
Case-II with $v_{\Delta}/v_{F}=1$ plus Case-III with $v_{\Delta}/v_{F}<1$.
The related numerical analysis of RG equations
in Fig.~\ref{fig_cofficient}(b3) exhibits Case-III is the dominant situation.
This henceforth corroborates the results at $l=l^*$ that fermion velocities
are forced to a weak anisotropy due to the contribution from the running
Yukawa coupling.

At last, we go to investigate the low-energy behaviors
of fermion velocities by virtue of the coupled RG flows~(\ref{Eq_z_vF})-(\ref{Eq_z_lambda})
nearby Type-$\tau_{z}$ QPT. The major results are presented in Fig.~\ref{fig_tau_z},
in which the distinctions between fixed-coupling and energy-dependent
cases are clearly exhibited. Studying from Fig.~\ref{fig_tau_z}(a),
$v_{F}$ rapidly drops down and vanishes at $l\approx l^*$ with
a fixed $\lambda=1$, but rather $v_{\Delta}$ progressively
descends and tends to a finite value~\cite{Wang2013PRB,Wang2015PLA,Wang2013NJP}.
While the Yukawa coupling $\lambda$ enters into
the RG equations, it becomes energy-dependent and
quickly climbs down with lowering the energy scale.
This brings significant effects to fermion velocities,
making $v_{\Delta}$ drop much more than that of $v_{F}$ despite
both of them smoothly decrease as the energy scale is decreased.
With respect to the ratio of fermion velocities at $l\ge l^*$
in Fig.~\ref{fig_tau_z}(b), we figure out that $v_{F}/v_{\Delta}$
bears similarities to $v_{\Delta}/v_{F}$ approaching the
Type-$\tau_{x}$ QPT illustrated in Fig.~\ref{fig_tau_z}(a).
In other words, the extreme anisotropy with $v_{F}/v_{\Delta}\rightarrow0$
at $\lambda=1$~\cite{Wang2013PRB} is sabotaged and replaced with a finite
anisotropy by the evolution of coupling $\lambda$.
By the same token, $v_{F}/v_{\Delta}$ for Type-$\tau_{z}$ QPT hereby
does not saturate at $l=l^*$ as shown in Inset A of Fig.~\ref{fig_tau_z}(b).
In this sense, we follow the previous tactic to identify
its final fate, which heavily hinges upon the coefficients
$\lambda^2$ and $C_{3}-C_{2}$ in Eqs.~(\ref{Eq_z_vF})-(\ref{Eq_z_lambda}).
In resemblance to the analysis for $v_{\Delta}/v_{F}$,
$C_3-C_2=0$ points to $v_F/v_\Delta=0$
and then three distinct fates are diagrammatically illustrated
in Fig.~\ref{fig_cofficient}(c2) including Case-I with $\lambda^2\neq0$,
Case-II with $v_F/v_{\Delta}=\lambda^2=0$ and Case-III with $v_F/v_{\Delta}\neq0$,
respectively. In the assistance of numerical evaluation,
Fig.~\ref{fig_cofficient}(c3) shows us that Case-III wins
the competition with $v_{F}/v_{\Delta}$ being governed by
a finite value. It therefore signals that the evolution
of coupling $\lambda$ drives the extreme anisotropy
$v_{F}/v_{\Delta}\rightarrow0$ into a finite anisotropy
at the lowest-energy limit.

\subsection{In the presence of fermion-fermion interactions}\label{Subsec_ff}

As aforementioned in Sec.~\ref{Sec_RG}, fermion-fermion interactions
enter into the coupled RG equations and then may play an important
role in the low-energy regime via intimately interacting with
quantum fluctuations of order parameters and fermion velocities.
Based upon the results in the absence of
fermion-fermion interactions, we are now in a suitable position
within this subsection to investigate how fermion-fermion interactions
impact the behaviors of fermion velocities
upon approaching the putative QCPs, which are
insufficiently taken into account in previous
efforts~\cite{Sachdev2008PRB,Wang2013PRB,
Kim-Kivelson2008PRB,Xu2008PRB,Wang2011PRB,
Wang2015PLA,Wang2013NJP,Liu2012PRB,She2015PRB}.

\begin{figure}
\centering
\includegraphics[width=4in]{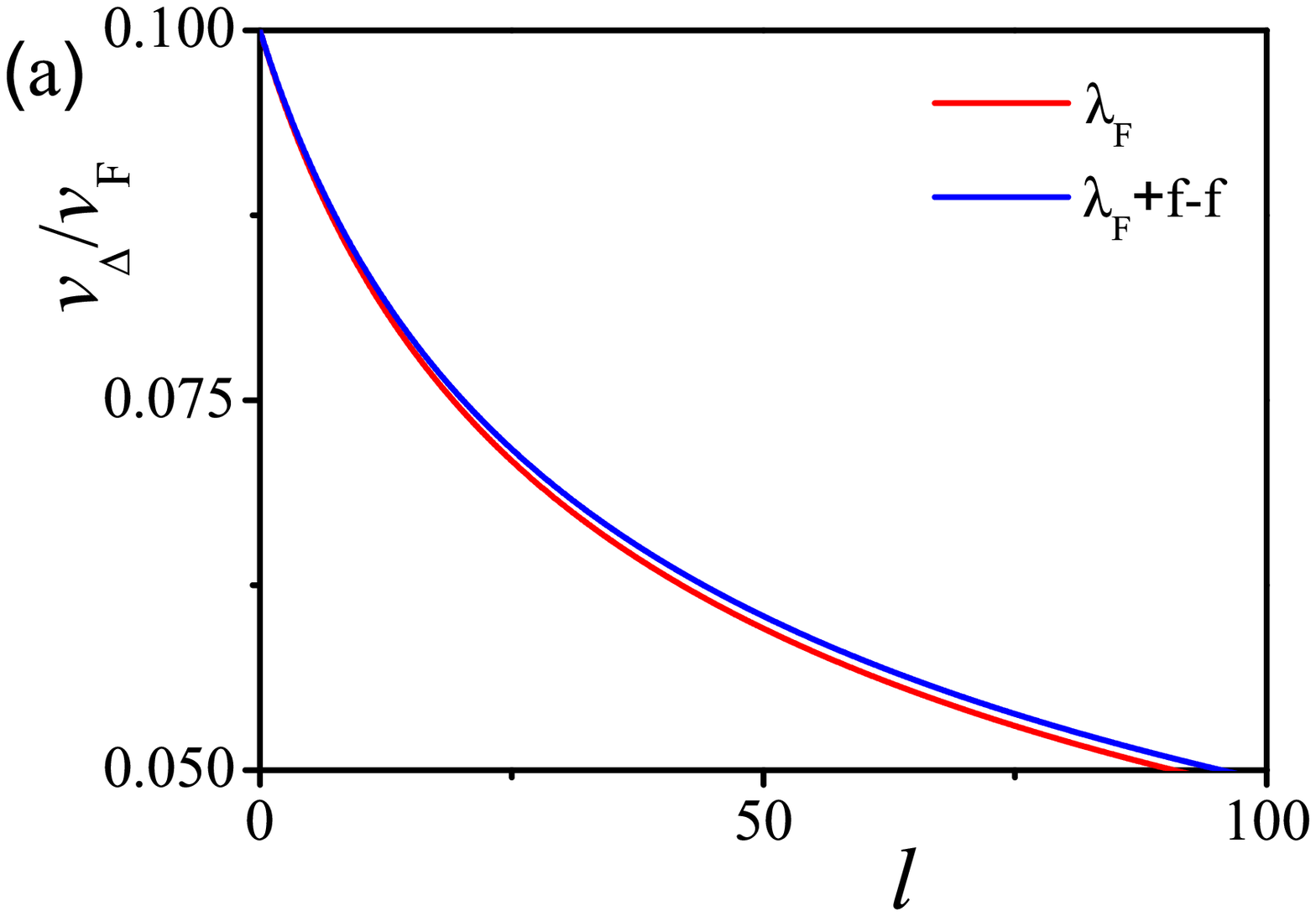}\vspace{-2.7cm}
\includegraphics[width=4in]{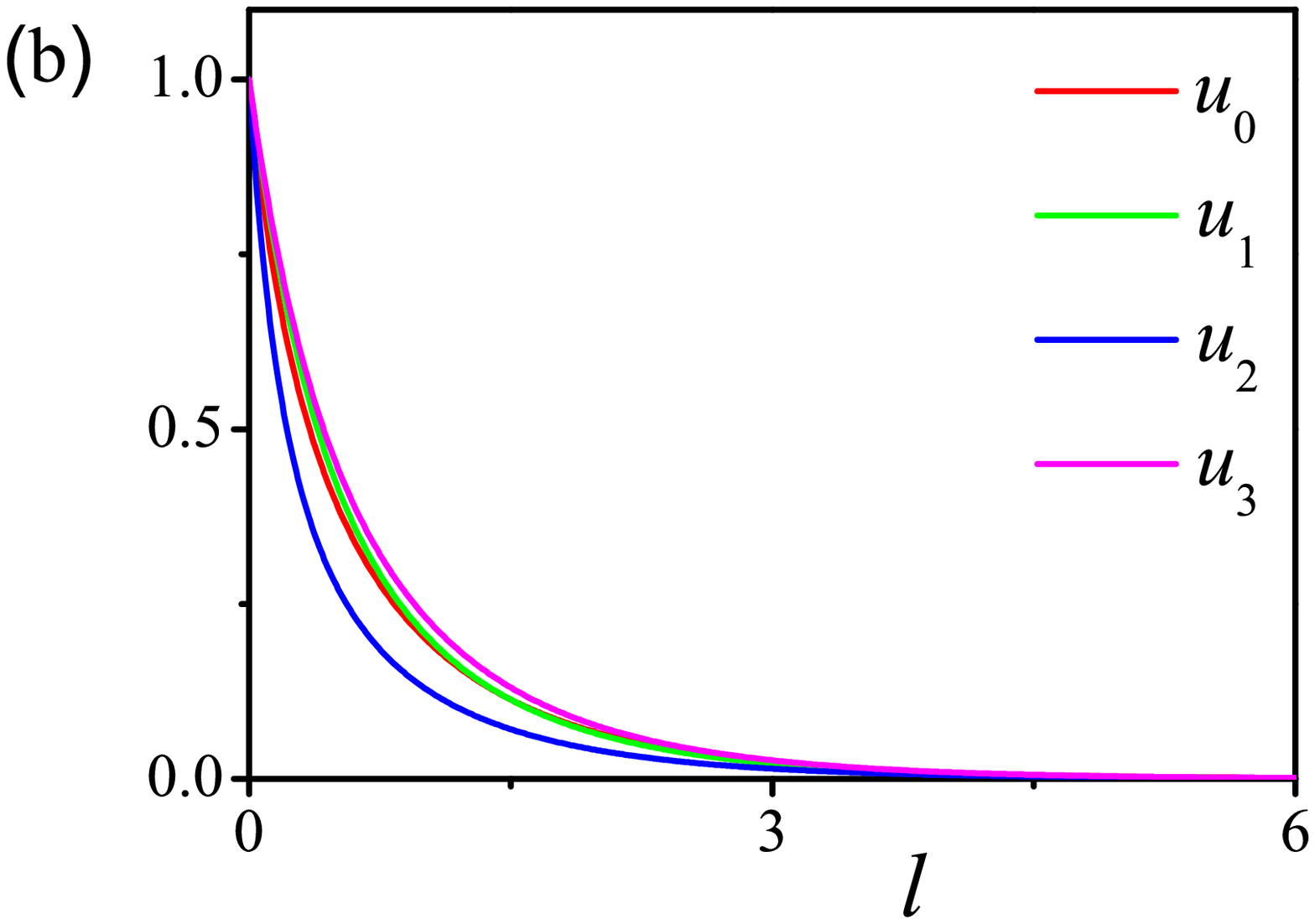}
\vspace{-2.7cm}
\caption{(Color online) Effects of fermion-fermion interactions for the Type-$\tau_{x}$
QPT on (a) the behaviors of $v_{\Delta}/v_{F}$ at an initial value $v_{\Delta0}/v_{F0}=0.1$  ($\lambda_{F}$ and $\lambda_F+$f-f correspond to the absence and presence of
fermion-fermion interactions, respectively), and (b) the energy-dependent
evolutions of fermion-fermion interaction strengths $u_{0,1,2,3}$
(the basic tendencies for Type-$\tau_{y,z}$ QPTs are similar
and hence not shown here).}\label{Fig_x_u}
\end{figure}

To achieve this goal, we have to study the coupled RG equations,
which consist of $v_F$, $v_\Delta$, and $\lambda$ as well as
fermion-fermion interactions characterized by $u_{i}$ with $i=0,1,2,3$.
To proceed, we at first consider the Yukawa fixed-coupling case~\cite{Sachdev2008PRB,Wang2013PRB,Xu2008PRB,Wang2011PRB,Kim-Kivelson2008PRB}.
Learning from the RG equations of fermion
velocities~(\ref{Eq_x_vD})-(\ref{Eq_x_vD/vF}),
we can infer that the fermion-fermion interactions $u_{i}$ cannot
directly affect $v_{F}$ and $v_{\Delta}$, but rather only indirectly
modify them via entangling with the interaction parameter $\lambda$.
In other words, the fermion velocities would receive the contributions from
fermion-fermion interactions once the coupling $\lambda$ flows under the
RG equations. This implies that the low-energy properties of fermion velocities
for the fixed-coupling situation are adequately robust against
fermion-fermion interactions.
Next, our focus is moved to the situation with the
energy-dependent evolution of Yukawa-coupling $\lambda$.
After carrying out the numerical analysis of coupled
RG equations~(\ref{Eq_x_vD})-(\ref{Eq_x_u_3}), we present the comparison between
the absence and presence of fermion-fermion interactions
in Fig.~\ref{Fig_x_u}(a) as approaching the
Type-$\tau_{x}$ QPT. It can be seen from Fig.~\ref{Fig_x_u}(a) that
the ratio of fermion velocities $v_{\Delta}/v_{F}$ under
fermion-fermion interactions shares the same downtrend with
its behaviors in the absence of fermion-fermion interactions.
However, one can unambiguously realize that fermion-fermion
interactions do bring considerable quantitative effects, which
are in favor of retarding the $v_{\Delta}/v_{F}$'s decrease
as the energy scale is lowered. These are consistent with the
fact exhibited in Fig.~\ref{Fig_x_u}(b) that the fermion-fermion
interactions $u_{i}$ with $i=0,1,2,3$ are all irrelevant to
one-loop level, which become less and less important
as the energy scale is decreased~\cite{Makhfudz2015AP}.
Paralleling above analysis to Type-$\tau_{y,z}$ QPTs gives rise to
the qualitative agreements with their Type-$\tau_{x}$'s counterpart.
As a consequence, the fermion-fermion interactions
would provide non-ignorable contributions
to the fermion velocities in the vicinity of a putative QPT although
they do not play a crucial role compared to the
quantum fluctuations of order parameters.

Before closing this section, a brief summary
is delivered as follows. With the help of coupled RG equations around
the putative QPTs, we systematically investigate the effects of quantum
fluctuations and fermion-fermion interactions on the low-energy behaviors of
fermion velocities and potential fixed points at the
lowest-energy limit. On one hand, we turn off the fermion-fermion interactions and then
notice that there exist two fixed points of $v_{\Delta}/v_{F}$
for Type-$\tau_{0}$ QPT, which are independent upon
the evolution of Yukawa coupling. In sharp comparison,
the fixed points for Type-$\tau_{x,y,z}$ QPTs derived at
a fixed-coupling  $\lambda=1$~\cite{Sachdev2008PRB, Wang2013PRB} are
seriously modified whilst the Yukawa coupling is involved in the
coupled RG equations. To be specific, the extreme anisotropy of
fermion velocities for Type-$\tau_{x,z}$ is broken and replaced
by some finite anisotropy. As to the Type-$\tau_{y}$ QPT,
the evolution of coupling $\lambda$ drives the isotropic system
into a finite anisotropic fixed point.
On the other hand, we find that the fates of fermion velocities are
principally robust against the fermion-fermion interactions although
certain quantitative effects are generated to retard
the tendencies flowing towards potential fixed points. Subsequently,
it is ready to examine the consequences of these
unusual behaviors of fermion velocities on the quantum criticality
of physical observables.

\section{Superfluid density and critical temperature }\label{Sec_rho-s}

Quantum criticality of fermion velocities around a putative QCP
is carefully studied and detailedly addressed in the previous section
~\ref{Sec_velocity} after simultaneously collecting the
quantum fluctuations of order parameters and fermion-fermion
interactions. In order to present these unique behaviors of fermion velocities
that are inconvenient to be detected directly, one can resort to
examining the low-energy physical observables
in that the fermion velocities plus their ratio $v_{\Delta}/v_{F}$ usually enter into
the physical quantities and play an important role in the low-energy regime~\cite{Orenstein2000Science,Lee1997PRL,Lee1993PRL,Durst2000PRB,
Mesot1999PRL,Vojta2009AP}. This therefore provides us
a useful routine to study the distinctions among different
QPTs and the very positions of QCPs.

For this purpose, we within this section concentrate on the properties of
superfluid density and critical temperature upon accessing the QCPs~\cite{Lee1997PRL,Xu2008PRB,Wang2013PRB,Wang2015PRB,She2015PRB,
Kim-Kivelson2008PRB,Wang2017PRB,Wang2007.14981}, which are two of
the most key quantities of superconductors.
In order to simplify our analysis, the effects of fermion-fermion interactions
are hereafter not considered since they are always subordinate to
the Yukawa couplings between nodal QPs and
order parameters~\cite{Makhfudz2015AP} and hence cannot alter the
basic results caused by the quantum fluctuations
as discussed in Sec.~\ref{Subsec_ff}. Rather, we primarily
try to examine how these two quantities behave under
distinct fates of fermion velocities with approaching the assumed QCPs
which are induced by the fermion-order parameter couplings and
explicitly presented in Sec.~\ref{Subsec_absence-ff}.

\subsection{Superfluid density and critical temperature nearby the QCP}\label{Sec_rho_s-Eqs}

Generally, the zero-temperature superfluid density of
$d$-wave superconductor in underdoped region depends linearly
on doping concentration $x$ and can be written as~\cite{Hardy1993PRL,Orenstein1990PRB}
\begin{eqnarray}
\rho^{s}(0)=\frac{x}{a^{2}},\label{Eq_rho_0}
\end{eqnarray}
where $a$ stands for the lattice spacing constant.
To proceed, it is inevitable that a certain amount of normal
nodal QPs would be thermally excited out from the SC condensate
at a finite temperature, which can efficiently deplete the
superfluid density~\cite{Lee1997PRL}. As a result, the
temperature-dependent superfluid density can be expressed as~\cite{Lee1997PRL}
\begin{eqnarray}
\rho^{s}(T)=\rho^{s}(0)-\rho^{n}(T),
\end{eqnarray}
where $\rho^{s}(T)$ and $\rho^{n}(T)$ serve as the superfluid
density and normal QPs density at $T>0$, respectively~\cite{Lee1997PRL,Liu2012PRB,
Wang2013PRB,Wang2015PRB}. In the non-interacting situation,
the normal QPs density exhibits a linear temperature
dependence~\cite{Hardy1993PRL} and takes the form
of~\cite{Lee1997PRL}
\begin{eqnarray}
\rho^{n}(T)=m\frac{2\ln2}{\pi}\frac{v_{F}}{v_{\Delta}}T,\label{Eq_rho_n}
\end{eqnarray}
with the parameter $m$ being the mass of nodal QP.

Hereby, it is of interest to address
Bo\v{z}ovi\'{c} \emph{et al.}~\cite{Bollinger2016Nature}
recently reported that the dependence of the zero-temperature superfluid density
on the critical superconducting temperature for the overdoped region
can change from linear to parabolic as the critical temperature is below
a very value about $12$~K.  However, we emphasize that above
formula~(\ref{Eq_rho_0})-(\ref{Eq_rho_n}) are believed to
capture the crucial information of the underdoped and
optimal regions as displayed in Fig.~\ref{fig1}.
In such scenario, the superfluid density decreases as the temperature is lifted and
thus the critical temperature can be
explicitly derived via assuming $\rho^{s}(T)=0$ at $T_c$,
\begin{eqnarray}
T_c=\frac{1}{2\ln2}\frac{v_\Delta}{v_F}\frac{x}{ma^2},
\end{eqnarray}
which is well consistent with the Uemura plot~\cite{Uemura1989PRL}.
This indicates $T_c$ is readily obtained for the region away from the QCPs,
in which the ratio of fermion velocities for noninteracting nodal QPs takes a constant,
for instance $v_{\Delta}/v_{F}\approx0.1$ for $\mathrm{YBa_{2}Cu_{3}O_{6+\delta}}$~\cite{Orenstein2000Science,Fournier2000PRB}.

\begin{figure}
\centering
\includegraphics[width=5in]{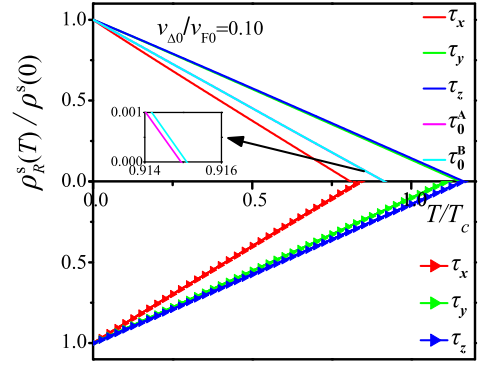}
\vspace{-3.76cm}
\caption{(Color online) The renormalized superfluid densities
and critical temperatures in the vicinity of Type-$\tau_{x,y,z,0}$
QCPs for both the fixed Yukawa coupling (bare lines) and
energy-dependent $\lambda$ (arrowed lines) cases at
a representative initial value $v_{\Delta 0}/v_{F0}=0.1$.
Hereby, $T_{c}$ denotes the critical temperature in the absence of any QCP
and the basic results are independent of the initial conditions as
depicted in Fig.~\ref{Fig_Tc_xyz-variation}.}\label{Fig_Tc_xyz0}
\end{figure}

In comparison, the involved physics is much more complicated but rather interesting
in the vicinity of certain QCP depicted in Fig.~\ref{fig1}. As systematically
addressed in Sec~\ref{Sec_velocity}, the fermion velocities $v_F$ and
$v_\Delta$ as well as other interaction parameters in the effective theory
with approaching the QCPs are heavily renormalized by ferocious quantum fluctuations
and become energy-dependent under the control of the coupled RG
equations in Sec.~\ref{Sec_RG-Eqs}. It is worth emphasizing that the ratio
of fermion velocities, which is directly related to the superfluid density as
delineated in Eq.~\ref{Eq_rho_n}, exhibits a cornucopia of energy-dependent behaviors
and flows towards several fixed points at the lowest-energy limit.
With these respects, in order to capture the effects of quantum criticality,
we take into account the renormalized fermion velocities
and follow the approach in Refs.~\cite{Lee1997PRL,Durst2000PRB,Liu2012PRB,Wang2013PRB,Wang2015PRB}
to construct the following renormalized normal
QPs density,
\begin{eqnarray}
\rho^{n}_{R}(T)=\frac{4m}{k_{B}T}\int^\Lambda\frac{d^{2}\mathbf{\mathbf{k}}}{(2\pi)^{2}}
\frac{v_{F}^{2}(k)e^{\frac{\sqrt{v_{F}^{2}(k)k_{x}^{2}
+v_{\Delta}^{2}(k)k_{y}^{2}}}{k_{B}T}}}
{\left(1+e^{\frac{\sqrt{v_{F}^{2}(k)k_{x}^{2}
+v_{\Delta}^{2}(k)k_{y}^{2}}}{k_{B}T}}\right)^{2}},\label{Eq_Tc_vF_vD_flow}
\end{eqnarray}
where $k_B$ denotes the Boltzmann constant and $v_{\Delta,F}$  are dictated
by associated RG equations in Sec.~\ref{Sec_RG-Eqs}. This henceforth
yields to the renormalized superfluid density
\begin{eqnarray}
\rho^{s}_{R}(T)=\rho^{s}(0)-\rho^{n}_{R}(T),\label{Eq_rho}
\end{eqnarray}
from which the renormalized critical temperature can be derived via
taking $\rho^{s}_{R}(T)=0$ at $T=T_c$.

As a consequence, Eq.~(\ref{Eq_Tc_vF_vD_flow}) together with Eq.~(\ref{Eq_rho})
signal that both superfluid density and critical temperature
are intimately associated with the energy-dependent fermion velocities,
which are governed by the coupled RG evolutions in Sec.~\ref{Sec_RG-Eqs} and
display many peculiar properties for all four sorts of QPTs
as presented in Sec.~\ref{Sec_velocity}. In the rest of this section, we
are about to pin down the fates of superfluid density and critical
temperature at the lowest-energy limit for all types of QPTs in Fig.~\ref{fig1}.

\begin{figure}
\centering
\includegraphics[width=3in]{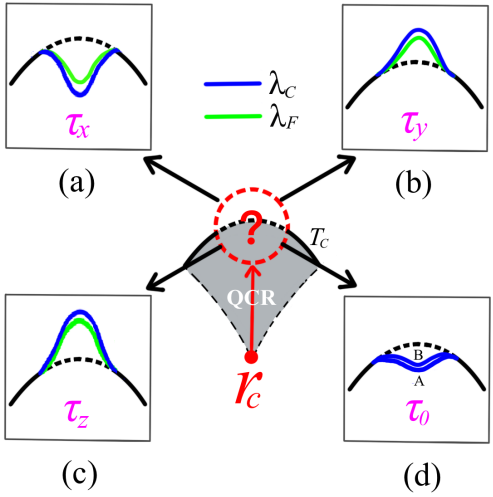}
\vspace{0.2cm}
\caption{(Color online) Schematic collections for the fates of
critical temperatures caused by quantum fluctuations around
the Type-$\tau_{x,y,z,0}$ QCPs as illustrated in Fig.~\ref{fig1}
in the absence ($\lambda_C$) and presence ($\lambda_F$) of evolution
of Yukawa coupling.}\label{Fig_Schem_Tc_xyz0}
\end{figure}

\begin{figure*}
\hspace{0cm}
\includegraphics[width=5.2in]{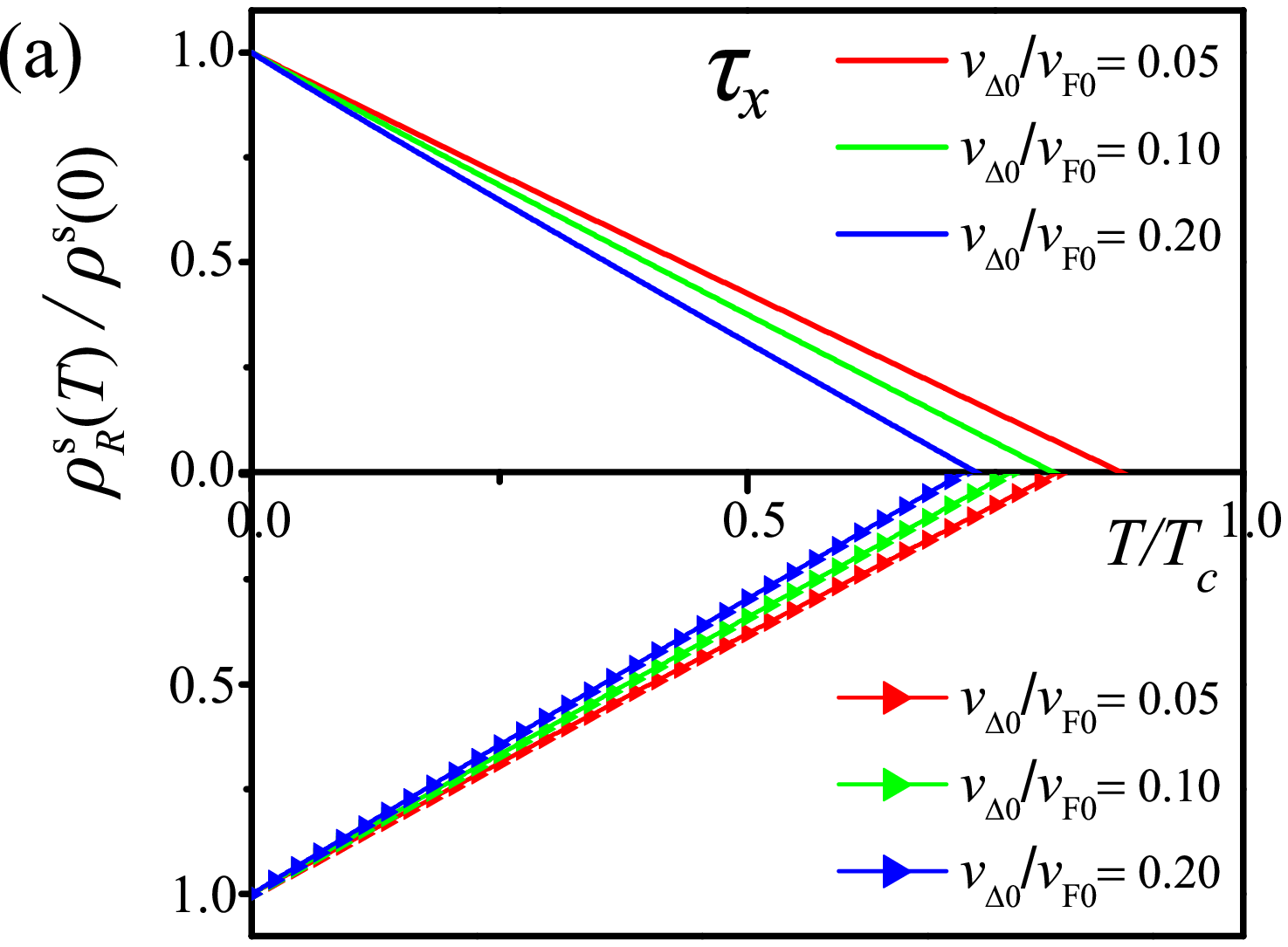}\hspace{-8.7cm}
\includegraphics[width=5.2in]{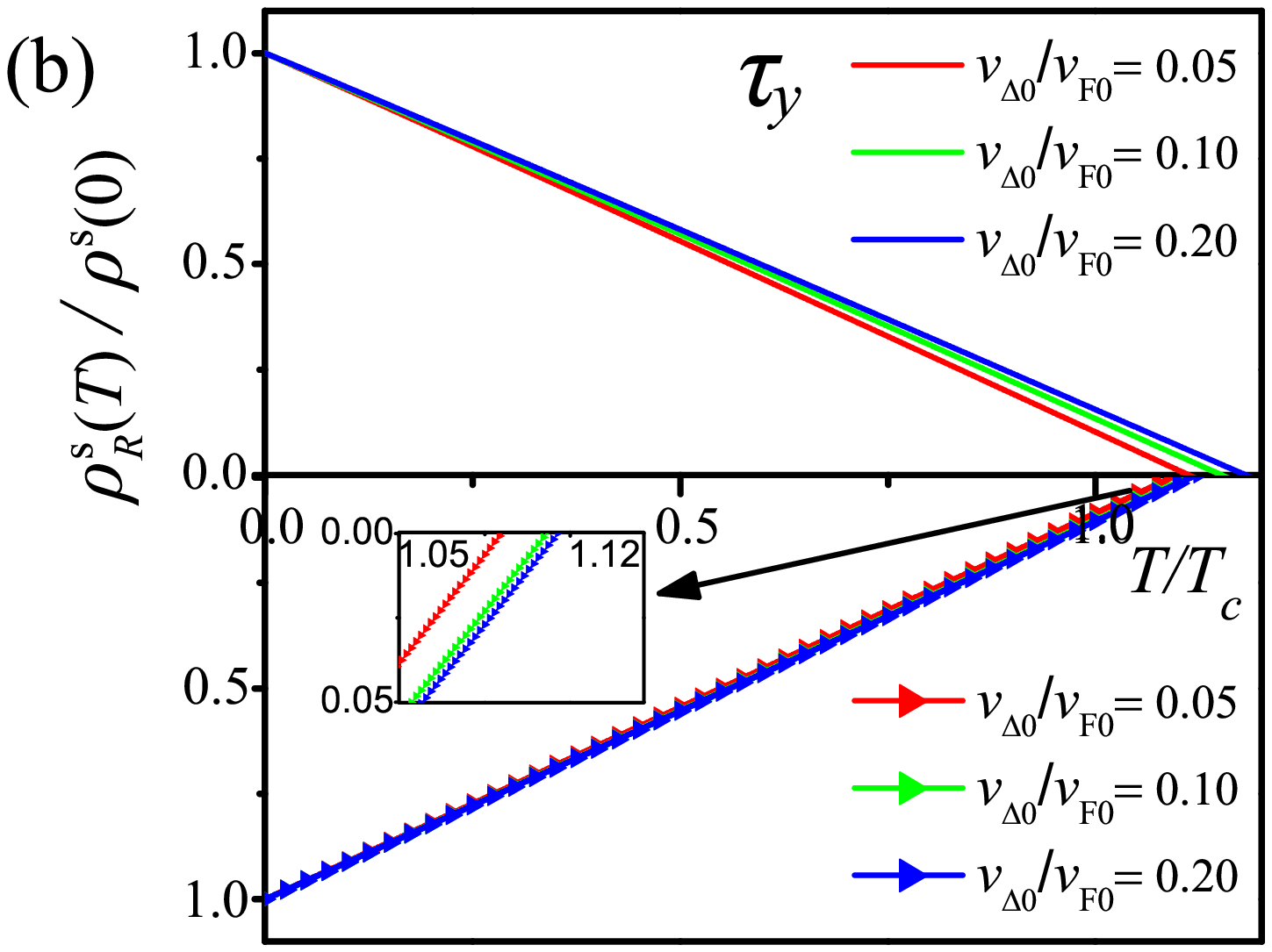}\vspace{-3.5cm}
\hspace{3.0cm}
\includegraphics[width=5.2in]{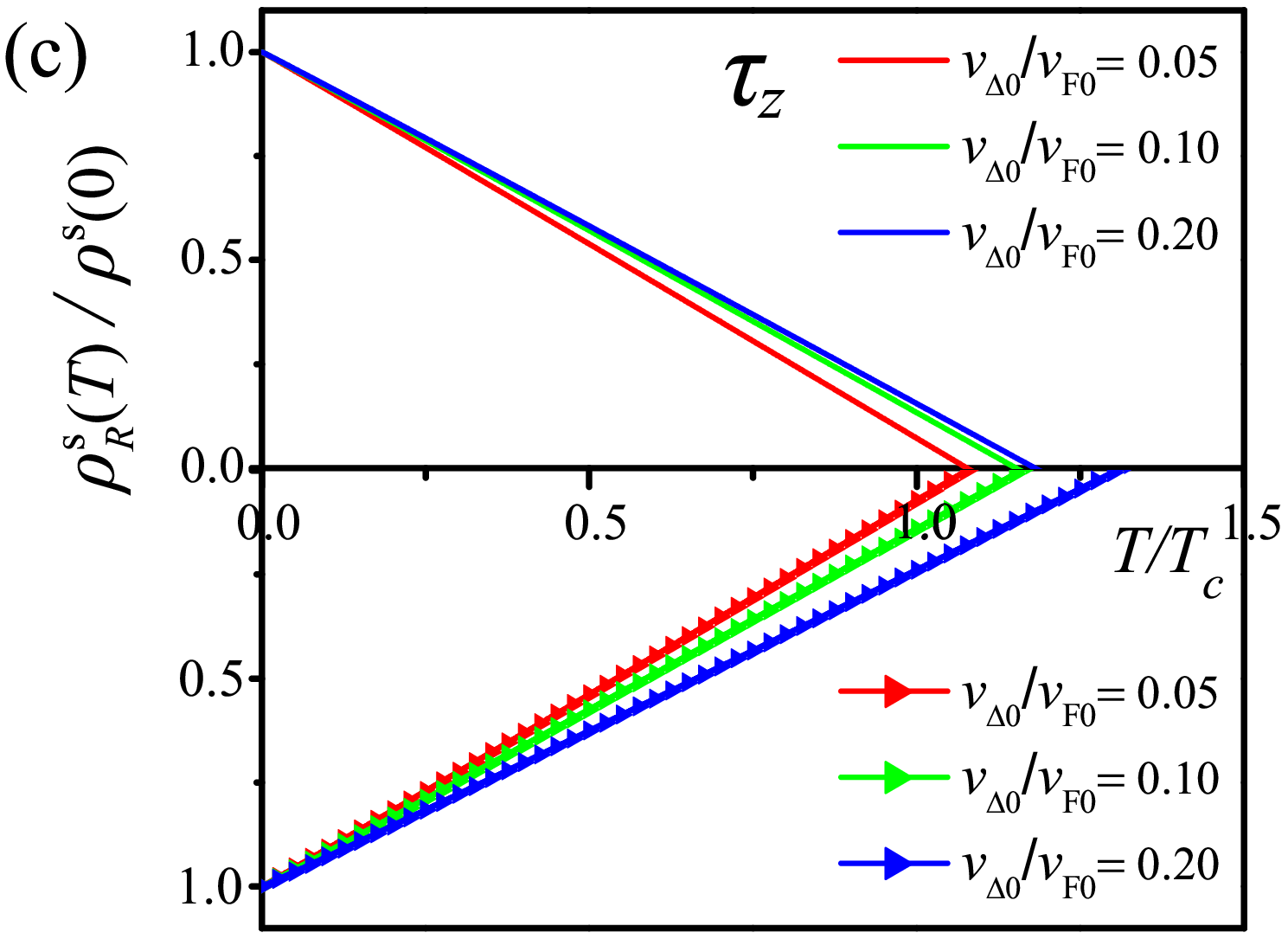}\hspace{-8.7cm}
\includegraphics[width=5.2in]{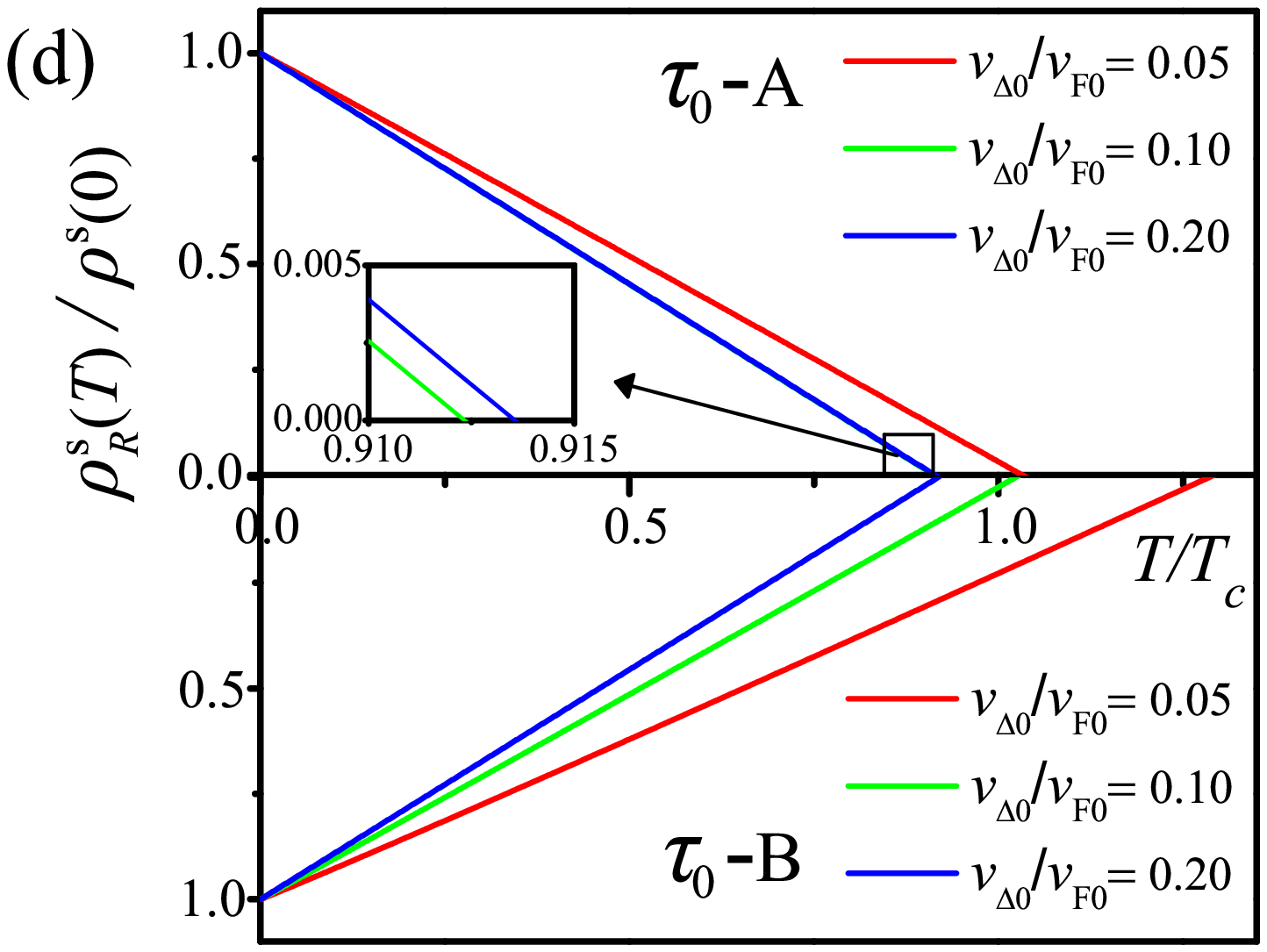}\vspace{-3cm}
\caption{(Color online) The stabilities of renormalized
superfluid densities and critical temperatures against the variation
of initial conditions upon approaching (a) Type-$\tau_{x}$,
(b) Type-$\tau_{y}$, (c) Type-$\tau_{z}$ and (d) Type-$\tau_{0}$
QPTs for both the fixed Yukawa coupling (bare lines) and
energy-dependent $\lambda$ (arrowed lines) circumstances.}\label{Fig_Tc_xyz-variation}
\end{figure*}

\subsection{Fates at $v_{\Delta0}/v_{F0}=0.1$}\label{Sec-fate-0.1}

On the basis of analysis in Sec.~\ref{Sec_rho_s-Eqs},
$v_{\Delta}/v_{F}$ plays a central role in determining the fates of
both the superfluid density and critical temperature as explicitly
displayed in Eq.~(\ref{Eq_Tc_vF_vD_flow}) and Eq.~(\ref{Eq_rho}).
In order to obtain the energy-dependent $v_{\Delta}/v_{F}$, it enables us to
fix its initial value and then carry out the numerical evaluation of
coupled RG equations in Sec.~\ref{Sec_RG-Eqs}.
Without loss of generality, we hereby place our primary focus on the
initial condition with $v_{\Delta0}/v_{F0}=0.1$ as such ratio appears
in most of high-$T_{c}$ superconductors~\cite{Orenstein2000Science,
Fournier2000PRB,Lee2006RMP}, and then discuss the stability
of basic results against the initial conditions in the following subsection.

To proceed, performing the numerical calculations of coupled RG
equations for all types of QPTs with such starting condition and inserting them
into Eqs.~(\ref{Eq_Tc_vF_vD_flow})-(\ref{Eq_rho}) give rise to
the critical behaviors of superfluid density and critical temperature for
distinct types of QPTs as collected in Fig.~\ref{Fig_Tc_xyz0}, in which
the $T_c$ denotes the critical temperature in the absence of a putative QCP.
It is noteworthy that the intersections of the curves $\rho_{R}^{s}(T)/\rho^{s}(0)$ with the horizontal axis $T/T_{c}$ characterize the renormalized critical temperatures after
incorporating the quantum criticality of the related QPTs. In addition,
given the low-energy tendencies of fermion velocities are of particular difference,
we utilize the $\bar{T}_{c}$ and $\tilde{T}_{c}$ to specify the critical temperatures
for a fixed Yukawa coupling ($\lambda=1$) and an evolution of $\lambda$, respectively.
The details are addressed as follows.

At first, we consider the Type-$\tau_{0}$ QPT.
In this case, the Yukawa coupling $\lambda$ mentioned
in Sec.~\ref{Subsubsec_fixed-lambda} is marginal to one-loop level
and henceforth it does not evolve with lowering the energy scales.
As a result, we only need to take into account the situation with $\lambda=1$.
It can be inferred from Fig.~\ref{Fig_Tc_xyz0}
that $\bar{T}^{0A}_{c}<\bar{T}^{0B}_{c}<T_{c}$, which is indicative of
the suppression of superconductivity for both the Type-$\tau_{0A}$
and Type-$\tau_{0B}$ components as schematically
illustrated in Fig.~\ref{Fig_Schem_Tc_xyz0}(d). Next, we turn to the Type-$\tau_{x}$ QPT.
In distinction to the Type-$\tau_{0}$ case, $v_{\Delta}/v_{F}$ with a running $\lambda$
exhibits very different behaviors compared to its fixed-coupling
($\lambda=1$) counterpart~\cite{Sachdev2008PRB} as presented in Sec.~\ref{subsub-flow-lambda}. In particular, the extreme
anisotropy of fermion velocities is broken due to the
evolution of Yukawa coupling as shown in Fig.~\ref{fig_tau_x}.
As a result, three sorts of critical temperatures are restricted
to follow $\bar{T}^{x}_{c}<\tilde{T}^{x}_{c}<T_{c}$ as depicted in
Fig.~\ref{Fig_Tc_xyz0}. In other words, despite of the suppression of
superconductivity, the flowing of $\lambda$ as portrayed in
Fig.~\ref{Fig_Schem_Tc_xyz0}(a) prefers to hinder the
decrease of critical temperature. At last, let us move to
the Type-$\tau_{y}$ and Type-$\tau_{z}$ QPTs,
in which the basic tendencies of critical temperatures are
analogous owing to the quantum fluctuations. At a fixed coupling
$\lambda=1$, fermion velocities are driven to
the isotropic situation for Type-$\tau_{y}$ QPT~\cite{Wang2013PRB}
but another extreme anisotropy with $v_{F}/v_{\Delta}\rightarrow0$
for Type-$\tau_{z}$ QPT~\cite{Sachdev2008PRB}.
Accordingly, Fig.~\ref{Fig_Tc_xyz0} presents that these cause a
little promotion for superconductivity with $T_{c}<\bar{T}^{y}_{c}
<\bar{T}^{z}_{c}$. In comparison, Sec.~\ref{subsub-flow-lambda} shows that
both the isotropic and extremely anisotropic fermion velocities are destroyed
by the evolution of coupling $\lambda$ but instead $v_{\Delta}/v_{F}$ are
attracted by some finite values at the lowest-energy limit. As a consequence, Fig.~\ref{Fig_Tc_xyz0} displays that critical temperatures of both cases
reduce. Although the relationship of $T_{c}<\tilde{T}^{y}_{c}
<\tilde{T}^{z}_{c}$ is preserved, the critical temperature of
Type-$\tau_{y}$ QPT falls a little more than that of Type-$\tau_{z}$ QPT as
illustrated in Fig.~\ref{Fig_Schem_Tc_xyz0}(b) and
Fig.~\ref{Fig_Schem_Tc_xyz0}(c).

To recapitulate, we come to a conclusion that the superconductivity
is enhanced nearby the Type-$\tau_{y}$ and Type-$\tau_{z}$ QPTs and conversely suppressed
in the proximity of the Type-$\tau_{0A}$, Type-$\tau_{0B}$ and Type-$\tau_{x}$ QPTs. The renormalized critical temperatures under these QPTs in Fig.~\ref{fig1} are followed by $\bar{T}^{z}_{c}>\bar{T}^{y}_{c}>T_{c}>\bar{T}^{0B}_{c}
>T^{0A}_{c}>\bar{T}^{x}_{c}$, which are schematically
summarized in Fig.~\ref{Fig_Schem_Tc_xyz0}.

\subsection{Stability of $T_{c}$ against $v_{\Delta0}/v_{F0}$}

For the sake of completeness, we are now in a suitable position to inspect the
stability of conclusions concerning the critical temperatures in Sec.~\ref{Sec-fate-0.1}
under the variation of initial condition $v_{\Delta0}/v_{F0}$ as approaching distinct
types of QPTs. There exist two points behind this issue as follows.
Although the final fixed points are considerably insensitive to the starting values of
fermion velocities as studied in Sec.~\ref{Sec_velocity}, one can learn from Eq.~(\ref{Eq_Tc_vF_vD_flow}) and Eq.~(\ref{Eq_rho})
that critical temperature depends not only upon the contributions from
the fixed point but also upon the whole low-energy regime.
In addition, although the $v_{\Delta0}/v_{F0}$
approximately equals $0.1$ in most of high-$T_{c}$ superconductors~\cite{Orenstein2000Science,Fournier2000PRB,Lee2006RMP},
this initial value is inevitable to be
affected by various uncontrollable and unexpected factors in real materials.

To proceed, we select three representative initial values
$v_{\Delta0}/v_{F0}=0.05, 0.1, 0.2$ to examine whether and how
the fates of critical temperatures nearby distinct types of QPTs
are renormalized by the initial conditions. In order to achieve
this end, we parallel the analogous procedures in Sec.~\ref{Sec-fate-0.1}
with the help of associated RG equations and then obtain the
main results collected in Fig.~\ref{Fig_Tc_xyz-variation}.
At the first sight, we figure out that the effects caused
by the variation of $v_{\Delta0}/v_{F0}$ on a
fixed-coupling $\lambda=1$ circumstance share
the qualitative results with that of the evolution of $\lambda$ case.
To be concrete, the critical temperatures around Type-$\tau_{x}$
and Type-$\tau_{0}$ QPTs are susceptible to the initial values of
$v_{\Delta}/v_{F}$ and present a little downtrends upon the increase
of $v_{\Delta0}/v_{F0}$ albeit the stability for suppression of
superconductivity displayed in Fig.~\ref{Fig_Tc_xyz-variation}(a)
and Fig.~\ref{Fig_Tc_xyz-variation}(d). In comparison, we can learn from Fig.~\ref{Fig_Tc_xyz-variation}(b) and Fig.~\ref{Fig_Tc_xyz-variation}(c)
for Type-$\tau_{y}$ and Type-$\tau_{z}$ QPTs that both of the critical
temperatures receives a certain mount of enhancements
with tuning up the value of $v_{\Delta0}/v_{F0}$. Meanwhile, the basic
restriction between the bare and renormalized critical temperature with $\tilde{T}_{c}>\bar{T}_{c}>T_{c}$ is insensitive to the initial condition.
This suggests that both the initial values of fermion velocities and
evolution of Yukawa coupling are subordinate to unusual behaviors of
the fermion velocities which are crucial to pin down the critical temperatures
around the QPT.

To be brief, the fates of critical temperatures are of particular robustness
against the initial values of fermion velocities in the proximity of the putative
QCPs. In other words, the low-energy properties of the fermion velocities that are
the external expressions of quantum criticality triggered by the QPTs play a more
significant role than initial condition in determining the superfluid density and
critical temperature. Accordingly, these distinct fates of critical temperatures schematically shown in Fig.~\ref{Fig_Schem_Tc_xyz0} are closely associated
with different sorts of QPTs and henceforth provide a helpful clue to experimentally
detect the very QPT and fix its location.

\section{Summary}\label{Sec_summary}

In summary, we study the low-energy fates of fermion velocities
and behaviors of superfluid density as well as critical temperature nearby
the putative QPTs in $d$-wave superconductors, which stem from
the topological change of nodal points~\cite{Vojta2000PRL,Vojta2000PRB,Vojta2000IJMPB}.
In order to facilitate the analysis, seven candidates of potential QPTs shown in Fig.~\ref{fig1} cluster into four effective categories, which are designated as the
Type-$\tau_{0}$, Type-$\tau_{x}$, Type-$\tau_{y}$ and Type-$\tau_{z}$ QPTs in Sec~\ref{Sec_eff-theory}~\cite{Vojta2000PRL,Wang2013PRB}.
By means of the momentum-shell RG approach~\cite{Shankar1994RMP,Wilson1975RMP,Polchinski1992},
all primary physical ingredients including quantum fluctuations
of order parameters, the couplings between order parameters and nodal QPs,
and fermion-fermion interactions can be equally captured and
encoded in a set of coupled RG equations after taking into account
one-loop corrections in Sec.~\ref{Sec_RG-Eqs}.
On the basis of these RG equations, we, with the help of both analytical
and numerical evaluations, achieve the main results concerning the
fixed points of fermion velocities and related physical quantities in the vicinity of
QPTs.

To be concrete, the fermion velocities exhibit a number of critical
properties caused by the effects of quantum fluctuations and
fermion-fermion interactions, which are expected to be in charge
of the low-energy fates around the underlying four types of QPTs.
At first, the focus is put on the fermion velocities.
Besides three distinct fixed points of fermion velocities obtained at
the fixed Yukawa coupling $\lambda=1$, including $v_{\Delta}/v_{F}\rightarrow0$ for Type-$\tau_{x}$~\cite{Sachdev2008PRB}, $v_{\Delta}/v_{F}\rightarrow1$ for Type-$\tau_{y}$~\cite{Wang2013PRB}, $v_{F}/v_{\Delta}\rightarrow0$ for Type-$\tau_{z}$~\cite{Wang2013PRB}, it is of particular interest to
point out that a series of new fixed points are generated due to the
interplay between evolution of Yukawa coupling $\lambda$ together with
other interaction parameters. As for the Type-$\tau_{0}$ QPT,
we notice that the ratio of fermion velocities is attracted by
either fixed point $(v_{\Delta}/v_{F})^*\approx0.0942$ or $(v_{\Delta}/v_{F})^*\approx0.3478$, which corresponds to
Type-$\tau_{0A}$ or Type-$\tau_{0B}$ component
and is insensitive to initial conditions. In comparison, the fixed points
for Type-$\tau_{x,y,z}$ QPTs at a fixed-coupling
$\lambda=1$~\cite{Sachdev2008PRB, Wang2013PRB} are
manifestly reshaped. Two kinds of extreme anisotropies of fermion
velocities for Type-$\tau_{x,z}$ QPTs are both broken and
replaced by finite anisotropies. Meanwhile, the isotropic fermion
velocities for Type-$\tau_{y}$ QPT are driven to a finite anisotropic
fixed point by the evolution of coupling $\lambda$. This indicates
that fermion velocities prefer to flow towards a finite anisotropy
as approaching a putative QPT. In addition to these results
caused by the evolution of Yukawa coupling $\lambda$, we examine
the effects of fermion-fermion interactions on fermion velocities as well,
which have not yet been considered seriously. Despite they
are subordinate to the quantum fluctuations of order parameters~\cite{Makhfudz2015AP},
it is noteworthy that the fermion-fermion interactions as shown
in Fig.~\ref{Fig_x_u} can bring non-ignorable quantitative
contributions to the fermion velocities in the vicinity
of a putative QPT. Next, both of superfluid density
and critical temperature, which are two of the most important observables
for superconductors, are carefully investigated
under the unconventional behaviors of fermion velocities around
all potential QPTs. Concretely, after combining the expressions
of these two observables that are dependent upon the fermion velocities
and the coupled RG equations of all interaction parameters,
we notice that the renormalized critical temperatures are restricted by $\bar{T}^{z}_{c}>\bar{T}^{y}_{c}>T_{c}>\bar{T}^{0B}_{c}
>T^{0A}_{c}>\bar{T}^{x}_{c}$ as schematically illustrated
in Fig.~\ref{Fig_Schem_Tc_xyz0} for all types of QPTs. In other words,
both Type-$\tau_{y}$ and Type-$\tau_{z}$ QPTs are in favor of the
superconductivity but rather the critical temperature is suppressed by
Type-$\tau_{0A}$, Type-$\tau_{0B}$ and Type-$\tau_{x}$ QPTs. In addition,
we check that the fates of critical temperatures are primarily determined
by unusual behaviors of the fermion velocities and considerably robust
against the initial values of fermion velocities.

Our results systematically account for the quantum criticality of
both fermion velocities and critical temperatures under the competition among
quantum fluctuations and interplay between nodal QPs and order parameters as well
as the fermion-fermion interactions near all potential QPTs in $d$-wave
superconductors. In particular, an underlying strategy is provided to
experimentally seek the putative QPTs and locate their very positions
by virtue of qualitatively distinct behaviors nearby different
types of QPTs. Additionally, this may offer an operable strategy to
classify the superconducting materials with distinct behaviors of critical temperatures.
What is more, these theoretical results may stimulate experimental
scientists to check and seek other potential critical physics
nearby these QCPs, as well as further explore the possible relationships between anomalous
properties in the normal state with $T>T_c$ and the quantum fluctuations.
To recapitulate, we anticipate that these instructive results would be helpful to improve
our understandings of the quantum criticality and structure of phase diagram
in the $d$-wave superconductors.

\section*{ACKNOWLEDGEMENTS}

X.Y.R. thanks J. -Q. Li and W. -H. Bian for the helpful discussions.
J.W. is partially supported by the National Natural
Science Foundation of China under Grant No. 11504360.

\section*{AUTHOR CONTRIBUTIONS}

J. W. initiated and supervised the project as well as performed the numerical analysis and wrote the manuscript with the assistance of the other two authors. X.Y. R. carried out the analytical calculations and plotted figures. Y. H. Z participated in the discussions and provided several useful suggestions.

\section*{ADDITIONAL INFORMATION}

\textbf{Competing interests:} The authors declare no Competing Financial or Non-Financial Interests.


\appendix

\section{One-loop corrections for Type-$\tau_y$, $\tau_z$, and $\tau_0$}\label{Appendix_correction}

\subsection{Self energy and vertex}\label{Appendix_Sigma-Gamma}

One-loop self energy as shown in Fig.~\ref{fig3}(b) receives the
corrections from the interplay between the nodal QPs and Type-$\mathcal{M}$
order parameter with $\mathcal{M}=\tau_{0,x,y,z}$ illustrated in Fig.~\ref{fig1}.
To be compact, we have just presented Type-$\tau_x$ in Eq.~\ref{Eq_Sigma-tau-x} of
Sec~\ref{Sec_correction}. In the following, the rest three types are collected after
integrating out the momentum shell within $b\Lambda-\Lambda$~
\cite{Sachdev2008PRB,Vafek2014PRB,Wang2017PRB,Wang2011PRB,Wang2013PRB,Vafek2012PRB,
She2010PRB, Kim-Kivelson2008PRB, She2015PRB, Roy-Sau2016PRB},
\begin{eqnarray}
\!\!\!\!\!\!\!\Sigma^{\tau_{y}}(\mathbf{k},\omega)\!\!\!
&=&\!\!\!\lambda^{2}\!\left[\mathcal{B}_{1}(-i\omega)
\!+\!\mathcal{B}_{2}v_{F}k_{x}\tau^{z}\!+\!\mathcal{B}_{3}
v_{\Delta}k_{y}\tau^{x}\right]\!l,\label{Eq_Sigma-y}\\
\!\!\!\!\!\!\!\Sigma^{\tau_{z}}(\mathbf{k},\omega)\!\!\!
&=&\!\!\!\lambda^{2}\!\left[\mathcal{C}_{1}(-i\omega)
\!+\mathcal{C}_{2}v_{F}k_{x}\tau^{z}\!+\mathcal{C}_{3}
v_{\Delta}k_{y}\tau^{x}\right]\!l,\label{Eq_Sigma-z}\\
\!\!\!\!\!\!\!\!\!\!\!\Sigma^{\tau_{0}}_A(\mathbf{k},\omega)\!\!\!\!
&=&\!\!\!\!\lambda^{2}\![\mathcal{D}^{A}_{1}(-i\omega)
\!+\!\mathcal{D}^{A}_{2}v_{F}k_{x}\tau^{z}\!
+\!\mathcal{D}^{A}_{3}v_{\Delta}k_{y}\tau^{x}]l,\label{Eq_Sigma-0-A}\\
\!\!\!\!\!\!\!\!\!\!\!\Sigma^{\tau_{0}}_B(\mathbf{k},\omega)\!\!\!\!
&=&\!\!\!\!\lambda^{2}\![\mathcal{D}^{B}_{1}(-i\omega)
\!+\!\mathcal{D}^{B}_{2}v_{F}k_{x}\tau^{z}\!
+\!\mathcal{D}^{B}_{3}v_{\Delta}k_{y}\tau^{x}]l,\label{Eq_Sigma-0-B}
\end{eqnarray}
where the indexes $A$ and $B$ denote the two components for Type-$\tau_0$.

With respect to the fermion-fermion interactions renormalized by
one-loop corrections as depicted in Fig.~\ref{Fig_1L-ff}, we
only provide the formal expression
in Eq.~(\ref{Eq_1L-ff-interaction}) of Sec.~\ref{Sec_correction}.
To remedy this, the details of the one-loop contributions are listed
as follows after practicing the strategy
in Refs.~\cite{Wang2017PRB,Wang2018JPCM,Wang2020PRB,Wang2021NPB,
Roy-Sau2016PRB,Mandal2018PRB, Roy2018PRX,
Roy-Saram2016PRB, Nandkishore2017PRB, Roy-Sau2017PRL,
Roy-Slager2018PRX, Roy2004.13043,Roy2021JHEP,
Roy2021PRB},
\begin{widetext}
\begin{eqnarray}
\Gamma^{\tau_{x}}_{u_{0}}
&=&u_{0}\int_{-\infty}^{\infty}
\frac{d\omega_{1}d\omega_{2}d\omega_{3}}{(2\pi)^{3}}\int^{b}
\frac{d^{2}\mathbf{k}_{1}d^{2}\mathbf{k}_{2}d^{2}\mathbf{k}_{3}}{(2\pi)^{6}}
\Psi^{\dag}_{\sigma}(\omega_{1},\mathbf{k}_{1})\tau_{0}
\Psi_{\sigma}(\omega_{2},\mathbf{k}_{2})
\Psi^{\dag}_{\sigma^{'}}(\omega_{3},\mathbf{k}_{3})\tau_{0}
\Psi_{\sigma^{'}}(\omega_{1}+\omega_{2}-\omega_{3},\mathbf{k}_{1}+\mathbf{k}_{2}
-\mathbf{k}_{3})\nonumber\\
&&\times\left[\frac{-(u_{1}u_{2}+u_{2}u_{3})}{4\pi u_{0}
v_{F}v_{\Delta}}+\frac{2\lambda^{2}}{3}\left(
\frac{u_{2}}{u_{0}}(\mathcal{A}_{3}-\mathcal{A}_{1})
-4\mathcal{A}_{1}\right)\right]l,\label{Eq_x0}\\
\Gamma^{\tau_{x}}_{u_{1}}
&=&u_{1}\int_{-\infty}^{\infty}
\frac{d\omega_{1}d\omega_{2}d\omega_{3}}{(2\pi)^{3}}\int^{b}
\frac{d^{2}\mathbf{k}_{1}d^{2}\mathbf{k}_{2}d^{2}\mathbf{k}_{3}}{(2\pi)^{6}}
\Psi^{\dag}_{\sigma}(\omega_{1},\mathbf{k}_{1})\tau_{1}
\Psi_{\sigma}(\omega_{2},\mathbf{k}_{2})
\Psi^{\dag}_{\sigma^{'}}(\omega_{3},\mathbf{k}_{3})\tau_{1}
\Psi_{\sigma^{'}}(\omega_{1}+\omega_{2}-\omega_{3},\mathbf{k}_{1}+\mathbf{k}_{2}
-\mathbf{k}_{3})\nonumber\\
&&\times\left[\frac{1}{4\pi v_{F}v_{\Delta}}(u_{0}-u_{1}-u_{2}-2u_{3}
+\frac{2u_{2}u_{3}}{u_{1}} )+\frac{2\lambda^{2}}{3}
\left(\frac{u_{3}}{u_{1}}(\mathcal{A}_{3}-\mathcal{A}_{1})
-4\mathcal{A}_{3}\right)\right]l,\label{Eq_x1}\\
\Gamma^{\tau_{x}}_{u_{2}}
&=&u_{2}\int_{-\infty}^{\infty}
\frac{d\omega_{1}d\omega_{2}d\omega_{3}}{(2\pi)^{3}}\int^{b}
\frac{d^{2}\mathbf{k}_{1}d^{2}\mathbf{k}_{2}d^{2}\mathbf{k}_{3}}{(2\pi)^{6}}
\Psi^{\dag}_{\sigma}(\omega_{1},\mathbf{k}_{1})\tau_{2}
\Psi_{\sigma}(\omega_{2},\mathbf{k}_{2})
\Psi^{\dag}_{\sigma^{'}}(\omega_{3},\mathbf{k}_{3})\tau_{2}
\Psi_{\sigma^{'}}(\omega_{1}+\omega_{2}-\omega_{3},\mathbf{k}_{1}+\mathbf{k}_{2}
-\mathbf{k}_{3})\nonumber\\
&&\times\left\{\frac{1}{4\pi v_{F}v_{\Delta}}\left[(2u_{0}-3u_{1}-2u_{2}-3u_{3})
+\frac{2u_{1}u_{3}
}{u_{2}}\right]
+\frac{2\lambda^{2}}{3}\left[4\mathcal{A}_{3}-5\mathcal{A}_{1}-5\mathcal{A}_{2}
+\frac{u_{3}}{u_{2}}(\mathcal{A}_{2}-\mathcal{A}_{3})
\right]\right\}l,\label{Eq_x2}\\
\Gamma^{\tau_{x}}_{u_{3}}
&=&u_{3}\int_{-\infty}^{\infty}
\frac{d\omega_{1}d\omega_{2}d\omega_{3}}{(2\pi)^{3}}\int^{b}
\frac{d^{2}\mathbf{k}_{1}d^{2}\mathbf{k}_{2}d^{2}\mathbf{k}_{3}}{(2\pi)^{6}}
\Psi^{\dag}_{\sigma}(\omega_{1},\mathbf{k}_{1})\tau_{3}
\Psi_{\sigma}(\omega_{2},\mathbf{k}_{2})
\Psi^{\dag}_{\sigma^{'}}(\omega_{3},\mathbf{k}_{3})\tau_{3}
\Psi_{\sigma^{'}}(\omega_{1}+\omega_{2}-\omega_{3},\mathbf{k}_{1}+\mathbf{k}_{2}
-\mathbf{k}_{3})\nonumber\\
&&\times\left\{\frac{1}{4\pi v_{F}v_{\Delta}}\left[(u_{0}-u_{3}-u_{1}-2u_{2})+\frac{2(
u_{1}u_{2})}{u_{3}}\right]
+\frac{2\lambda^{2}}{3}\left[
\frac{u_{2}}{u_{3}}(\mathcal{A}_{2}-\mathcal{A}_{3})
-\mathcal{A}_{1}-5\mathcal{A}_{2}\right]\right\}l,\label{Eq_x3}\\
\Gamma^{\tau_{y}}_{u_{0}}
&=&u_{0}\int_{-\infty}^{\infty}
\frac{d\omega_{1}d\omega_{2}d\omega_{3}}{(2\pi)^{3}}\int^{b}
\frac{d^{2}\mathbf{k}_{1}d^{2}\mathbf{k}_{2}d^{2}\mathbf{k}_{3}}{(2\pi)^{6}}
\Psi^{\dag}_{\sigma}(\omega_{1},\mathbf{k}_{1})\tau_{0}
\Psi_{\sigma}(\omega_{2},\mathbf{k}_{2})
\Psi^{\dag}_{\sigma^{'}}(\omega_{3},\mathbf{k}_{3})\tau_{0}
\Psi_{\sigma^{'}}(\omega_{1}+\omega_{2}-\omega_{3},\mathbf{k}_{1}+\mathbf{k}_{2}
-\mathbf{k}_{3})\nonumber\\
&&\times\left[\frac{-(u_{1}u_{2}+u_{2}u_{3})}{4\pi u_{0}v_{F}v_{\Delta}}
+\frac{2\lambda^{2}}{3}\left(-4\mathcal{B}_{1}
-\frac{u_{1}}{u_{0}}(\mathcal{B}_{1}+\mathcal{B}_{2})
+\frac{u_{2}}{u_{0}}(\mathcal{B}_{2}+\mathcal{B}_{3})
-\frac{u_{3}}{u_{0}}(\mathcal{B}_{1}+\mathcal{B}_{2})\right)\right]l,\label{Eq_y0}\\
\Gamma^{\tau_{y}}_{u_{1}}
&=&u_{1}\int_{-\infty}^{\infty}
\frac{d\omega_{1}d\omega_{2}d\omega_{3}}{(2\pi)^{3}}\int^{b}
\frac{d^{2}\mathbf{k}_{1}d^{2}\mathbf{k}_{2}d^{2}\mathbf{k}_{3}}{(2\pi)^{6}}
\Psi^{\dag}_{\sigma}(\omega_{1},\mathbf{k}_{1})\tau_{1}
\Psi_{\sigma}(\omega_{2},\mathbf{k}_{2})
\Psi^{\dag}_{\sigma^{'}}(\omega_{3},\mathbf{k}_{3})\tau_{1}
\Psi_{\sigma^{'}}(\omega_{1}+\omega_{2}-\omega_{3},\mathbf{k}_{1}+\mathbf{k}_{2}
-\mathbf{k}_{3})\nonumber\\
&&\times\left[\frac{1}{4\pi v_{F}v_{\Delta}}(u_{0}-u_{1}-u_{2}-2u_{3}
+\frac{2u_{2}u_{3}}{u_{1}} )+\frac{2\lambda^{2}}{3}
\left(\frac{u_{3}}{u_{1}}(\mathcal{B}_{2}+\mathcal{B}_{3})
-4\mathcal{B}_{3}\right)\right]l,\label{Eq_y1}\\
\Gamma^{\tau_{y}}_{u_{2}}
&=&u_{2}\int_{-\infty}^{\infty}
\frac{d\omega_{1}d\omega_{2}d\omega_{3}}{(2\pi)^{3}}\int^{b}
\frac{d^{2}\mathbf{k}_{1}d^{2}\mathbf{k}_{2}d^{2}\mathbf{k}_{3}}{(2\pi)^{6}}
\Psi^{\dag}_{\sigma}(\omega_{1},\mathbf{k}_{1})\tau_{2}
\Psi_{\sigma}(\omega_{2},\mathbf{k}_{2})
\Psi^{\dag}_{\sigma^{'}}(\omega_{3},\mathbf{k}_{3})\tau_{2}
\Psi_{\sigma^{'}}(\omega_{1}+\omega_{2}-\omega_{3},\mathbf{k}_{1}+\mathbf{k}_{2}
-\mathbf{k}_{3})\nonumber\\
&&\!\!\times\!\left\{\frac{1}{4\pi v_{F}v_{\Delta}}
\left[(2u_{0}\!-\!3u_{1}\!-\!2u_{2}\!-\!3u_{3})
+\frac{2u_{1}u_{3}
}{u_{2}}\!\right]
\!+\!\frac{2\lambda^{2}}{3}\left[4(\mathcal{B}_{1}\!+\!\mathcal{B}_{2}\!+\!\mathcal{B}_{3})
-\frac{u_{1}}{u_{2}}(\mathcal{B}_{1}\!+\!\mathcal{B}_{2})
-\frac{u_{3}}{u_{2}}(\mathcal{B}_{1}\!+\!\mathcal{B}_{3})\right]\right\}l,\label{Eq_y2}\\
\Gamma^{\tau_{y}}_{u_{3}}
&=&u_{3}\int_{-\infty}^{\infty}
\frac{d\omega_{1}d\omega_{2}d\omega_{3}}{(2\pi)^{3}}\int^{b}
\frac{d^{2}\mathbf{k}_{1}d^{2}\mathbf{k}_{2}d^{2}\mathbf{k}_{3}}{(2\pi)^{6}}
\Psi^{\dag}_{\sigma}(\omega_{1},\mathbf{k}_{1})\tau_{3}
\Psi_{\sigma}(\omega_{2},\mathbf{k}_{2})
\Psi^{\dag}_{\sigma^{'}}(\omega_{3},\mathbf{k}_{3})\tau_{3}
\Psi_{\sigma^{'}}(\omega_{1}+\omega_{2}-\omega_{3},\mathbf{k}_{1}+\mathbf{k}_{2}
-\mathbf{k}_{3})\nonumber\\
&&\times\left\{\frac{1}{4\pi v_{F}v_{\Delta}}\left[(u_{0}-u_{3}-u_{1}-2u_{2})+\frac{2(
u_{1}u_{2})}{u_{3}}
\right]
+\frac{2\lambda^{2}}{3}\left[
\frac{u_{1}}{u_{3}}(\mathcal{B}_{2}+\mathcal{B}_{3})
-4\mathcal{B}_{2}\right]\right\}l,\label{Eq_y3}\\
\Gamma^{\tau_{z}}_{u_{0}}
&=&u_{0}\int_{-\infty}^{\infty}
\frac{d\omega_{1}d\omega_{2}d\omega_{3}}{(2\pi)^{3}}\int^{b}
\frac{d^{2}\mathbf{k}_{1}d^{2}\mathbf{k}_{2}d^{2}\mathbf{k}_{3}}{(2\pi)^{6}}
\Psi^{\dag}_{\sigma}(\omega_{1},\mathbf{k}_{1})\tau_{0}
\Psi_{\sigma}(\omega_{2},\mathbf{k}_{2})
\Psi^{\dag}_{\sigma^{'}}(\omega_{3},\mathbf{k}_{3})\tau_{0}
\Psi_{\sigma^{'}}(\omega_{1}+\omega_{2}-\omega_{3},\mathbf{k}_{1}+\mathbf{k}_{2}
-\mathbf{k}_{3})\nonumber\\
&&\times\left[\frac{-(u_{1}u_{2}+u_{2}u_{3})}{4\pi u_{0}v_{F}
v_{\Delta}}+\frac{2\lambda^{2}}{3}\left(
\frac{u_{2}}{u_{0}}(\mathcal{C}_{2}-\mathcal{C}_{1})
-4\mathcal{C}_{1}\right)\right]l,\label{Eq_z0}\\
\Gamma^{\tau_{z}}_{u_{1}}
&=&u_{1}\int_{-\infty}^{\infty}
\frac{d\omega_{1}d\omega_{2}d\omega_{3}}{(2\pi)^{3}}\int^{b}
\frac{d^{2}\mathbf{k}_{1}d^{2}\mathbf{k}_{2}d^{2}\mathbf{k}_{3}}{(2\pi)^{6}}
\Psi^{\dag}_{\sigma}(\omega_{1},\mathbf{k}_{1})\tau_{1}
\Psi_{\sigma}(\omega_{2},\mathbf{k}_{2})
\Psi^{\dag}_{\sigma^{'}}(\omega_{3},\mathbf{k}_{3})\tau_{1}
\Psi_{\sigma^{'}}(\omega_{1}+\omega_{2}-\omega_{3},\mathbf{k}_{1}+\mathbf{k}_{2}
-\mathbf{k}_{3})\nonumber\\
&&\times\left[\frac{1}{4\pi v_{F}v_{\Delta}}(u_{0}-u_{1}-u_{2}-2u_{3}
+\frac{2u_{2}u_{3}}{u_{1}} )+\frac{2\lambda^{2}}{3}
\left(\frac{u_{2}}{u_{1}}(\mathcal{C}_{3}-\mathcal{C}_{2})
-\mathcal{C}_{1}-5\mathcal{C}_{3}\right)\right]l,\label{Eq_z1}\\
\Gamma^{\tau_{z}}_{u_{2}}
&=&u_{2}\int_{-\infty}^{\infty}
\frac{d\omega_{1}d\omega_{2}d\omega_{3}}{(2\pi)^{3}}\int^{b}
\frac{d^{2}\mathbf{k}_{1}d^{2}\mathbf{k}_{2}d^{2}\mathbf{k}_{3}}{(2\pi)^{6}}
\Psi^{\dag}_{\sigma}(\omega_{1},\mathbf{k}_{1})\tau_{2}
\Psi_{\sigma}(\omega_{2},\mathbf{k}_{2})
\Psi^{\dag}_{\sigma^{'}}(\omega_{3},\mathbf{k}_{3})\tau_{2}
\Psi_{\sigma^{'}}(\omega_{1}+\omega_{2}-\omega_{3},\mathbf{k}_{1}+\mathbf{k}_{2}
-\mathbf{k}_{3})\nonumber\\
&&\times\left\{\frac{1}{4\pi v_{F}v_{\Delta}}\left[(2u_{0}-3u_{1}-2u_{2}-3u_{3})
+\frac{2u_{1}u_{3}
}{u_{2}}\right]
+\frac{2\lambda^{2}}{3}\left[4\mathcal{C}_{2}-5\mathcal{C}_{1}-5\mathcal{C}_{3}
+\frac{u_{1}}{u_{2}}(\mathcal{C}_{3}-\mathcal{C}_{2})
\right]\right\}l,\label{Eq_z2}\\
\Gamma^{\tau_{z}}_{u_{3}}
&=&u_{3}\int_{-\infty}^{\infty}
\frac{d\omega_{1}d\omega_{2}d\omega_{3}}{(2\pi)^{3}}\int^{b}
\frac{d^{2}\mathbf{k}_{1}d^{2}\mathbf{k}_{2}d^{2}\mathbf{k}_{3}}{(2\pi)^{6}}
\Psi^{\dag}_{\sigma}(\omega_{1},\mathbf{k}_{1})\tau_{3}
\Psi_{\sigma}(\omega_{2},\mathbf{k}_{2})
\Psi^{\dag}_{\sigma^{'}}(\omega_{3},\mathbf{k}_{3})\tau_{3}
\Psi_{\sigma^{'}}(\omega_{1}+\omega_{2}-\omega_{3},\mathbf{k}_{1}+\mathbf{k}_{2}
-\mathbf{k}_{3})\nonumber\\
&&\times\left\{\frac{1}{4\pi v_{F}v_{\Delta}}\left[(u_{0}-u_{3}-u_{1}-2u_{2})
+\frac{2(u_{1}u_{2})}{u_{3}}\right]
+\frac{2\lambda^{2}}{3}\left[
\frac{u_{1}}{u_{3}}(\mathcal{C}_{2}-\mathcal{C}_{1})
-4\mathcal{C}_{2}\right]\right\}l,\label{Eq_z3}
\end{eqnarray}
where Eqs.~(\ref{Eq_x0})-(\ref{Eq_x3}) are linked to
Type-$\tau_{x}$, Eqs.~(\ref{Eq_y0})-(\ref{Eq_y3}) to
Type-$\tau_{y}$ and Eqs.~(\ref{Eq_z0})-(\ref{Eq_z3}) to $\tau_{z}$, respectively.

\begin{figure}
\centering
\includegraphics[width=0.66in]{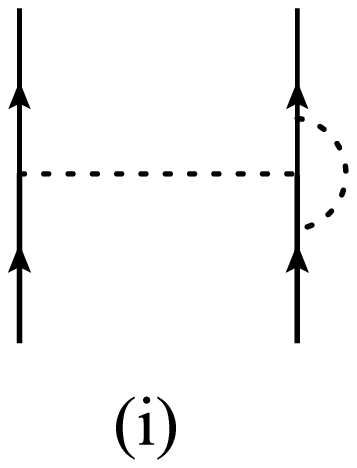}\hspace{0.9cm}
\includegraphics[width=0.66in]{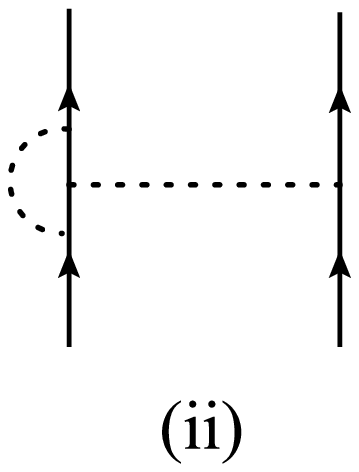}\hspace{0.9cm}
\includegraphics[width=0.65in]{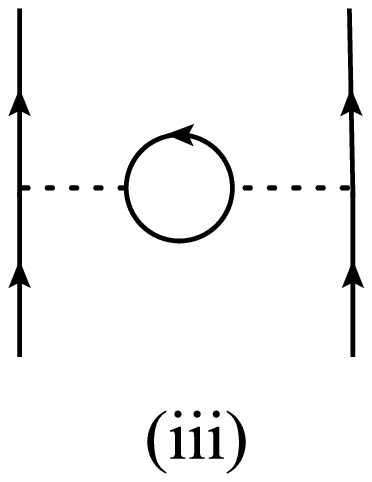}\hspace{0.9cm}
\includegraphics[width=0.6in]{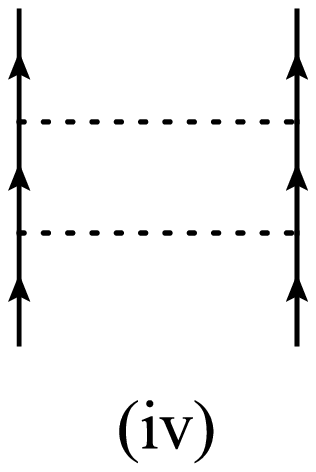}\hspace{0.9cm}
\includegraphics[width=0.6in]{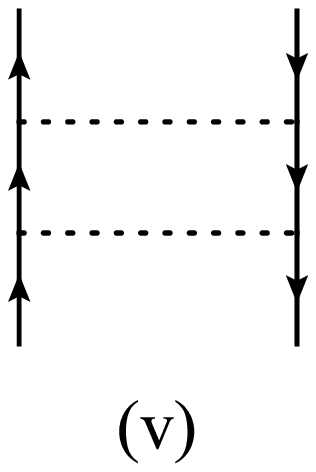}\vspace{0.5cm}\\
\includegraphics[width=0.7in]{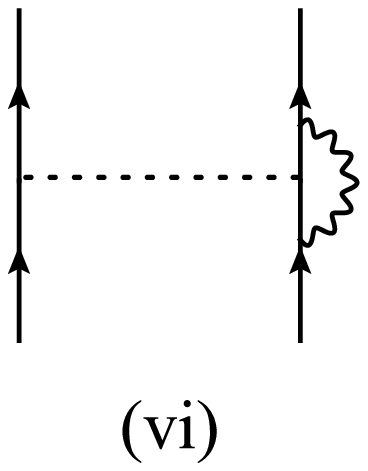}\hspace{0.9cm}
\includegraphics[width=0.68in]{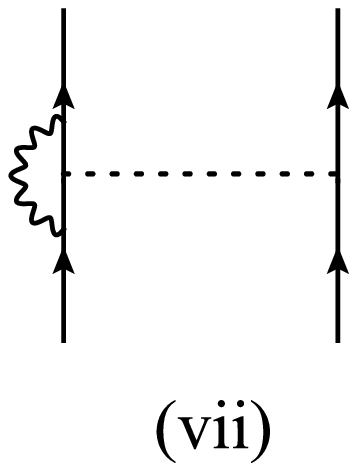}\hspace{0.9cm}
\includegraphics[width=0.63in]{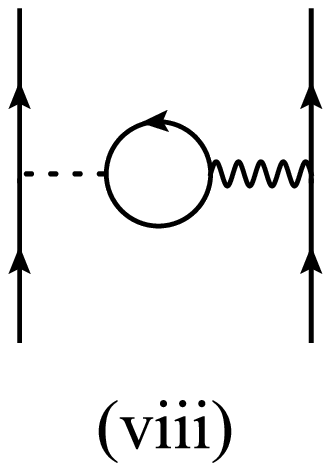}\hspace{0.9cm}
\includegraphics[width=0.6in]{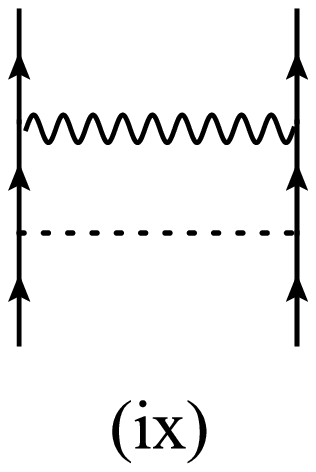}\hspace{0.9cm}
\includegraphics[width=0.6in]{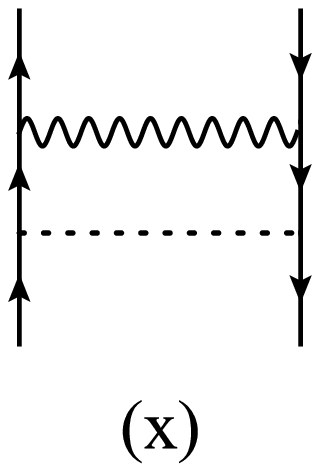}\\
\vspace{0.3cm}
\caption{One-loop corrections to the fermion-fermion
interactions due to fermion-fermion interactions (i)-(v) and Yukawa coupling between
nodal fermion and order parameter (vi)-(x). The
solid, dashed and wavy lines represent the fermion propagator,
four-fermion interaction and order parameter, respectively.}\label{Fig_1L-ff}
\end{figure}

\subsection{Designated coefficients}\label{Appendix_Coefficients}

All related coefficients appearing in both above equations and elsewhere
are designated by
\begin{eqnarray}
\mathcal{A}_{1}&=&\frac{2(v_{\Delta}/v_{F})}{N_{f}\pi^{3}}
\int_{-\infty}^{\infty}dx\int_{0}^{2\pi}d\theta
\frac{x^{2}-\cos^{2}\theta-(v_{\Delta}/v_{F})^{2}
\sin^{2}\theta}{[x^{2}+\cos^{2}\theta+(v_{\Delta}/v_{F})^{2}
\sin^{2}\theta]^{2}}\mathcal{G}_{\mathrm{I}}(x,\theta),\\
\mathcal{A}_{2}&=&\frac{2(v_{\Delta}/v_{F})}{N_{f}\pi^{3}}
\int_{-\infty}^{\infty}dx\int_{0}^{2\pi}d\theta
\frac{[-x^{2}+\cos^{2}\theta-(v_{\Delta}/v_{F})^{2}
\sin^{2}\theta]}{[x^{2}+\cos^{2}\theta+(v_{\Delta}/v_{F})^{2}
\sin^{2}\theta]^{2}}\mathcal{G}_{\mathrm{I}}(x,\theta),\\
\mathcal{A}_{3}&=&\frac{2(v_{\Delta}/v_{F})}{N_{f}\pi^{3}}
\int_{-\infty}^{\infty}dx\int_{0}^{2\pi}d\theta
\frac{[x^{2}+\cos^{2}\theta-(v_{\Delta}/v_{F})^{2}\sin^{2}\theta]}
{[x^{2}+\cos^{2}\theta+(v_{\Delta}/v_{F})^{2}
\sin^{2}\theta]^{2}}\mathcal{G}_{\mathrm{I}}(x,\theta),\\
\mathcal{B}_{1}&=&\frac{2(v_{\Delta}/v_{F})}{N_{f}\pi^{3}}
\int_{-\infty}^{\infty}dx\int_{0}^{2\pi}d\theta
\frac{x^{2}-\cos^{2}\theta-(v_{\Delta}/v_{F})^{2}
\sin^{2}\theta}{[x^{2}+\cos^{2}\theta+(v_{\Delta}/v_{F})^{2}
\sin^{2}\theta]^{2}}\mathcal{G}_{\mathrm{II}}(x,\theta),\\
\mathcal{B}_{2}&=&\frac{2(v_{\Delta}/v_{F})}{N_{f}\pi^{3}}
\int_{-\infty}^{\infty}dx\int_{0}^{2\pi}d\theta
\frac{[-x^{2}+\cos^{2}\theta-(v_{\Delta}/v_{F})^{2}
\sin^{2}\theta]}{[x^{2}+\cos^{2}\theta+(v_{\Delta}/v_{F})^{2}
\sin^{2}\theta]^{2}}\mathcal{G}_{\mathrm{II}}(x,\theta),\\
\mathcal{B}_{3}&=&\frac{2(v_{\Delta}/v_{F})}{N_{f}\pi^{3}}
\int_{-\infty}^{\infty}dx\int_{0}^{2\pi}d\theta
\frac{[-x^{2}-\cos^{2}\theta+(v_{\Delta}/v_{F})^{2}\sin^{2}\theta]}
{[x^{2}+\cos^{2}\theta+(v_{\Delta}/v_{F})^{2}
\sin^{2}\theta]^{2}}\mathcal{G}_{\mathrm{II}}(x,\theta),\\
\mathcal{C}_{1}&=&\frac{2(v_{F}/v_{\Delta})}{N_{f}\pi^{3}}
\int_{-\infty}^{\infty}dx\int_{0}^{2\pi}d\theta
\frac{x^{2}-\cos^{2}\theta-(v_{\Delta}/v_{F})^{2}
\sin^{2}\theta}{[x^{2}+\cos^{2}\theta+(v_{\Delta}/v_{F})^{2}
\sin^{2}\theta]^{2}}\mathcal{G}_{\mathrm{III}}(x,\theta),\\
\mathcal{C}_{2}&=&\frac{2(v_{F}/v_{\Delta})}{N_{f}\pi^{3}}
\int_{-\infty}^{\infty}dx\int_{0}^{2\pi}d\theta
\frac{[x^{2}-\cos^{2}\theta+(v_{\Delta}/v_{F})^{2}
\sin^{2}\theta]}{[x^{2}+\cos^{2}\theta+(v_{\Delta}/v_{F})^{2}
\sin^{2}\theta]^{2}}\mathcal{G}_{\mathrm{III}}(x,\theta),\\
\mathcal{C}_{3}&=&\frac{2(v_{F}/v_{\Delta})}{N_{f}\pi^{3}}
\int_{-\infty}^{\infty}dx\int_{0}^{2\pi}d\theta
\frac{[-x^{2}-\cos^{2}\theta+(v_{\Delta}/v_{F})^{2}\sin^{2}\theta]}
{[x^{2}+\cos^{2}\theta+(v_{\Delta}/v_{F})^{2}
\sin^{2}\theta]^{2}}\mathcal{G}_{\mathrm{III}}(x,\theta),\\
\mathcal{D}^{A}_{1}\!&=&\!\frac{2(v_{\Delta}/v_{F})}{N_{f}\pi^{3}}
\int_{-\infty}^{\infty}dx\int_{0}^{2\pi}d\theta
\frac{x^{2}-\cos^{2}\theta-(v_{\Delta}/v_{F})^{2}
\sin^{2}\theta}{[x^{2}+\cos^{2}\theta+(v_{\Delta}/v_{F})^{2}
\sin^{2}\theta]^{2}}\mathcal{G}_{\mathrm{IVA}}(x,\theta),\\
\mathcal{D}^{A}_{2}\!\!&=&\!\!\frac{2(v_{\Delta}/v_{F})}{N_{f}\pi^{3}}
\int_{-\infty}^{\infty}dx\int_{0}^{2\pi}d\theta
\frac{[x^{2}-\cos^{2}\theta+(v_{\Delta}/v_{F})^{2}
\sin^{2}\theta]}{[x^{2}+\cos^{2}\theta+(v_{\Delta}/v_{F})^{2}
\sin^{2}\theta]^{2}}\mathcal{G}_{\mathrm{IVA}}(x,\theta),\\
\mathcal{D}^{A}_{3}\!\!&=&\!\!\frac{2(v_{\Delta}/v_{F})}{N_{f}\pi^{3}}
\int_{-\infty}^{\infty}dx\int_{0}^{2\pi}d\theta
\frac{[x^{2}+\cos^{2}\theta-(v_{\Delta}/v_{F})^{2}\sin^{2}\theta]}
{[x^{2}+\cos^{2}\theta+(v_{\Delta}/v_{F})^{2}
\sin^{2}\theta]^{2}}\mathcal{G}_{\mathrm{IVA}}(x,\theta),\\
\mathcal{D}^{B}_{1}&=&\frac{2(v_{\Delta}/v_{F})}{N_{f}\pi^{3}}
\int_{-\infty}^{\infty}dx\int_{0}^{2\pi}d\theta
\frac{x^{2}-\sin^{2}\theta-(v_{\Delta}/v_{F})^{2}
\cos^{2}\theta}{[x^{2}+\sin^{2}\theta+(v_{\Delta}/v_{F})^{2}
\cos^{2}\theta]^{2}}\mathcal{G}_{\mathrm{IVB}}(x,\theta),\\
\mathcal{D}^{B}_{2}&=&\frac{2(v_{\Delta}/v_{F})}{N_{f}\pi^{3}}
\int_{-\infty}^{\infty}dx\int_{0}^{2\pi}d\theta
\frac{[x^{2}-\sin^{2}\theta+(v_{\Delta}/v_{F})^{2}
\cos^{2}\theta]}{[x^{2}+\sin^{2}\theta+(v_{\Delta}/v_{F})^{2}
\cos^{2}\theta]^{2}}\mathcal{G}_{\mathrm{IVB}}(x,\theta),\\
\mathcal{D}^{B}_{3}&=&\frac{2(v_{\Delta}/v_{F})}{N_{f}\pi^{3}}
\int_{-\infty}^{\infty}dx\int_{0}^{2\pi}d\theta
\frac{[x^{2}+\sin^{2}\theta-(v_{\Delta}/v_{F})^{2}\cos^{2}\theta]}
{[x^{2}+\sin^{2}\theta+(v_{\Delta}/v_{F})^{2}
\cos^{2}\theta]^{2}}\mathcal{G}_{\mathrm{IVB}}(x,\theta),
\end{eqnarray}
where the associated functions $\mathcal{G}_{\mathrm{I}}$, $\mathcal{G}_{\mathrm{II}}$,
$\mathcal{G}_{\mathrm{III}}$, $\mathcal{G}_{\mathrm{IVA}}$, and
$\mathcal{G}_{\mathrm{IVB}}$ are nominated as
\begin{eqnarray}
\mathcal{G}_{\mathrm{I}}^{-1}&=&\frac{x^{2}+\cos^{2}\theta}
{\sqrt{x^{2}+\cos^{2}\theta
+(v_{\Delta}/v_{F})^{2}\sin^{2}\theta}}
+\frac{x^{2}+\sin^{2}\theta}{\sqrt{x^{2}
+\sin^{2}\theta+(v_{\Delta}/v_{F})^{2}\cos^{2}\theta}},\\
\mathcal{G}_{\mathrm{II}}^{-1} &=& \sqrt
{x^2+\cos^2\theta+(v_\Delta/v_F)^2 \sin^2\theta}
+ \sqrt{x^2 +\sin^2\theta+(v_\Delta/v_F)^2\cos^2\theta},\\
\mathcal{G}_{\mathrm{III}}^{-1} &=& \frac{x^2+(v_\Delta/v_F)^2\sin^2\theta}{\sqrt
{x^2+\cos^2\theta+(v_\Delta/v_F)^2 \sin^2\theta}}
+ \frac{x^2+(v_\Delta/v_F)^2\cos^2\theta}{\sqrt{x^2 +\sin^2\theta+(v_\Delta/v_F)^2\cos^2\theta}},\\
\mathcal{G}_{\mathrm{IVA}}^{-1}
&=&-\sqrt{x^{2}+\cos^{2}\theta+(v_{\Delta}/v_{F})^{2}\sin^{2}\theta},\\
\mathcal{G}_{\mathrm{IVB}}^{-1}&=&-\sqrt{x^{2}+\sin^{2}\theta
+(v_{\Delta}/v_{F})^{2}\cos^{2}\theta}.
\end{eqnarray}

\section{Coupled RG equations for Type-$\tau_y$, $\tau_z$,
and $\tau_0$}\label{Appendix_RG}

Besides the coupled RG equations for Type-$\tau_x$
phase transition exhibited in Eqs.~(\ref{Eq_x_vD})-(\ref{Eq_x_u_3}),
we perform the standard procedures of momentum-shell RG approach~\cite{Shankar1994RMP,Wilson1975RMP,Polchinski1992}
and then deliver the corresponding RG evolutions for other types
as follows,
\begin{eqnarray}
\frac{dv_{F}}{dl}&=&\lambda^{2}(\mathcal{B}_{1}-\mathcal{B}_{2})v_{F},\label{Eq_y_vF}\\
\frac{dv_{\Delta}}{dl}&=&\lambda^{2}(\mathcal{B}_{1}-\mathcal{B}_{3})v_{\Delta},\\
\frac{d\frac{v_{\Delta}}{v_{F}}}{dl}&=&\lambda^{2}(\mathcal{B}_{2}-\mathcal{B}_{3})
\frac{v_{\Delta}}{v_{F}},\label{Eq_y_vD/vF}\\
\frac{d\lambda}{dl}&=&\left[2\mathcal{B}_{1}+\mathcal{B}_{2}+\mathcal{B}_{3}+\frac{[(u_{1})^{2}
+(u_{3})^{2}-(u_{0})^{2}-(u_{2})^{2}]}{4\pi v_{F}v_{\Delta}}\right]\lambda^{3},\label{Eq_y_lambda}\\
\frac{du_{0}}{dl}
&=&\left\{-1+2\mathcal{B}_1-\frac{(u_{1}u_{2}
+u_{2}u_{3})}{4\pi v_{F}v_{\Delta}u_{0}}+\frac{2\lambda^{2}}{3}\left[-4\mathcal{B}_{1}
-\frac{u_{1}}{u_{0}}(\mathcal{B}_{1}+\mathcal{B}_{3})
-\frac{u_{3}}{u_{0}}(\mathcal{B}_{1}+\mathcal{B}_{2})\right]\right\}u_{0},\\
\frac{du_{1}}{dl}
&=&\left\{-1+2\mathcal{B}_1
+\frac{1}{4\pi v_{F}v_{\Delta}}(u_{0}-u_{1}-u_{2}-2u_{3}
+\frac{2u_{2}u_{3}}{u_{1}} )+\frac{2\lambda^{2}}{3}\left[
\frac{u_{3}}{u_{1}}(\mathcal{B}_{2}+\mathcal{B}_{3})
-4\mathcal{B}_{3}\right]\right\}u_{1},\\
\frac{du_{2}}{dl}
\!\!&=&\!\!\left\{-1\!+\!2\mathcal{B}_1
+\frac{1}{4\pi v_{F}v_{\Delta}}\left[(2u_{0}\!-\!3u_{1}\!
-\!2u_{2}\!-\!3u_{3})+\frac{2u_{1}u_{3}
}{u_{2}}\right]
+\frac{2\lambda^{2}}{3}\left[4(\mathcal{B}_{1}+\mathcal{B}_{2}+\mathcal{B}_{3})
-\frac{u_{1}}{u_{2}}(\mathcal{B}_{1}+\mathcal{B}_{2})\right.\right.\nonumber\\
&&\left.\left.-\frac{u_{3}}{u_{2}}(\mathcal{B}_{1}+\mathcal{B}_{3})\right]\right\}u_{2},\\
\frac{du_{3}}{dl}
&=&\left\{-1+2\mathcal{B}_1
+\frac{1}{4\pi v_{F}v_{\Delta}}\left[(u_{0}-u_{3}-u_{1}-2u_{2})
+\frac{2u_{1}u_{2}}{u_{3}}
\right]
+\frac{2\lambda^{2}}{3}\left[
\frac{u_{1}}{u_{3}}(\mathcal{B}_{2}+\mathcal{B}_{3})
-4\mathcal{B}_{2}\right]\right\}u_{3}\label{Eq_y_u3},
\end{eqnarray}
for Type-$\tau_{y}$,
\begin{eqnarray}
\frac{dv_{F}}{dl}&=&\lambda^{2}(\mathcal{C}_{1}-\mathcal{C}_{2})v_{F},\label{Eq_z_vF}\\
\frac{dv_{\Delta}}{dl}&=&\lambda^{2}(\mathcal{C}_{1}-\mathcal{C}_{3})v_{\Delta},\\
\frac{d\frac{v_{F}}{v_{\Delta}}}{dl}
&=&\lambda^{2}(\mathcal{C}_{3}-\mathcal{C}_{2})\frac{v_{F}}{v_{\Delta}},\label{Eq_z_vD/vF}\\
\frac{d\lambda}{dl}&=&\left[\mathcal{C}_{1}-\mathcal{C}_{2}+\frac{[(u_{1})^{2}
+(u_{2})^{2}-(u_{0})^{2}-(u_{3})^{2}]}{8\pi v_{F}v_{\Delta}}\right]\lambda^{3},\label{Eq_z_lambda}\\
\frac{du_{0}}{dl}
&=&\left\{-1+2\mathcal{C}_1-\frac{(u_{1}u_{2}
+u_{2}u_{3})}{4\pi v_{F}v_{\Delta}u_{0}}+\frac{2\lambda^{2}}{3}\left[
\frac{u_{2}}{u_{0}}(\mathcal{C}_{2}-\mathcal{C}_{1})-4\mathcal{C}_{1}\right]\right\}u_{0},\\
\frac{du_{1}}{dl}
&=&\left\{-1+2\mathcal{C}_1
+\frac{1}{4\pi v_{F}v_{\Delta}}(u_{0}-u_{1}-u_{2}-2u_{3}
+\frac{2u_{2}u_{3}}{u_{1}} )+\frac{2\lambda^{2}}{3}\left[
\frac{u_{2}}{u_{1}}(\mathcal{C}_{3}-\mathcal{C}_{2})-\mathcal{C}_{1}
-5\mathcal{C}_{3}\right]\right\}u_{1},\\
\frac{du_{2}}{dl}
&=&\left\{-1+2\mathcal{C}_1
+\frac{1}{4\pi v_{F}v_{\Delta}}\left[(2u_{0}-3u_{1}-2u_{2}-3u_{3})+\frac{2u_{1}u_{3}
}{u_{2}}\right]
+\frac{2\lambda^{2}}{3}\left[4\mathcal{C}_{2}-5\mathcal{C}_{1}
-5\mathcal{C}_{3}\right.\right.\nonumber\\
&&\left.\left.
+\frac{u_{1}}{u_{2}}(\mathcal{C}_{3}-\mathcal{C}_{2})\right]\right\}u_{2},\\
\frac{du_{3}}{dl}
&=&\left\{-1+2\mathcal{C}_1
+\frac{1}{4\pi v_{F}v_{\Delta}}\left[(u_{0}-u_{3}-u_{1}-2u_{2})
+\frac{2u_{1}u_{2}}{u_{3}}
\right]
+\frac{2\lambda^{2}}{3}\left[
\frac{u_{1}}{u_{3}}(\mathcal{C}_{2}-\mathcal{C}_{1})
-4\mathcal{C}_{2}\right]\right\}u_{3}\label{Eq_z_u3},
\end{eqnarray}
\end{widetext}
for Type-$\tau_{z}$, and
\begin{eqnarray}
\frac{dv_{F}}{dl}&=&\lambda^{2}(\mathcal{D}_{1}^{A,B}
-\mathcal{D}_{2}^{A,B})v_{F}\label{Eq_0_vF},\\
\frac{dv_{\Delta}}{dl}&=&\lambda^{2}(\mathcal{D}_{1}^{A,B}
-\mathcal{D}_{3}^{A,B})v_{\Delta},\\
\frac{d\frac{v_{\Delta}}{v_{F}}}{dl}
&=&\lambda^{2}(\mathcal{D}_{2}^{A,B}-\mathcal{D}_{3}^{A,B})
\frac{v_{\Delta}}{v_{F}},\\
\frac{d\lambda}{dl}&=&(\mathcal{D}_{1}^{A,B}
-\mathcal{D}_{1}^{A,B})\lambda^{3}=0\label{Eq_0_lambda},
\end{eqnarray}
for Type-$\tau_{0}$ phase transitions, respectively.
It is worth pointing out that the Yukawa coupling $\lambda$ is
still marginal to the one-loop level and hence does not flow with
the decrease of energy scale. Given the fermion-fermion interactions
can only indirectly influence the fermion velocities and accompanied
physical implications via modifying such Yukawa coupling, we
henceforth can safely skip the effects caused by
fermion-fermion interactions, in other words neglecting
the one-loop RG equations of the fermion-fermion interactions.

\vspace{0.5cm}



\begin{thebibliography}{10}

\bibitem{Sigrist1991RMP}
M. Sigrist and K. Ueda,
\href{https://https://journals.aps.org/rmp/abstract/10.1103/RevModPhys.63.239}
{Rev. Mod. Phys. {\bf 63}, 239 (1991)}.

\bibitem{Dagotto1994RMP}
E. Dagotto,
\href{https://journals.aps.org/rmp/abstract/10.1103/RevModPhys.66.763}
{Rev. Mod. Phys. 66, 763 (1994)}.

\bibitem{Kivelson1995Nature}
V. J. Emery and S. A. Kivelson,
\href{https://www.nature.com/articles/374434a0}
{Nature, {\bf 374}, 30 (1995)}.

\bibitem{Sigrist1995RMP}
M. Sigrist and T. M. Rice,
\href{https://journals.aps.org/rmp/abstract/10.1103/RevModPhys.67.503}
{Rev. Mod. Phys. {\bf 67}, 503 (1995)}.

\bibitem{Tinkham1996Book}
M. Tinkham, \emph{Introduction to Superconductivity, Dover
Books on Physics Series}, Dover Publications, (1996).

\bibitem{Anderson1997Book}
P. W. Anderson, \emph{The Theory of Superconductivity in the High-Tc
Cuprate Superconductors}, Princeton University Press, (1997).

\bibitem{Kivelson1998Nature}
S. A. Kivelson, E. Fradkin, and V. J. Emery,
\href{https://www.nature.com/articles/31177}
{Nature (London), {\bf 393}, 550 (1998)}.

\bibitem{Sachdev2000Science}
S. Sachdev,
\href{https://science.sciencemag.org/content/288/5465/475}
{Science {\bf 288}, 475 (2000)}.

\bibitem{Dagotto2005Science}
E. Dagotto,
\href{https://science.sciencemag.org/content/309/5732/257}
{Science {\bf 309}, 257 (2005)}.

\bibitem{Sachdev2011PT}
S. Sachdev and B. Keimer,
\href{https://physicstoday.scitation.org/doi/10.1063/1.3554314}
{Phys. Today {\bf 64(2)}, 29 (2011)}.

\bibitem{Vojta2000PRL}
M. Vojta, Y. Zhang, and S. Sachdev,
\href{https://journals.aps.org/prl/abstract/10.1103/PhysRevLett.85.4940}
{Phys. Rev. Lett. {\bf 85}, 4090 (2000)}.

\bibitem{Vojta2000PRB}
M. Vojta, Y. Zhang, and S. Sachdev,
\href{https://journals.aps.org/prb/abstract/10.1103/PhysRevB.62.6721}
{Phys. Rev. B {\bf 62}, 6721 (2000)}.

\bibitem{Vojta2000IJMPB}
M. Vojta, Y. Zhang, and S. Sachdev,
\href{https://www.worldscientific.com/doi/abs/10.1142/S0217979200004271}
{Int. J. Mod. Phys. B {\bf 14}, 3719 (2000)}.

\bibitem{Sachdev2003RMP}
S. Sachdev,
\href{https://journals.aps.org/rmp/abstract/10.1103/RevModPhys.75.913}
{Rev. Mod. Phys. {\bf 75}, 913 (2003)}.

\bibitem{Kivelson2003RMP_DFS}
S. A. Kivelson, I. P. Bindloss, E. Fradkin, V. Oganesyan, J. M. Tranquada, A. Kapitulnik, and C. Howald,
\href{https://journals.aps.org/rmp/abstract/10.1103/RevModPhys.75.1201}
{Rev. Mod. Phys. {\bf 75}, 1201 (2003)}.

\bibitem{Lee2006RMP}
P. A. Lee, N. Nagaosa, and X. -G. Wen,
\href{https://journals.aps.org/rmp/abstract/10.1103/RevModPhys.78.17}
{Rev. Mod. Phys. {\bf 78}, 17 (2006)}.

\bibitem{Sachdev2008PRB}
Y. Huh and S. Sachdev,
\href{https://journals.aps.org/prb/abstract/10.1103/PhysRevB.78.064512}
{Phys. Rev. B {\bf 78}, 064512 (2008)}.

\bibitem{Wang2011PRB}
J. Wang, G. Z. Liu, and H. Kleinert,
\href{https://journals.aps.org/prb/abstract/10.1103/PhysRevB.83.214503}
{Phys. Rev. B {\bf 83}, 214503 (2011)}.

\bibitem{Kim-Kivelson2008PRB}
E. A. Kim, M. J. Lawler, P. Oreto, S. Sachdev, E. Fradkin, and S. A. Kivelson,
\href{https://journals.aps.org/prb/abstract/10.1103/PhysRevB.77.184514}
{Phys. Rev. B {\bf 77}, 184514 (2008)}.

\bibitem{Xu2008PRB}
C. Xu, Y. Qi, and S. Sachdev,
\href{https://journals.aps.org/prb/abstract/10.1103/PhysRevB.78.134507}
{Phys. Rev. B {\bf 78}, 134507 (2008)}.

\bibitem{She2010PRB}
J. H. She, J. Zaanen, A. R. Bishop, and A. V. Balatsky,
\href{https://journals.aps.org/prb/abstract/10.1103/PhysRevB.82.165128}
{Phys. Rev. B {\bf 82}, 165128 (2010)}.

\bibitem{She2015PRB}
J. H. She, M. J. Lawler, and E. A. Kim,
\href{https://journals.aps.org/prb/abstract/10.1103/PhysRevB.92.035112}
{Phys. Rev. B {\bf 92}, 035112 (2015)}.

\bibitem{Fradkin2012NPhys}
E. Fradkin and S. A. Kivelson,
\href{https://www.nature.com/articles/nphys2498}
{Nature Phys {\bf 8}, 864 (2012)}.


\bibitem{Kivelson2014PNAS}
H. Watanabe, A. Vishwanath, and S. A. Kivelson,
\href{https://www.pnas.org/content/111/46/16314}
{PNAS {\bf 111}, 16314 (2014)}.

\bibitem{Fradkin2015RMP}
E. Fradkin, S. A. Kivelson, and J. M. Tranquada,
\href{https://journals.aps.org/rmp/abstract/10.1103/RevModPhys.87.457}
{Rev. Mod. Phys. {\bf 87}, 457 (2015)}.

\bibitem{Phillips2020NPhys}
P. W. Phillips, L. Yeo, and E. W. Huang,
\href{https://www.nature.com/articles/s41567-020-0988-4}
{Nat. Phys. {\bf 16}, 1175 (2020)}.

\bibitem{Larkin2005Book}
A. Larkin and A. Varlamov, \emph{Theory of fluctuations in superconductors},
Oxford University Press (New York), (2005).

\bibitem{Ding1996Nature}
H. Ding, T. Yokoya, J. C. Campuzano, T. Takahashi, M. Randeria, M. Norman,
T. Mochiku, and J. Giapintzakis,
\href{https://www.nature.com/articles/382051a0}
{Nature (London), {\bf 382}, 51 (1996)}.

\bibitem{Loeser1996Science}
A. G. Loeser, Z. -X. Shen, D. S. Desau, D. S. Marshall, C. H. Park, P. Fournier, and
A. Kapitulnik,
\href{https://science.sciencemag.org/content/273/5273/325}
{Science {\bf 273}, 325 (1996)}.

\bibitem{Valla1999Science}
T. Valla, A. Fedorov, P. Johnson, B. Wells, S. Hulbert, Q. Li,
G. Gu, and N. Koshizuka,
\href{https://science.sciencemag.org/content/285/5436/2110}
{Science {\bf 285}, 2110 (1999)}.

\bibitem{Orenstein2000Science}
J. Orenstein and A. J. Millis,
\href{https://science.sciencemag.org/content/288/5465/468}
{Science {\bf 288}, 468 (2000)}.

\bibitem{Yoshida2003PRL}
T. Yoshida, X. J. Zhou, T. Sasagawa, W. L. Yang, P. V. Bogdanov,
A. Lanzara, Z. Hussain, T. Mizokawa, A. Fujimori, H. Eisaki,
Z. -X. Shen, T. Kakeshita, and S. Uchida,
\href{https://journals.aps.org/prl/abstract/10.1103/PhysRevLett.91.027001}
{Phys. Rev. Lett. {\bf 91}, 027001 (2003)}.

\bibitem{Vojta2003RPP}
M. Vojta,
\href{https://iopscience.iop.org/article/10.1088/0034-4885/66/12/R01}
{Rep. Prog. Phys. {\bf 66}, 2069 (2003).}

\bibitem{Coleman2005Nature}
P. Coleman and A. J. Schofield,
\href{https://www.nature.com/articles/nature03279}
{Nature, {\bf 433}, 20 (2005)}.

\bibitem{Sachdev2011Book}
S. Sachdev, \emph{Quantum Phase Transitions}, 2nd edn., Cambridge University Press, Cambridge, (2011).

\bibitem{Paaske2001PRL}
D. V. Khveshchenko and J. Paaske,
\href{https://journals.aps.org/prl/abstract/10.1103/PhysRevLett.86.4672}
{Phys. Rev. Lett. {\bf 86}, 4672 (2001)}.

\bibitem{Sachdev2009PRB}
L. Fritz and S. Sachdev,
\href{https://journals.aps.org/prb/abstract/10.1103/PhysRevB.80.144503}
{Phys. Rev. B {\bf 80}, 144503 (2009)}.

\bibitem{Liu2012PRB}
G. -Z. Liu, J. -R. Wang, and J. Wang,
\href{https://journals.aps.org/prb/abstract/10.1103/PhysRevB.85.174525}
{Phys. Rev. B {\bf 85}, 174525 (2012)}.

\bibitem{Liu2013NJP}
J. -R. Wang and G. -Z. Liu,
\href{https://iopscience.iop.org/article/10.1088/1367-2630/15/6/063007}
{New J. Phys. {\bf 15}, 063007 (2013)}.

\bibitem{Moon2010PRB}
E. G. Moon and S. Sachdev,
\href{https://journals.aps.org/prb/abstract/10.1103/PhysRevB.82.104516}
{Phys. Rev. B {\bf 82}, 104516 (2010)}.

\bibitem{Moon2012PRB}
E. G. Moon and S. Sachdev,
\href{https://journals.aps.org/prb/abstract/10.1103/PhysRevB.85.184511}
{Phys. Rev. B {\bf 85}, 184511 (2012)}.

\bibitem{Moon2016PRB}
Y. Huh, E. -G. Moon, and Y. B. Kim,
\href{https://journals.aps.org/prb/abstract/10.1103/PhysRevB.93.035138}
{Phys. Rev. B {\bf 93}, 035138 (2016)}.

\bibitem{Moon2016SRep}
E. -G. Moon,
\href{https://www.nature.com/articles/srep31051}
{Sci. Rep. {\bf 6}, 31051 (2016)}.

\bibitem{Wang-EM2014PRB}
J. Wang, A. Eberlein, and W. Metzner,
\href{https://journals.aps.org/prb/abstract/10.1103/PhysRevB.89.121116}
{Phys. Rev. B {\bf 89}, 121116(R) (2014)}.

\bibitem{Lee1993PRL}
P. A. Lee,
\href{https://journals.aps.org/prl/abstract/10.1103/PhysRevLett.71.1887}
{Phys. Rev. Lett. {\bf 71}, 1887 (1993)}.

\bibitem{Castellani1997ZPB}
C. Castellani, C. D. Castro, and M. Grilli,
\href{https://link.springer.com/article/10.1007%2Fs002570050347}
{Z. Phys. B: Condens. Matter {\bf 103}, 137 (1997)}.

\bibitem{She2011PRL}
S. -X. Yang, H. Fotso, S. -Q. Su, D. Galanakis, E. Khatami, J. -H. She,
J. Moreno, J. Zaanen, and M. Jarrell,
\href{https://journals.aps.org/prl/abstract/10.1103/PhysRevLett.106.047004}
{Phys. Rev. Lett. {\bf 106}, 047004 (2011)}.

\bibitem{Wang2013PRB}
J. Wang,
\href{https://journals.aps.org/prb/abstract/10.1103/PhysRevB.87.054511}
{Phys. Rev. B {\bf 87}, 054511 (2013)}.

\bibitem{Keimer2008Science}
V. Hinkov, D. Haug, B. Fauque, P. Bourges, Y. Sidis, A. Ivanov, C.
Bernhard, C. T. Lin, and B. Keimer,
\href{https://science.sciencemag.org/content/319/5863/597}
{Science {\bf 319}, 597 (2008)}.

\bibitem{Durst2000PRB}
A. C. Durst and P. A. Lee,
\href{https://journals.aps.org/prb/abstract/10.1103/PhysRevB.62.1270}
{Phys. Rev. B {\bf 62}, 1270 (2000)}.

\bibitem{Lee1997PRL}
P. A. Lee and X. -G. Wen,
\href{https://journals.aps.org/prl/abstract/10.1103/PhysRevLett.78.4111}
{Phys. Rev. Lett. {\bf 78}, 4111 (1997)}.

\bibitem{Mesot1999PRL}
J. Mesot, M. R. Norman, H. Ding, M. Randeria, J. C. Campuzano,
A. Paramekanti, H. M. Fretwell, A. Kaminski, T. Takeuchi, T.
Yokoya, T. Sato, T. Takahashi, T. Mochiku,
and K. Kadowaki,
\href{https://journals.aps.org/prl/abstract/10.1103/PhysRevLett.83.840}
{Phys. Rev. Lett. {\bf 83}, 840 (1999)}.

\bibitem{Vojta2009AP}
M. Vojta,
\href{https://www.tandfonline.com/doi/abs/10.1080/00018730903122242}
{Adv. Phys. {\bf 58}, 699 (2009)}.

\bibitem{Metzner2000PRL}
C. J. Halboth and W. Metzner,
\href{https://journals.aps.org/prl/abstract/10.1103/PhysRevLett.85.5162}
{Phys. Rev. Lett. {\bf 85}, 5162 (2000)}.

\bibitem{Kivelson2001PRB}
V. Oganesyan, S. A. Kivelson, and E. Fradkin,
\href{https://journals.aps.org/prb/abstract/10.1103/PhysRevB.64.195109}
{Phys. Rev. B {\bf 64}, 195109 (2001)}.

\bibitem{Sachdev2002PRB}
Y. Zhang, E. Demler, and S. Sachdev,
\href{https://journals.aps.org/prb/abstract/10.1103/PhysRevB.66.094501}
{Phys. Rev. B {\bf 66}, 094501 (2002)}.

\bibitem{Metzner2007PRB}
H. Yamase and W. Metzner,
\href{https://journals.aps.org/prb/abstract/10.1103/PhysRevB.75.155117}
{Phys. Rev. B {\bf 75}, 155117 (2007)}.


\bibitem{Kivelson2009PRB}
S. Raghu, A. Paramekanti, E. -A. Kim, R. A. Borzi, S. A. Grigera, A. P. Mackenzie, and S. A. Kivelson,
\href{https://journals.aps.org/prb/abstract/10.1103/PhysRevB.79.214402}
{Phys. Rev. B {\bf 79}, 214402 (2009)}.

\bibitem{Kim2010Nature}
M. J. Lawler, K. Fujita, J. Lee, A. R. Schmidt, Y. Kohsaka, K. C. Kim, H. Eisaki, S.
Uchida, J. C. Davis, J. P. Sethna, and E. -A. Kim,
\href{https://www.nature.com/articles/nature09169}
{Nature {\bf 466}, 347 (2010)}.

\bibitem{Sachdev2010PRB}
E. G. Moon and S. Sachdev,
\href{https://journals.aps.org/prb/abstract/10.1103/PhysRevB.82.104516}
{Phys. Rev. B {\bf 82}, 104516 (2010)}.

\bibitem{Kim2010PRB}
E. -A. Kim and M. J. Lawler,
\href{https://journals.aps.org/prb/abstract/10.1103/PhysRevB.81.132501}
{Phys. Rev. B {\bf 81}, 132501 (2010)}.

\bibitem{Fradkin2010ARCMP}
E. Fradkin, S. A. Kivelson, M. J. Lawler, J. P. Eisenstein, and
A. P. Mackenzie,
\href{https://www.annualreviews.org/doi/10.1146/annurev-conmatphys-070909-103925}
{Annu. Rev. Condens. Matter Phys. {\bf 1}, 153 (2010)}.

\bibitem{Wang2013NJP}
J. Wang and G. Z. Liu,
\href{https://iopscience.iop.org/article/10.1088/1367-2630/15/7/073039}
{New J. Phys. {\bf 15}, 073039 (2013)}.

\bibitem{Wang2015PLA}
J. Wang,
\href{https://www.sciencedirect.com/science/article/abs/pii/S0375960115004582?via%3Dihub}
{Phys. Let. A {\bf 379}, 1917 (2015)}.


\bibitem{Roy2004.13043}
B. Roy,
\href{https://arxiv.org/abs/2004.13043}
{arXiv :2004.13043, (2020)}.

\bibitem{Chubukov2010PRB}
S. Maiti and A. V. Chubukov,
\href{https://journals.aps.org/prb/abstract/10.1103/PhysRevB.82.214515}
{Phys. Rev. B {\bf 82}, 214515 (2010)}.

\bibitem{Vafek2010PRB}
O. Vafek,
\href{https://journals.aps.org/prb/abstract/10.1103/PhysRevB.82.205106}
{Phys. Rev. B {\bf 82}, 205106 (2010)}.

\bibitem{Vafek2012PRB}
V. Cvetkovi\'{c}, R. E. Throckmorton, and O. Vafek,
\href{https://journals.aps.org/prb/abstract/10.1103/PhysRevB.86.075467}
{Phys. Rev. B {\bf 86}, 075467 (2012)}.

\bibitem{Vafek2014PRB}
J. M. Murray and O. Vafek,
\href{https://journals.aps.org/prb/abstract/10.1103/PhysRevB.89.201110}
{Phys. Rev. B {\bf 89}, 201110 (2014)}.

\bibitem{Chubukov2012NPhys_chiral_SC}
R. Nandkishore, L. S. Levitov, and A. V. Chubukov,
\href{https://www.nature.com/articles/nphys2208}
{Nature Phys, {\bf 8}, 158 (2012)}.

\bibitem{Nandkishore2013PRB}
R. Nandkishore, J. Maciejko, D. A. Huse, and S. L. Sondhi,
\href{https://journals.aps.org/prb/abstract/10.1103/PhysRevB.87.174511}
{Phys. Rev. B {\bf 87}, 174511 (2013)}.

\bibitem{Khodas2016PRX}
A. V. Chubukov, M. Khodas, and R. M. Fernandes,
\href{https://journals.aps.org/prx/abstract/10.1103/PhysRevX.6.041045}
{Phys. Rev. X {\bf 6}, 041045 (2016)}.

\bibitem{Nandkishore2016NJP_RG-shell}
S. Sur and R. Nandkishore,
\href{https://iopscience.iop.org/article/10.1088/1367-2630/18/11/115006}
{New J. Phys. {\bf 18}, 115006 (2016)}.

\bibitem{Roy2016SR}
B. Roy, V. Juricic, and S. D. Sarma,
\href{https://www.nature.com/articles/srep32446}
{Sci Rep {\bf 6}, 32446 (2016)}.

\bibitem{Roy-Sau2016PRB}
B. Roy, P. Goswami, and J. D. Sau,
\href{https://journals.aps.org/prb/abstract/10.1103/PhysRevB.94.041101}
{Phys. Rev. B {\bf 94}, 041101(R) (2016)}.


\bibitem{Roy-Saram2016PRB}
B. Roy and S. D. Sarma,
\href{https://journals.aps.org/prb/abstract/10.1103/PhysRevB.94.115137}
{Phys. Rev. B {\bf 94}, 115137 (2016)}.

\bibitem{Nandkishore2017PRB}
R. M. Nandkishore and S. A. Parameswaran,
\href{https://journals.aps.org/prb/abstract/10.1103/PhysRevB.95.205106}
{Phys. Rev. B {\bf 95}, 205106 (2017)}.

\bibitem{Roy2017PRB}
B. Roy,
\href{https://journals.aps.org/prb/abstract/10.1103/PhysRevB.96.041113}
{Phys. Rev. B {\bf 96}, 041113 (2017)}.

\bibitem{Roy-Sau2017PRL}
B. Roy, Y. Alavirad, and J. D. Sau,
\href{https://journals.aps.org/prl/abstract/10.1103/PhysRevLett.118.227002}
{Phys. Rev. Lett. {\bf 118}, 227002 (2017)}.

\bibitem{Wang2017PRB}
J. Wang, C. Ortix, J. van den Brink, and D. V. Efremov,
\href{https://journals.aps.org/prb/abstract/10.1103/PhysRevB.96.201104}
{Phys. Rev. B {\bf 96}, 201104(R) (2017)}.

\bibitem{Wang2017PRB-2}
J. Wang, G. -Z. Liu, D. V. Efremov, and J. van den Brink,
\href{https://journals.aps.org/prb/abstract/10.1103/PhysRevB.95.024511}
{Phys. Rev. B {\bf 95}, 024511 (2017)}.

\bibitem{Wang2018JPCM}
J. Wang,
\href{https://iopscience.iop.org/article/10.1088/1361-648X/aaa8ce}
{J. Phys. Condens. Matter {\bf 30}, 125401 (2018)}.

\bibitem{Wang2019JPCM}
Y. -M. Dong, D. -X. Zheng, and J. Wang,
\href{https://iopscience.iop.org/article/10.1088/1361-648X/ab142d}
{J. Phys. Condens. Matter {\bf 31}, 275601 (2019)}.

\bibitem{Roy-Slager2018PRX}
B. Roy, R. J. Slager, and V. Juri\u{c}i\'{c},
\href{https://journals.aps.org/prx/abstract/10.1103/PhysRevX.8.031076}
{Phys. Rev. X {\bf 8}, 031076 (2018)}.

\bibitem{Roy2018PRX}
B. Roy and M. S. Foster,
\href{https://journals.aps.org/prx/abstract/10.1103/PhysRevX.8.011049}
{Phys. Rev. X {\bf 8}, 011049 (2018)}.

\bibitem{Mandal2018PRB}
I. Mandal and R. M. Nandkishore,
\href{https://journals.aps.org/prb/abstract/10.1103/PhysRevB.97.125121}
{Phys. Rev. B {\bf 97}, 125121 (2018)}.

\bibitem{Roy2019PRL}
S. Sur and B. Roy,
\href{https://journals.aps.org/prl/abstract/10.1103/PhysRevLett.123.207601}
{Phys. Rev. Lett. {\bf 123}, 207601 (2019)}.

\bibitem{Wang2020NPB}
J. Wang,
\href{https://www.sciencedirect.com/science/article/pii/S0550321320303151}
{Nucl. Phys. B {\bf 961}, 115230 (2020)}.

\bibitem{Hui2020EPJB}
Y. H. Zhai and J. Wang,
\href{https://link.springer.com/article/10.1140%2Fepjb%2Fe2020-10049-x}
{Eur. Phys. J. B  {\bf 93}, 86 (2020)}.

\bibitem{Wang2020PRB}
Y. M. Dong, Y. H. Zhai, D. X. Zheng, and J. Wang,
\href{https://journals.aps.org/prb/abstract/10.1103/PhysRevB.102.134204}
{Phys. Rev. B {\bf 102}, 134204 (2020)}.

\bibitem{Wang2021NPB}
Y. H. Zhai and J. Wang,
\href{https://www.sciencedirect.com/science/article/pii/S0550321321000687}
{Nucl. Phys. B {\bf 966}, 115371 (2021)}.

\bibitem{Roy2021PRB}
A. L. Szab\'{o} and B. Roy,
\href{https://journals.aps.org/prb/abstract/10.1103/PhysRevB.103.205135}
{Phys. Rev. B {\bf 103}, 205135 ( 2021)}.

\bibitem{Roy2021JHEP}
A. L. Szab\'{o} and B. Roy,
\href{https://link.springer.com/article/10.1007%2FJHEP01%282021%29004}
{JHEP {\bf 01}, 004 (2021)}.

\bibitem{Moon2017PRB}
S. E. Han, G. Y. Cho, and E. -G. Moon,
\href{https://journals.aps.org/prb/abstract/10.1103/PhysRevB.95.094502}
{Phys. Rev. B {\bf 95}, 094502 (2017)}.

\bibitem{Herbut2016JHEP}
B. Roy, V. Juri\u{c}i\'{c}, and I. F. Herbut,
\href{https://link.springer.com/article/10.1007/JHEP04%282016%29018}
{JHEP {\bf 04}, 018 (2016)}.

\bibitem{Moon2016SRep-2}
G. Y. Cho and E. -G. Moon,
\href{https://www.nature.com/articles/srep19198}
{Sci. Rep. {\bf 6}, 19198 (2016)}.

\bibitem{Yao2017PRB}
S. -K. Jian and H. Yao,
\href{https://journals.aps.org/prb/abstract/10.1103/PhysRevB.96.155112}
{Phys. Rev. B {\bf 96}, 155112 (2017)}.

\bibitem{Herbut2018Science}
H. -K Tang, J. N. Leaw, J. N. B. Rodrigues, I. F. Herbut, P. Sengupta,
F. F. Assaad, and S. Adam,
\href{https://science.sciencemag.org/content/361/6402/570}
{Science {\bf 361}, 570 (2018)}.

\bibitem{Yao2021PRB}
S. -X. Zhang, S. -K. Jian, and H. Yao,
\href{https://journals.aps.org/prb/abstract/10.1103/PhysRevB.103.165129}
{Phys. Rev. B {\bf 103}, 165129 (2021)}.

\bibitem{Wilson1975RMP}
K. G. Wilson,
\href{https://journals.aps.org/rmp/abstract/10.1103/RevModPhys.47.773}
{Rev. Mod. Phys. {\bf 47}, 773 (1975)}.

\bibitem{Polchinski1992}
J. Polchinski,
\href{https://arxiv.org/abs/hep-th/9210046}
{arXiv:hep-th/9210046 (unpublished)}.

\bibitem{Shankar1994RMP}
R. Shankar,
\href{https://journals.aps.org/rmp/abstract/10.1103/RevModPhys.66.129}
{Rev. Mod. Phys. {\bf 66}, 129 (1994)}.

\bibitem{Ong1995PRL} 
K. Krishana, J. M. Harris, and N. P. Ong,
\href{https://journals.aps.org/prl/abstract/10.1103/PhysRevLett.75.3529}
{Phys. Rev. Lett. {\bf 75}, 3529 (1995).}

\bibitem{Makhfudz2015AP}
I. Makhfudz,
\href{https://www.sciencedirect.com/science/article/abs/pii/S000349161500189X?via%3Dihub}
{Annals of Physics {\bf 360}, 113 (2015)}.

\bibitem{Wang2015PRB}
J. Wang and G. -Z. Liu,
\href{https://journals.aps.org/prb/abstract/10.1103/PhysRevB.92.184510}
{Phys. Rev. B {\bf 92}, 184510 (2015)}.


\bibitem{Wang2007.14981}
J. Wang, \href{https://arxiv.org/abs/2007.14981}
{arXiv: 2007.14981v1 (2020)}.

\bibitem{Orenstein1990PRB}
J. Orenstein, G. A. Thomas, A. J. Millis, S. L. Cooper, D. H. Rapkine, T. Timusk, L. F. Schneemeyer, and J. V. Waszczak,
\href{https://journals.aps.org/prb/abstract/10.1103/PhysRevB.42.6342}
{Phys. Rev. B {\bf 42}, 6342 (1990)}.

\bibitem{Hardy1993PRL}
W. N. Hardy, D. A. Bonn, D. C. Morgan, R. Liang, and K. Zhang,
\href{https://journals.aps.org/prl/abstract/10.1103/PhysRevLett.70.3999}
{Phys. Rev. Lett. {\bf 70}, 3999 (1993)}.

\bibitem{Bollinger2016Nature}
I. Bo\v{z}ovi\'{c}, X. He, J. Wu, and A. T. Bollinger,
\href{https://www.nature.com/articles/nature19061}
{Nature {\bf 536}, 309 (2016)}.

\bibitem{Uemura1989PRL}
Y. J. Uemura, G. M. Luke, B. J. Sternlieb, J. H. Brewer, J. F. Carolan,
W. N. Hardy, R. Kadono, J. R. Kempton, R. F. Kief, S. R. Kreitzman,
P. Mulhern, T. M. Riseman, D. L. Williams, B. X. Yang, S. Uchida,
H. Takagi, J. Gopalakrishnan, A. W. Sleight, M. A. Subramanian,
C. L. Chien, M. Z. Cieplak, G. Xiao, V. Y. Lee, B. W. Statt,
C. E. Stronach, W. J. Kossler, X. H. Yu,
\href{https://journals.aps.org/prl/abstract/10.1103/PhysRevLett.62.2317}
{Phys. Rev. Lett. {\bf 62}, 2317(1989)}.

\bibitem{Fournier2000PRB}
M. Chiao, R. W. Hill, C. Lupien, L. Taillefer, P. Lambert, R. Gagnon,
and P. Fournier,
\href{https://journals.aps.org/prb/abstract/10.1103/PhysRevB.62.3554}
{Phys. Rev. B {\bf 62}, 3554 (2000)}.














\end{thebibliography}
\end{document}